\documentclass[fleqn,usenatbib]{mnras}

\usepackage{newtxtext,newtxmath}
\usepackage{bm,subcaption}
\usepackage{algorithm} 
\usepackage{algpseudocode}

\usepackage[T1]{fontenc}

\DeclareRobustCommand{\VAN}[3]{#2}
\let\VANthebibliography\thebibliography
\def\thebibliography{\DeclareRobustCommand{\VAN}[3]{##3}\VANthebibliography}

\usepackage{graphicx}	
\usepackage{amsmath}	
\usepackage{comment}
\usepackage[table]{xcolor}
\usepackage{booktabs}
\usepackage{multirow}

\newcommand{\Mpc}{\,\text{Mpc}}
\newcommand{\norm}[1]{\left|\left| #1 \right|\right|}
\newcommand{\grad}{\mathtt{grad}\,}
\newcommand{\Diag}{\mathtt{Diag}}

\renewcommand{\hat}{\widehat}

\usepackage{array}
\newcommand{\PreserveBackslash}[1]{\let\temp=\\#1\let\\=\temp}
\newcolumntype{C}[1]{>{\PreserveBackslash\centering}p{#1}}
\newcolumntype{R}[1]{>{\PreserveBackslash\raggedleft}p{#1}}
\newcolumntype{L}[1]{>{\PreserveBackslash\raggedright}p{#1}}

\allowdisplaybreaks

\title[\texttt{SCONCE}: Spherical and Conic Cosmic Web Finder]{\texttt{SCONCE}: A cosmic web finder for spherical and conic geometries}

\author[Yikun Zhang \textit{et al.}]{
	Yikun Zhang,$^{1}$\thanks{E-mail: yikun@uw.edu}
	Rafael S. de Souza,$^{2}$\thanks{E-mail: drsouza@shao.ac.cn}
	and Yen-Chi Chen$^{1}$\thanks{E-mail: yenchic@uw.edu}
	\\
	$^{1}$Department of Statistics, University of Washington, Seattle, WA 98195, United States\\
	$^{2}$Key Laboratory for Research in Galaxies and Cosmology, Shanghai Astronomical Observatory, Chinese Academy of Sciences, Shanghai 200030, China\\
}

\date{Accepted XXX. Received YYY; in original form ZZZ}

\pubyear{2022}

\begin{document}
	\label{firstpage}
	\pagerange{\pageref{firstpage}--\pageref{lastpage}}
	\maketitle
	
	\begin{abstract}
        The latticework structure known as the cosmic web provides a valuable insight into the assembly history of large-scale structures. Despite the variety of methods to identify the cosmic web structures, they mostly rely on the assumption that galaxies are embedded in a Euclidean geometric space. Here we present a novel cosmic web identifier called \texttt{SCONCE} (\textbf{S}pherical and \textbf{CON}ic \textbf{C}osmic w\textbf{E}b finder) that inherently considers the 2D (RA,DEC) spherical or the 3D (RA,DEC,\emph{z}) conic geometry. The proposed algorithms in \texttt{SCONCE} generalize the well-known subspace constrained mean shift (\texttt{SCMS}) method and primarily address the predominant filament detection problem. They are intrinsic to the spherical/conic geometry and invariant to data rotations.
        We further test the efficacy of our method with an artificial cross-shaped filament example and apply it to the SDSS galaxy catalogue, revealing that the 2D spherical version of our algorithms is robust even in regions of high declination. Finally, using N-body simulations from Illustris, we show that the 3D conic version of our algorithms is more robust in detecting filaments than the standard \texttt{SCMS} method under the redshift distortions caused by the peculiar velocities of halos. 
        Our cosmic web finder is packaged in \texttt{python} as \texttt{SCONCE-SCMS} and has been made publicly available.
	\end{abstract}
	
	
	
	
	\section{Introduction}
	The cosmic web, a latticework structure of magnificent proportions, represents one of the most prominent physical patterns in the Universe~\citep{Bond1996}. Its network nature is a byproduct of the hierarchical growth of large-scale structures of the Universe, which have been corroborated by observations of large-scale surveys \citep{LSST2019,DR17-2021} and cosmological N-body simulations \citep{Millennium2005, Illustris-I2014}. The structure is composed of a network of galaxy groups and clusters, filaments, sheets, and the large regions of near-emptiness known as cosmic voids \citep{Zel1982}. 
	In particular, the filamentary part of the cosmic web, as traced by galaxies, has been observed by several galaxy surveys, including the Sloan Digital Sky Survey \citep[SDSS;][]{York2000}, the two degree Field Galaxy Redshift Survey \citep[2dFGRS;][]{Colless2003}, the Cosmic Evolution Survey \citep[COSMOS;][]{Scoville_2007}, the VIMOS VLT deep survey \citep{Le2005}, the 6dF Galaxy Survey \citep[6dFGS; ][]{Jones2009}, the Galaxy and Mass Assembly \citep[GAMA; ][]{Driver2011},  VIPERS \citep{Guzzo2014} and SAMI \citep{Bryant2015}.

	As the baryonic matter distribution correlates -- to a certain extent -- with the distribution of dark matter, galaxy properties are expected to correlate to their environment and, as such, to the local properties of the cosmic web. Filaments are known to be correlated with the stellar masses, intrinsic alignments, and luminosity of nearby galaxies \citep{zhang2013alignments,clampitt2016detection,poudel2017effect,chen2017detectSDSSIII,chen2019detect-align}. For instance, the filament between clusters Abell 0399 and 0401 has been shown to host quiescent galaxies and hot gas and to emit in radio~\citep{bonjean2018gas,govoni2019radio}. \citet{Darvish2014} suggested that mild galaxy-galaxy interactions play a role in increasing the fraction of H$_\alpha$ star-forming galaxies in filaments. Using spectroscopic data from the main galaxy sample of the SDSS data release 7 \citep{Abazajian2009}, \citet{Kraljic2020} found that galaxy proprieties such as stellar mass, star formation activity, and morphology display some degree of dependence on the connectivity of the cosmic web.

	Detection of filaments in a large volume has recently been carried out by~\citet{malavasi2020}, using galaxy samples from SDSS \citep{York2000}, while~\citet{galarraga2020} applies a similar approach to hydrodynamical simulations to derive the expected statistical properties of filaments. There are several reasons why we are mainly interested in identifying one-dimensional cosmic filaments. Methodologically, detecting the filaments is relatively more robust than identifying cosmic sheets, whose tenuous structures may have low signal-to-noise ratios. Furthermore, filaments connect those zero-dimensional galaxy clusters and circumvent the empty cosmic voids. Any precise filament detection method can infer the locations of cosmic voids by outlining their boundaries, which provides rich resources to infer the cosmology \citep{PreciseCosm2021}. Diving deeper into this vein, filaments also provide information about the nature of dark matter halos \citep{spin2007,zhang2009spin}.

	Despite the undoubtedly role of cosmic filaments in galaxy formation, identifying them on either observed or simulated galaxy distributions is challenging due to their intricate geometrical patterns \citep[see, e.g.][for a detailed review]{cautun2014evolution,tracing2018}. Some filament finders in the literature span from investigating the Hessian matrix of the matter/galaxy density function: Skeleton \citep{novikov2006skeleton}, \texttt{Multiscale Morphology Filter} \citep{aragon2007multiscale},  \texttt{NEXUS} \citep{cautun2013nexus}, \texttt{COWS} \citep{COWS2022} to 
	directly segmenting the density function via some topological principles: \texttt{SpineWeb} \citep{aragon2010spine},  \texttt{DisPerSE} \citep{sousbie2011I,sousbie2011II}.

	In this work, we present a novel cosmic web finder that is adaptive to the spherical and light-cone geometries by generalizing the classical subspace constrained mean shift (\texttt{SCMS}) algorithm \citep{ozertem2011locally} and its prototypical mean shift algorithm \citep{MS2002}. \texttt{SCMS} has been used in astronomy to identify cosmic filaments as density ridges  \citep{Chen2015methods,chen2015investigating, Chen2016Catalog, Hendel2019, Moews2021,novel_cat2021}. Our first generalization extends upon these past implementations and is more accurate in recovering cosmic web structures on the celestial sphere indicated by the (RA,DEC) coordinate system, especially in regions of higher declination. Our second generalization further extends our new \texttt{SCMS} method to a 3D light cone specified by the (RA,DEC,$z$) coordinate system. In essence, the \textbf{S}pherical and \textbf{CON}ic \textbf{C}osmic w\textbf{E}b finder (\texttt{SCONCE}) estimates the matter/galaxy density function on the celestial sphere or the 3D light cone based on the Cartesian representation of (RA,DEC). Such a density function on the sphere (or 3D light cone) is known as the directional (or directional-linear) density in Statistics \citep{mardia2009directional,ley2017modern,pewsey2021recent}. For a given estimated matter density function on the sphere or spherical cone, \texttt{SCONCE} recovers the filaments as the density ridges of the estimated density via adaptive gradient ascent iterations \citep{DirMS2020, DirSCMS2021,DirLinProd2021}. 
	
	This paper is organized as follows. \autoref{Sec:model} reviews the standard \texttt{SCMS} algorithm and describes our extensions on the spherical and conic geometries. Applications to mock and real-world astronomical data are presented at \autoref{Sec:Application}, including comparisons with the other \texttt{SCMS} typed and \texttt{DisPerSE} algorithms. Finally, \autoref{sec:conclusion} provides our summary and discussion. Throughout this work, we assume a Wilkinson Microwave Anisotropy Probe (WMAP)-9 cosmology ($\Omega_m = 0.2726$, $\Omega_b = 0.0456$, $\Omega_{\Lambda} = 0.7274$, and $h = 0.704$; \citealt{WMAP9_2013}).

	\section{Methodology}
	\label{Sec:model}
	
   This section describes the general aspects of the standard \texttt{SCMS} algorithm in the Euclidean space and our extended \texttt{SCMS} algorithms in \texttt{SCONCE}. The extensions enable better characterizations of the density ridges (\emph{i.e.}, our filament model) and their associated nodes according to a collection of discrete observations on the celestial sphere 
	\begin{align*}
	\mathbb{S}^2 &=\left\{\bm{x}\in \mathbb{R}^3:\norm{\bm{x}}_2 =1 \right\} \simeq \\
	&\left\{(\alpha, \delta) \in [0^{\circ}, 360)\times [-90^{\circ}, 90^{\circ}]: \alpha \text{ is RA and } \delta \text{ is DEC}\right\}, 
	\end{align*} 
	or the 3D light cone $\mathbb{S}^2\times \mathbb{R}$. Here, $\simeq$ stands for the equivalent representation, $\mathbb{R}$ is the real line, and $\norm{\cdot}_2$ is the Euclidean norm. 
	\texttt{SCONCE} has been made publicly available\footnote{The code is available at \url{https://pypi.org/project/sconce-scms/}.} and can be installed via pip command.

	\subsection{{\tt SCMS} Algorithm on the Euclidean Space $\mathbb{R}^d$}
	\label{Sec:SCMS_std}
	
	Introduced by~\citet{ozertem2011locally}, the standard \texttt{SCMS} algorithm is part of the taxonomy of statistical methods dealing with the estimation of local principal curves, more widely known as density ridges \citep{eberly1996ridges,NonparRidges2014,chen2015asymptotic} in the flat Euclidean space $\mathbb{R}^d$. Suppose we start with a mesh of points placed in equidistant steps across the data space. In that case, the algorithm seeks to establish density ridges in an iterative process resembling the gradient ascent method. The procedure can be visualized as a cloud of points shifting closer to the nearest underlying structure at each iteration. 
	
	\subsubsection{Filament Model in $\mathbb{R}^d$}
	Formally speaking, a density ridge is the maximization of the local density in the normal direction defined by the Hessian matrix. Let $\nabla p(\bm{x})$ be the gradient of a probability density function $p$ on a $d$-dimensional Euclidean space $\mathbb{R}^d$, and $H(\bm{x}) \equiv \nabla\nabla p(\bm{x})$ be its Hessian matrix of second derivatives, whose orthonormal eigenvectors  $\bm{v}_1(\bm{x}),...,\bm{v}_d(\bm{x})$ are associated with a descending order of eigenvalues $\lambda_1(\bm{x}) \geq \cdots \geq \lambda_d(\bm{x})$. The density ridge of $p$ is defined as the collection of points satisfying
	\begin{eqnarray}
	\label{ridge}
	R(p) = \left\{\bm{x}\in \mathbb{R}^d: V_E(\bm{x})^T \nabla p(\bm{x})=\bm{0}, \lambda_2(\bm{x}) < 0 \right\},
	\end{eqnarray}
	where $V_E(\bm{x}) = \left[\bm{v}_2(\bm{x}),...,\bm{v}_d(\bm{x}) \right] \in \mathbb{R}^{d\times (d-1)}$ has its columns composed by all but the first eigenvector of the Hessian $H(\bm{x})$. In principle, the density function $p$ is distinctively curved along the space defined by the columns of $V_E(\bm{x})$, within which $R(p)$ consists of its local maxima. The density ridge $R(p)$ serves as a theoretical cosmic filament model in $\mathbb{R}^d$ with an arbitrary dimension $d$.
	
	\subsubsection{Filament Estimation in $\mathbb{R}^d$: The Standard {\tt SCMS} Algorithm}
	
	\begin{figure*}
		\captionsetup[subfigure]{justification=centering}
		\centering
		\begin{subfigure}[t]{.32\textwidth}
			\centering
			\includegraphics[width=\linewidth]{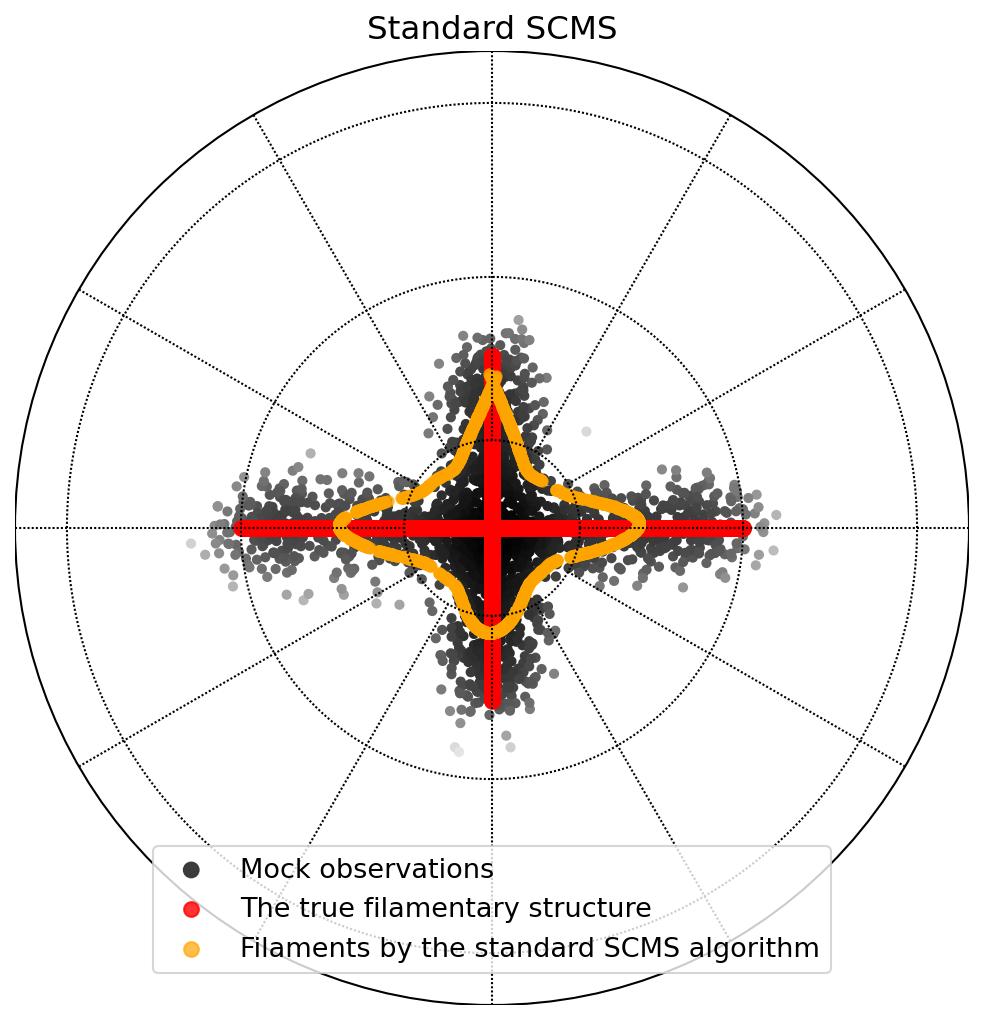}
		\end{subfigure}
		\begin{subfigure}[t]{.32\textwidth}
			\centering
			\includegraphics[width=\linewidth]{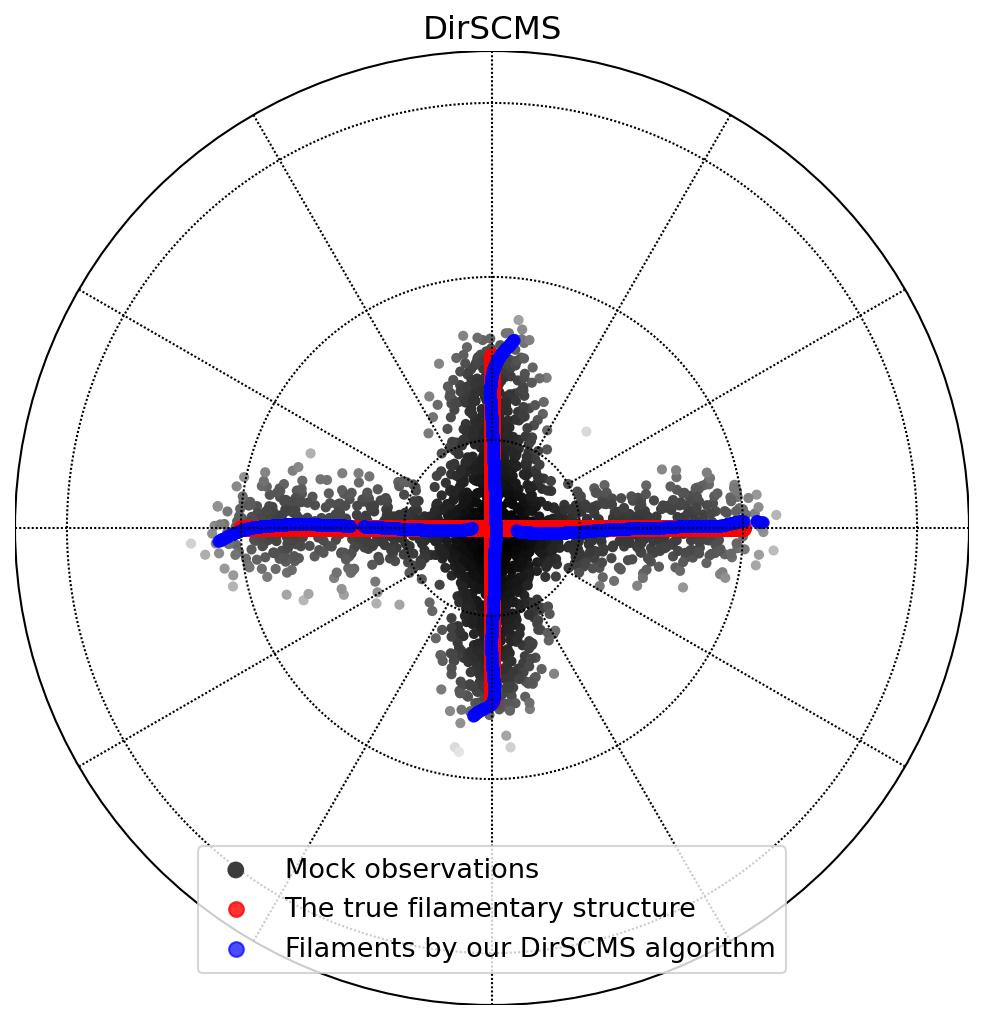}
		\end{subfigure}
		\hfil
		\begin{subfigure}[t]{.32\textwidth}
			\centering
			\includegraphics[width=\linewidth]{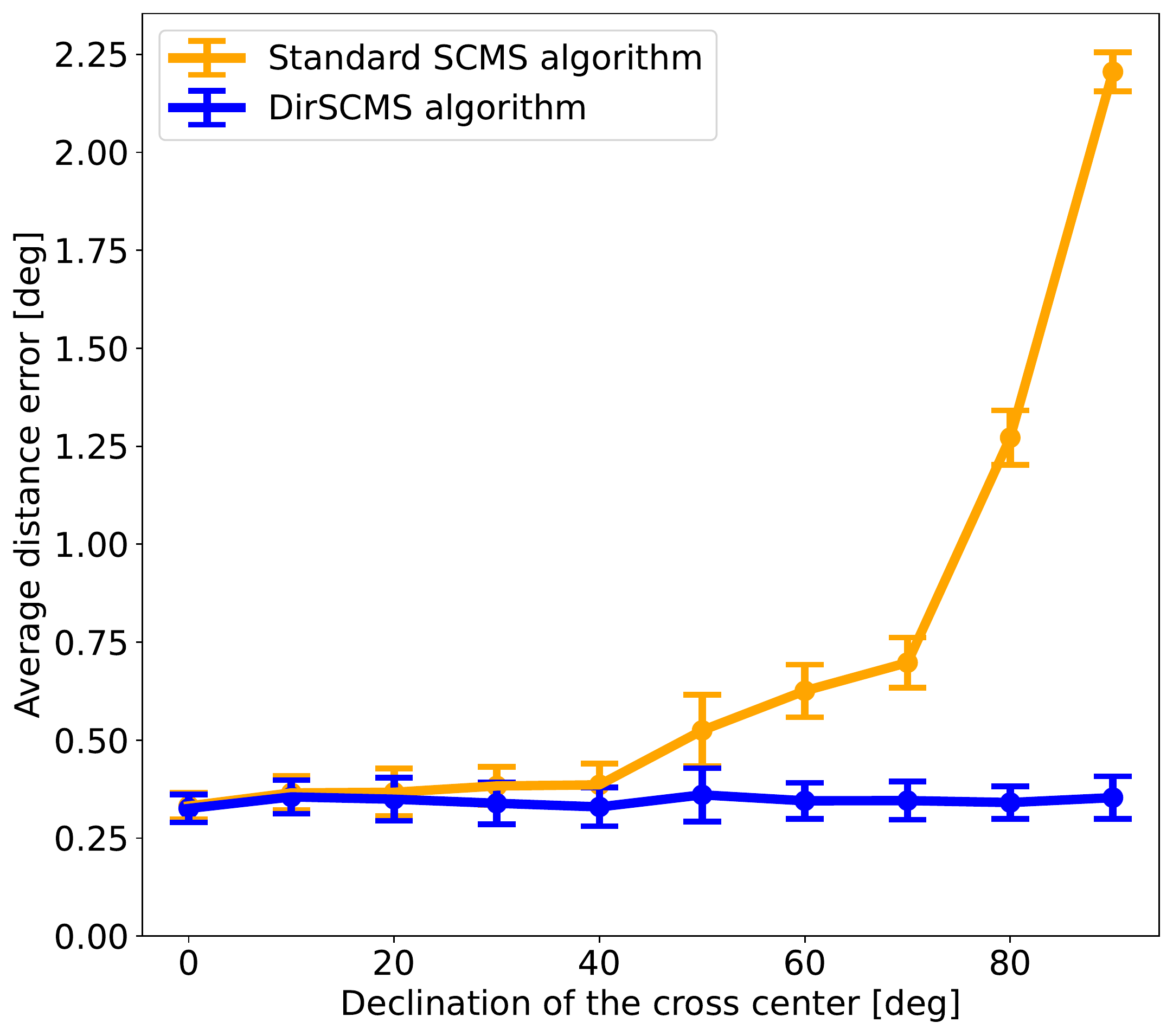}
		\end{subfigure}
		\caption{Comparisons between the standard \texttt{SCMS} and our extended \texttt{DirSCMS} algorithms applied to the mock dataset on $\mathbb{S}^2$. The \emph{left} and \emph{middle} panels depict the mock observations, true cross-shaped filament, and the estimated filaments yielded by the standard \texttt{SCMS} and our \texttt{DirSCMS} algorithms, respectively. The \emph{right} panel presents the average distance errors from the estimated filaments yielded by the standard \texttt{SCMS} and our \texttt{DirSCMS} algorithms to the true filaments as we varies the declination values on which the true filament centers.}
		\label{fig:ridges_Eu_Dir}
	\end{figure*}
	
	To approximate the theoretical density ridge \eqref{ridge}, a common strategy is to use the density ridge $\hat{R}\equiv R(\hat{p})$ defined by the kernel density estimator (KDE) $\hat{p}$. 
	Given some observations $\bm{X}_1,...,\bm{X}_n \subset \mathbb{R}^d$ from $p$, the KDE $\hat{p}$ estimates the density function as:
	\begin{eqnarray}
	\label{KDE}
	\hat{p}(\bm{x}) = \frac{1}{nb^d} \sum_{i=1}^n K\left(\norm{\frac{\bm{x} - \bm{X}_i}{b}}_2^2 \right),
	\end{eqnarray}
	where $K(\cdot)$ is a kernel function\footnote{For instance, the Gaussian kernel $K(r) = \frac{1}{(2\pi)^{d/2}} \exp\left(-\frac{r}{2}\right)$.} and $b$ is the smoothing bandwidth parameter \citep[see, e.g.,][for more details about the choices of $K(\cdot)$ and $b$]{Scott2015,YC2017KDE}. Commonly, a larger value of $b$ leads to a smoother $\hat{p}$ and a less intricate density ridge $R(\hat{p})$. The standard \texttt{SCMS} algorithm iterates an initial mesh of points along the direction of the (estimated) projected gradient $\hat{V}_E(\bm{x}) \hat{V}_E(\bm{x})^T \nabla \hat{p}(\bm{x})$ defined by the KDE $\hat{p}$ until convergence (see, e.g., Algorithm 1 in \cite{DirSCMS2021} for the full formulation). Unlike the ordinary gradient ascent method, which requires some delicate choices of the step size parameter, the \texttt{SCMS} and its prototypical mean shift algorithms embrace an adaptive step size with respect to the data and bandwidth parameter $b$. As long as $b$ is reasonably small, the mean shift and \texttt{SCMS} algorithms are guaranteed to converge \citep{MS1995,MS2007pf,MSconsist2016,SCMSconv2013,DirSCMS2021}. The set of converged points under the \texttt{SCMS} algorithm becomes a discrete sample from the density ridge $R(\hat{p})$.

    Despite its broad applicability in detecting filamentary structures \citep{chen2015investigating,Fernandez2020,Velcohere2020}, the standard \texttt{SCMS} algorithm has shortcomings when the curvature where the data live is non-negligible, which may produce highly biased estimates of the cosmic web. To showcase this effect, we create a simple experiment. Starting from a cross-shaped filamentary structure with its center at the North pole of the sphere $\mathbb{S}^2$, we generate a mock dataset consisting of 2,000 mock observations from the true filament with independent Gaussian noises of zero mean and 0.05 standard deviation (see \autoref{fig:ridges_Eu_Dir}). 
    We then apply both the standard \texttt{SCMS} and our extended \texttt{DirSCMS} method (\autoref{Sec:DirSCMS}) in order to recover the original cross-shaped structure. A visual inspection of \autoref{fig:ridges_Eu_Dir} reveals that the standard \texttt{SCMS} algorithm fails to approximate the true filamentary structure near the North pole (\emph{i.e.}, the center of the cross). On the contrary, our \texttt{DirSCMS} method manages to recover the entire cross-shaped filament.\footnote{We selected the bandwidth parameters for the standard \texttt{SCMS} and our \texttt{DirSCMS} algorithms via Equations~\eqref{bw_Eu} and \eqref{bw_Dir} with $A_0=1.25$ and $B_0=1$. The results in \autoref{fig:ridges_Eu_Dir} are similar under other choices of the parameters.}
    To further quantify how the location of the true filament would influence each algorithm's estimate, we generate the same experiment several times but with different declination values for the center of the true filament, spanning from $0^{\circ}$ to $90^{\circ}$. For each declination, we compute the average distance errors from the estimated filaments detected by the standard \texttt{SCMS} and our \texttt{DirSCMS} algorithms to the true cross-shaped filament on $\mathbb{S}^2$, respectively. Results are shown on the right panel of \autoref{fig:ridges_Eu_Dir}. The error bars indicate the standard deviations of the average distance errors, which are obtained by repeating the data sampling and filament detection procedure 20 times. This toy experiment suggests that our \texttt{DirSCMS} algorithm outperforms the standard \texttt{SCMS} algorithm when the filamentary structures are located at the regions with declination higher than $50^{\circ}$. More importantly, the estimation bias of our \texttt{DirSCMS} algorithm is invariant under translations on $\mathbb{S}^2$.

	\subsection{{\tt SCMS} Algorithm on the Celestial Sphere $\mathbb{S}^2$}
	\label{Sec:SCMS_sph}
	
	To address the estimation bias of the standard \texttt{SCMS} algorithm in the regions of high declination on the (unit) celestial sphere $\mathbb{S}^2$, we introduce our first \texttt{SCMS} extension to tackle the spherical geometry.
	We consider $n$ observations with their angular coordinates as $(\alpha_i,\delta_i) \in [0,360^{\circ})\times [-90^{\circ}, 90^{\circ}]$ for $i=1,...,n$, where $\alpha_i$ is the right ascension (RA) of the $i$-th object and $\delta_i$ is its declination (DEC). 
	
	Note that both \cite{Moews2021} and \cite{novel_cat2021} have considered estimating the matter/galaxy density function (as well as its gradient and Hessian fields) under the geodesic distance on the (RA,DEC) angular coordinate system of $\mathbb{S}^2$, by leveraging either the haversine formula \citep{inman1849navigation} or the Hierarchical Equal Area isoLatitude Pixelisation (HEALPix; \citealt{healpix2005}) scheme. While their approaches reduce the bias in density estimation on the sphere, the subsequent \texttt{SCMS} iterations fail to converge to the (true) filamentary structures at the regions of high declination. Moreover, given the anisotropy of the (RA,DEC) angular coordinate system between regions of low and high declination, the \texttt{SCMS} algorithm based on this coordinate system is not completely invariant to the data rotation on $\mathbb{S}^2$. 
	
	\begin{figure*}
		\captionsetup[subfigure]{justification=centering}
		\centering
		\begin{subfigure}[t]{.24\textwidth}
			\centering
			\includegraphics[width=\linewidth]{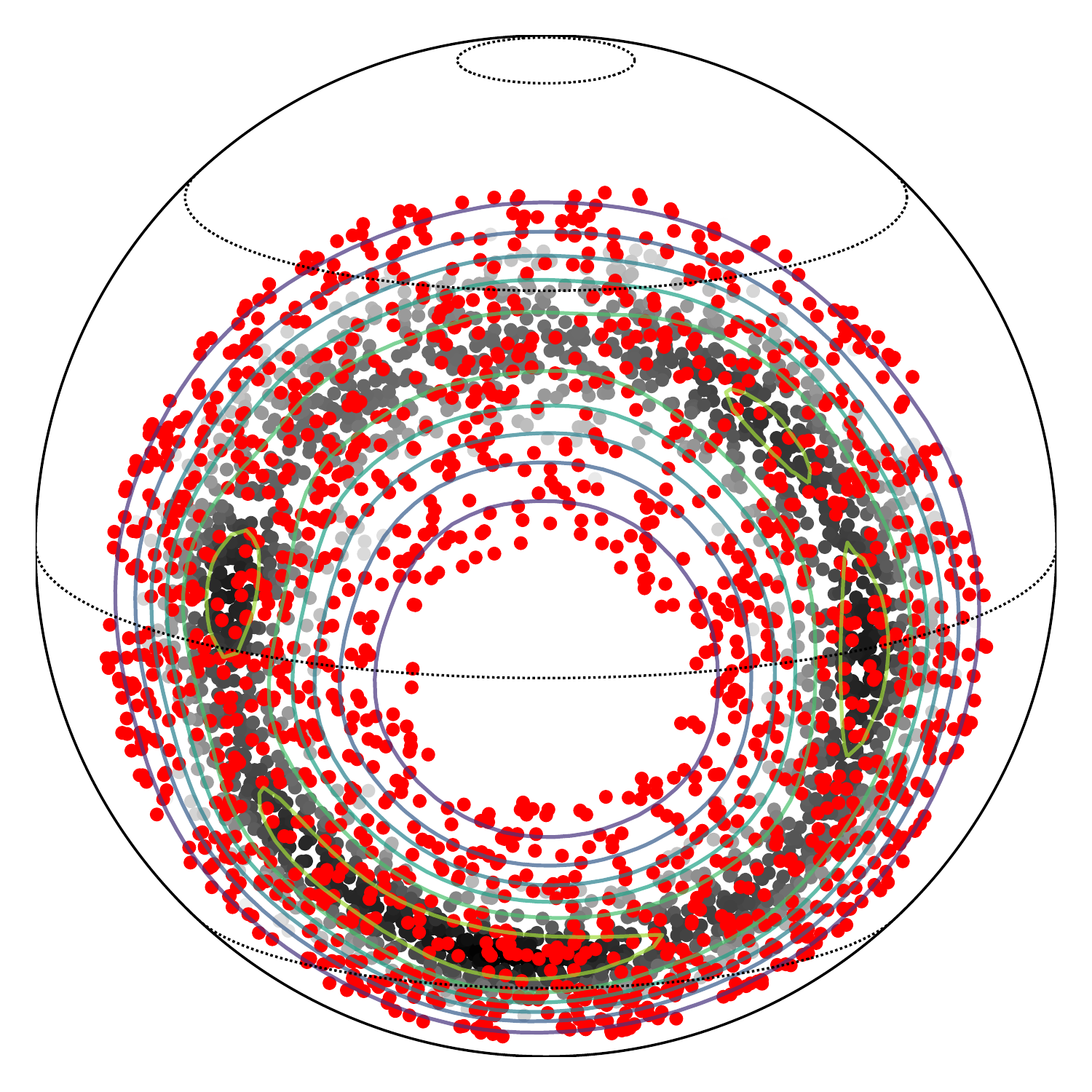}
		\end{subfigure}
		\hfil
		\begin{subfigure}[t]{.24\textwidth}
			\centering
			\includegraphics[width=\linewidth]{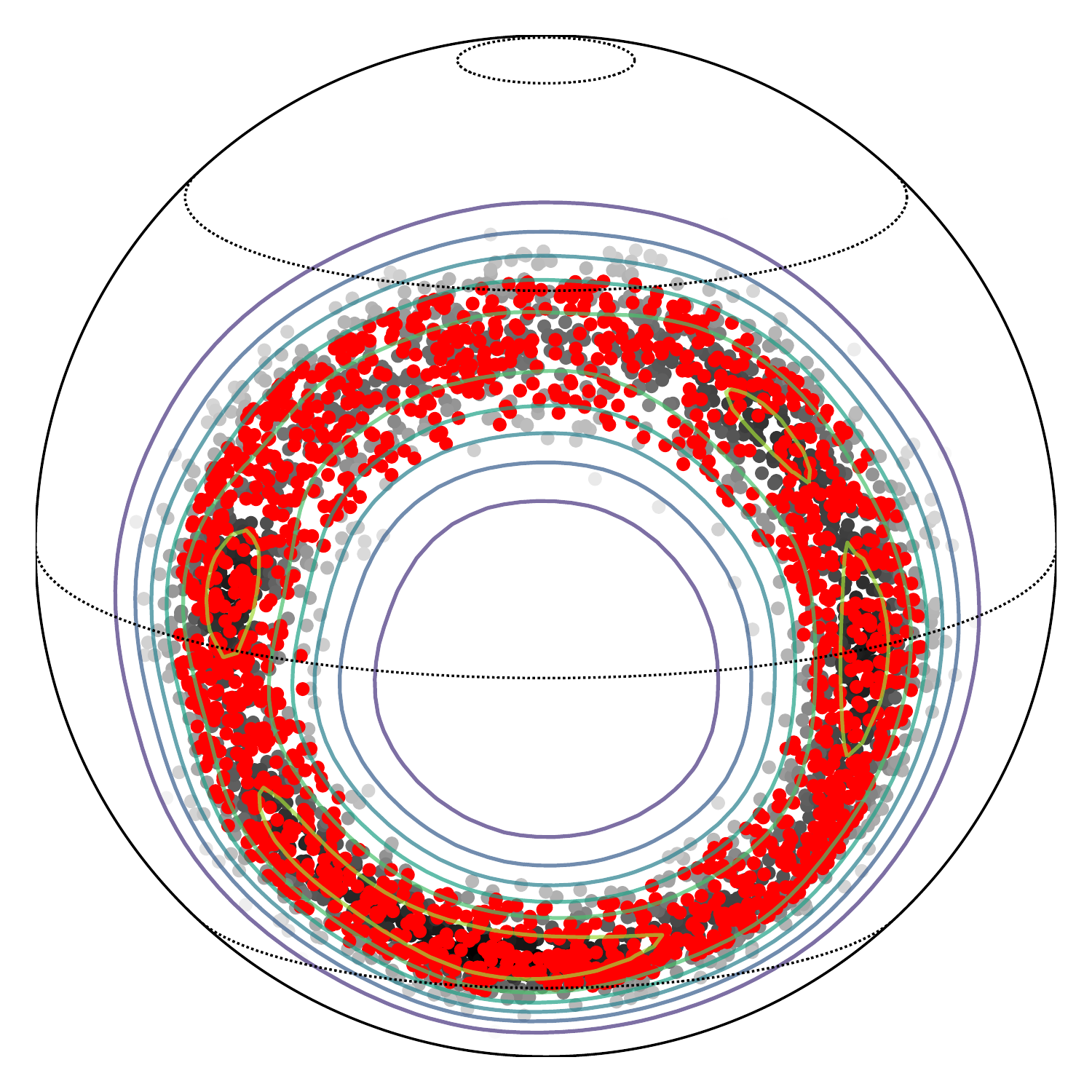}
		\end{subfigure}
		\hfil
		\begin{subfigure}[t]{.24\textwidth}
			\centering
			\includegraphics[width=\linewidth]{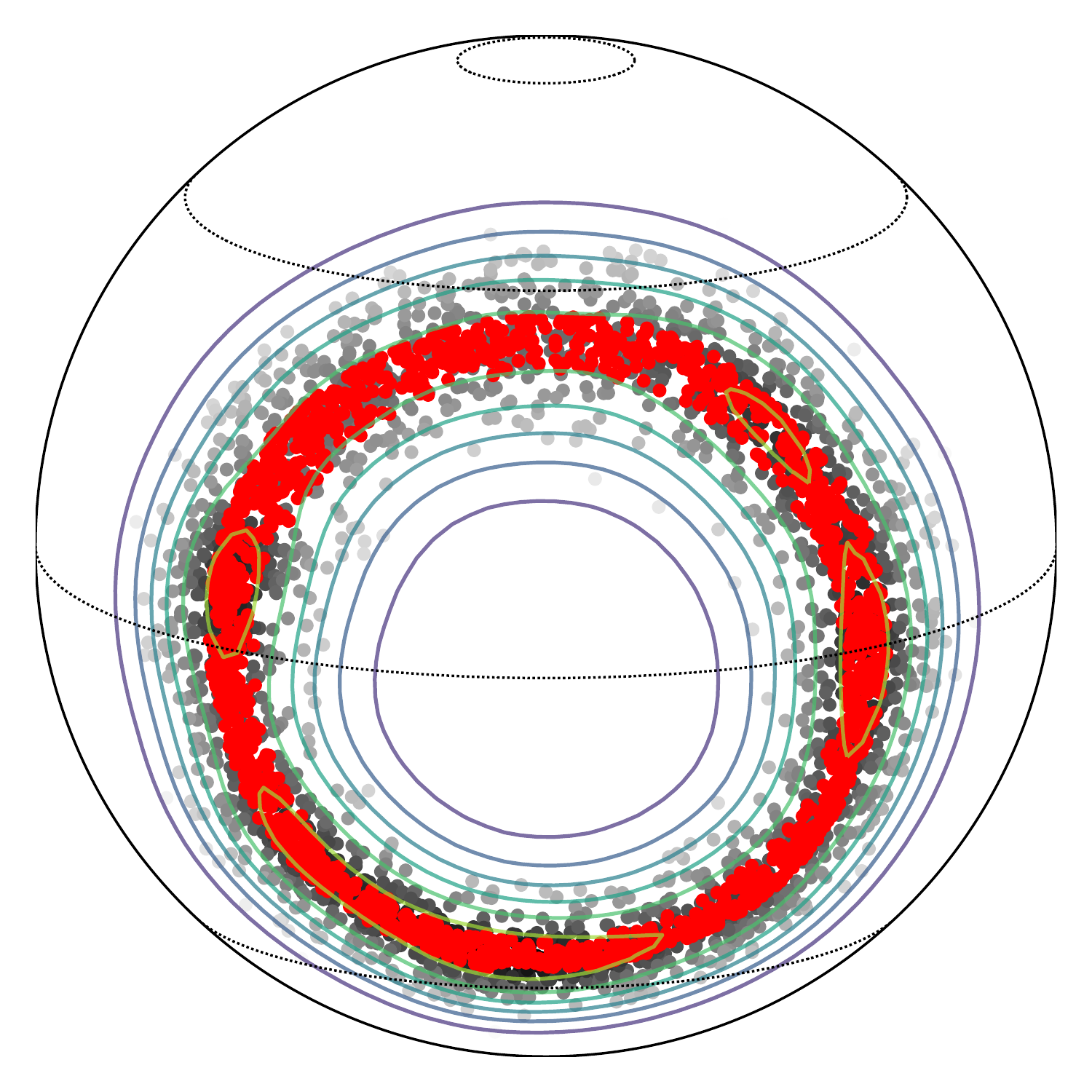}
		\end{subfigure}
		\hfil
		\begin{subfigure}[t]{.24\textwidth}
			\centering
			\includegraphics[width=\linewidth]{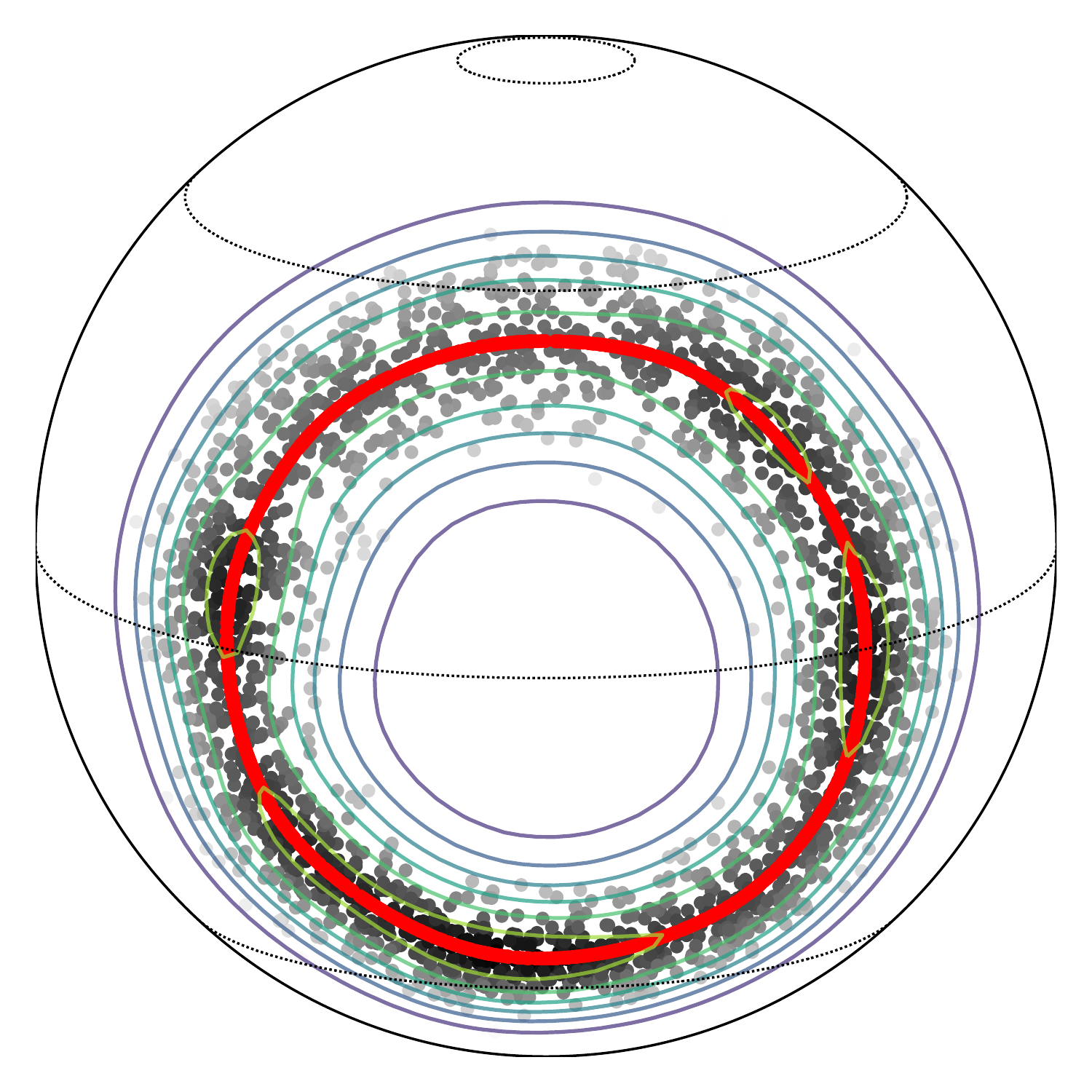}
		\end{subfigure}
		\caption{Step-by-step illustrations of our \texttt{DirSCMS} algorithm on $\mathbb{S}^2$. We sample some uniform mesh points (red dots) covering the input observations (gray dots) and apply our \texttt{DirSCMS} algorithm to the mesh points. The panels from left to right present the \texttt{DirSCMS} iteration at step 0, 1, 2, and 8. The contour lines in each panel indicate the directional KDE \eqref{DirKDE} applied to the input observations under their Cartesian coordinates $\bm{X}_1,...,\bm{X}_n \subset \mathbb{S}^2$.}
		\label{fig:DirSCMS}
	\end{figure*}
	
	\subsubsection{Filament Model on $\mathbb{S}^2$}
	
	Different from how \cite{Moews2021} and \cite{novel_cat2021} handle the spherical geometry, we convert the (RA,DEC) coordinates of the $n$ observations to their Cartersian representations $\bm{X}_1,...,\bm{X}_n$ on the (unit) sphere $\mathbb{S}^2$ as:
	\begin{eqnarray}
	\label{change_of_coordinate}
	\bm{X}_i=\left(\cos\delta_i \cos \alpha_i, \cos\delta_i \sin \alpha_i, \sin\delta_i \right) \quad \text{ for } i=1,...,n.
	\end{eqnarray}
	We assume that $\bm{X}_1,...,\bm{X}_n$ are random observations sampled from a (directional) density function $f$ on $\mathbb{S}^2$. The filaments are modeled as the one-dimensional (directional) density ridge of $f$ on $\mathbb{S}^2$. Mathematically, a (directional) density ridge $R(f)$ is defined through the Riemannian gradient $\grad f(\bm{x})$ and Riemannian Hessian matrix $\mathcal{H} f(\bm{x})$ on $\mathbb{S}^2$; see Appendix~\ref{App:Riem_terms} for their detailed expressions. Let $\bm{v}_1(\bm{x}),\bm{v}_2(\bm{x})$ be the orthonormal eigenvectors of $\mathcal{H} f(\bm{x})$ that lie within the tangent space $T_{\bm{x}}$ at $\bm{x}\in \mathbb{S}^2$ and $\lambda_1(\bm{x})\geq \lambda_2(\bm{x})$ be their associated eigenvalues.\footnote{Notice that $\bm{v}_1(\bm{x}),\bm{v}_2(\bm{x})$ span the two-dimensional tangent space $T_{\bm{x}}$, and the Riemannian Hessian matrix $\mathcal{H} f(\bm{x})$, by definition, has another unit eigenvector $\bm{x}$ that is orthogonal to $T_{\bm{x}}$ associated with eigenvalue 0.} Then, the (directional) density ridge $R(f)$ on $\mathbb{S}^2$ is the set of points satisfying
	\begin{eqnarray}
	\label{DirRidges}
	R(f) = \left\{\bm{x}\in \mathbb{S}^2: \bm{v}_2(\bm{x})^T \grad f(\bm{x}) = 0, \lambda_2(\bm{x}) < 0 \right\}.
	\end{eqnarray}
	
	In contrast to the classical definition of density ridges in flat Euclidean spaces \citep{eberly1996ridges,NonparRidges2014,chen2015asymptotic}, our (directional) density ridge $R(f)$ is defined through the (intrinsic/Riemannian) gradient and Hessian of $f$ within the tangent spaces of $\mathbb{S}^2$. As a result, this filament model on $\mathbb{S}^2$ is, by nature, adaptive to the spherical geometry and invariant to any rotation (\emph{i.e.}, isotropic) in the (RA,DEC) space.

	\subsubsection{Filament Estimation on $\mathbb{S}^2$: Directional {\tt SCMS} Algorithm}
	\label{Sec:DirSCMS}
	
	To identify the density ridge \eqref{DirRidges} from $n$ observations on $\mathbb{S}^2$ with Cartesian coordinates $\left\{\bm{X}_1,...,\bm{X}_n \right\} \subset \mathbb{S}^2$, we first estimate the (directional) density function $f$ through the directional KDE \citep{hall1987kernel,Bai1988,zhao2001central,garcia2013exact} as:
	\begin{eqnarray}
	\label{DirKDE}
	\hat{f}(\bm{x}) = \frac{C_L(b)}{n} \sum_{i=1}^n L\left(\frac{1-\bm{x}^T\bm{X}_i}{b^2} \right),
	\end{eqnarray}
	where $L(\cdot)$ is a directional kernel function (\emph{e.g.}, the von Mises kernel $L(r)=e^{-r}$), $b$ is the smoothing bandwidth parameter, and $C_L(b)$ is a normalizing constant ensuring that the integration of $\hat{f}$ on $\mathbb{S}^2$ is one. 
    The choice of smoothing bandwidth parameter $b$ is a critical factor in determining the performance of the estimator \eqref{DirKDE} and its density ridge filament model. Several statistical papers have been addressing the bandwidth selection problem \citep{hall1987kernel,Autobw2008,Oliveira2012,garcia2013exact,NonpDirHDR2020}, so we omit a detailed discussion here.  
	
	Given the directional KDE \eqref{DirKDE}, our practical filament model on $\mathbb{S}^2$ is given by a plug-in estimator of the theoretical one $R(f)$ as:
	\begin{eqnarray}
	\label{DirRidges_est}
	R(\hat{f}) = \left\{\bm{x}\in \mathbb{S}^2: \hat{\bm{v}}_2(\bm{x})^T \grad \hat{f}(\bm{x}) = 0, \hat{\lambda}_2(\bm{x}) < 0 \right\},
	\end{eqnarray}
	where $\hat{\bm{v}}_2(\bm{x})$ is the unit eigenvector of the estimated Riemannian matrix $\mathcal{H} \hat{f}(\bm{x})$ within the tangent space $T_{\bm{x}}$ associated with the smallest eigenvalue $\hat{\lambda}_2(\bm{x})$; see Appendix~\ref{App:DirSCMS_detail} for the expression for $\mathcal{H} \hat{f}(\bm{x})$. While the standard \texttt{SCMS} algorithm is incapable of identifying the (directional) density function and its ridge on $\mathbb{S}^2$ (recall Figure~\ref{fig:ridges_Eu_Dir}), we propose our directional SCMS (\texttt{DirSCMS}) algorithm with an iterative formula as:
	\begin{equation}
	\label{DirSCMS_iter}
	\tilde{\bm{x}}^{(t+1)} \gets \bm{x}^{(t)} - \hat{\bm{v}}_2(\bm{x}^{(t)}) \hat{\bm{v}}_2(\bm{x}^{(t)})^T  \left[\frac{\sum_{i=1}^n \bm{X}_i L'\left(\frac{1-\bm{X}_i^T \bm{x}^{(t)}}{b^2} \right)}{\norm{\sum_{i=1}^n \bm{X}_i L'\left(\frac{1-\bm{X}_i^T \bm{x}^{(t)}}{b^2} \right)}}_2 \right], 
	\end{equation}
	and $\bm{x}^{(t+1)} \gets \frac{\tilde{\bm{x}}^{(t+1)}}{\norm{\tilde{\bm{x}}^{(t+1)}}_2}$ for $t=0,1,...$, recalling that $\left\{\bm{X}_1,...,\bm{X}_n\right\}$ consists of Cartesian coordinates of the observations on $\mathbb{S}^2$. At a high level, our \texttt{DirSCMS} algorithm updates a given point $\bm{x}^{(t)}$ on $\mathbb{S}^2$ along the direction of the (estimated) projected Riemannian gradient $\hat{\bm{v}}_2(\bm{x}^{(t)}) \hat{\bm{v}}_2(\bm{x}^{(t)})^T \grad \hat{f}(\bm{x}^{(t)})$. It leads to an iterative sequence $\big\{\bm{x}^{(t)}\big\}_{t=0}^{\infty}\subset \mathbb{S}^2$ that converges to the estimated density ridge (or our practical filament model) $R(\hat{f})$; see \autoref{fig:DirSCMS} for graphical illustrations. More results about the convergence of our \texttt{DirSCMS} algorithm can be found in Section 4.2 in \cite{DirSCMS2021}. In Appendix~\ref{Sec:GenDirEst}, we describe a more general formulation of the \texttt{DirSCMS} algorithm on $\mathbb{S}^2$ that takes into account the stellar properties of the observational data in its iteration. Additionally, we discuss how to quantify the uncertainty levels of filamentary points on the estimated density ridge $R(\hat{f})$ with our \texttt{DirSCMS} algorithm and bootstrap techniques in Appendix~\ref{Sec:UncertMeasure}. Given that different bootstrap schemes give rise to linearly correlated uncertainty measures for the estimated filaments (see Appendix~\ref{App:boot_comp}), we will implement the nonparametric bootstrap with the number of bootstrapping times as $B=100$ for measuring the uncertainty of any estimated filament in this paper.

	\subsection{{\tt SCMS} Algorithm on the Light Cone $\mathbb{S}^2 \times \mathbb{R}$}
	\label{Sec:SCMS_cone}
	
	In the previous subsection, we have offered our first extended \texttt{SCMS} algorithm to detect cosmic filaments under the 2D spherical geometry. Although this extension is convenient for the local analysis of 2D spherical cosmic web structures within some thin redshift slices, it will inevitably miss some light-of-sight filaments in the large-scale structure of our universe. Therefore, we provide the second extension of the \texttt{SCMS} algorithm to handle the filament finding task in the 3D light cone $\mathbb{S}^2\times \mathbb{R}$.
	
	\subsubsection{Filament Model on $\mathbb{S}^2\times \mathbb{R}$}
	
	Suppose that the data comprise $n$ observations $(\bm{X}_1,Z_1),...,(\bm{X}_n,Z_n)$ sampled from a (directional-linear) density function $f_{dl}(\bm{x},z)$ in the 3D light cone (or equivalently, the 3D redshift space) $\mathbb{S}^2\times \mathbb{R}$, where $\left\{\bm{X}_i\right\}_{i=1}^n$ encodes their Cartesian coordinates on the (unit) celestial sphere $\mathbb{S}^2$ and $\left\{Z_i\right\}_{i=1}^n$ are their redshift values. 
	The traditional avenue to detect cosmic filaments in the 3D redshift space is to convert the redshifts into comoving distances so that each observation has its unique representation in the Euclidean space $\mathbb{R}^3$ \citep{Tempel2014catal}. However, this conversion relies on a specific cosmological model and needs to handle the finger-of-god effects or other redshift distortions in designing the conversion formula. Due to these two factors, the resulting cosmic web structures tend not to be very flexible in downstream analyses.
	
	As a result, we consider estimating the matter/galaxy density function $f_{dl}(\bm{x},z)$ and its filamentary structures directly from the data $\left\{(\bm{X}_i,Z_i) \right\}_{i=1}^n$ on the 3D light cone $\mathbb{S}^2\times \mathbb{R}$. In particular, the density function is estimated by the directional-linear KDE \citep{DirLinear2013,garcia2015central} as:
	\begin{eqnarray}
	\label{DLKDE}
	\hat{f}_{dl}(\bm{x},z) = \frac{C_L(b_1)}{nb_2} \sum_{i=1}^n L\left(\frac{1-\bm{x}^T\bm{X}_i}{b_1^2} \right) K\left(\norm{\frac{z-Z_i}{b_2}}_2^2 \right),
	\end{eqnarray}
	where $L(\cdot)$ and $K(\cdot)$ are the directional and linear kernel functions while $b_1$ and $b_2$ are the smoothing bandwidth parameters for directional and linear data components. 
	In our practical applications, we take $L(r)=e^{-r}$ as the directional (von Mises) kernel and $K(r) = \frac{1}{\sqrt{2\pi}} e^{-\frac{r}{2}}$ as the linear (Gaussian) kernel.
	
	As before, the directional-linear KDE \eqref{DLKDE} provides a plug-in estimator $R(\hat{f}_{dl})$ of the density ridge (or the theoretical filament model) of the underlying density function $f_{dl}$ defined as:
	\begin{align}
	\label{DL_ridge}
	\begin{split}
	R(f_{dl}) = \Big\{(\bm{x},z) \in \mathbb{S}^2\times \mathbb{R}: & V_{dl}(\bm{x},z)^T \grad f_{dl}(\bm{x},z) = \bm{0},\\ 
	&\lambda_{dl,2}(\bm{x},z)<0\Big\},
	\end{split}
	\end{align}
	where $V_{dl}(\bm{x},z) = \left[\bm{v}_{dl,2}(\bm{x},z),\bm{v}_{dl,3}(\bm{x},z) \right] \in \mathbb{R}^{4\times 2}$ has its columns as the othornormal eigenvectors inside the tangent space $T_{(\bm{x},z)}$ of $\mathbb{S}^2\times \mathbb{R}$ associated with the last two eigenvalues $\lambda_{dl,2}(\bm{x},z) \geq \lambda_{dl,3}(\bm{x},z)$ of the Riemannian Hessian $\mathcal{H} f_{dl}(\bm{x},z) \in \mathbb{R}^{4\times 4}$; see Appendix~\ref{App:Riem_terms} for the expressions of $\grad f_{dl}(\bm{x},z)$ and $\mathcal{H} f_{dl}(\bm{x},z)$.
	
	\subsubsection{Filament Estimation on $\mathbb{S}^2\times \mathbb{R}$: Directional-Linear {\tt SCMS} Algorithm}
	
	While the generalization of KDE and density ridges (or our cosmic filament model) from the density function $f$ on the celestial sphere $\mathbb{S}^2$ to $f_{dl}$ on the 3D light cone $\mathbb{S}^2\times \mathbb{R}$ is not very challenging, the extension of the \texttt{SCMS} algorithm to the directional-linear data $\left\{(\bm{X}_i,Z_i)\right\}_{i=1}^n$ requires special attention. A naive generalization from the mean shift algorithm to its \texttt{SCMS} counterpart as how the standard \texttt{SCMS} method does \citep{ozertem2011locally} will lead to an incorrect estimate of $R(\hat{f}_{dl})$; see the related discussion in Section 4 of \cite{DirLinProd2021}. Here, we formulate the correct \texttt{SCMS} iteration formula at point $(\bm{x}^{(t)},z^{(t)}) \in \mathbb{S}^2 \times \mathbb{R}$ as:
	\begin{align}
	\label{DirLinSCMS}
	\begin{split}
	\begin{pmatrix}
	\tilde{\bm{x}}^{(t+1)}\\
	z^{(t+1)}
	\end{pmatrix} &\gets 
  \begin{pmatrix}
	\bm{x}^{(t)}\\
	z^{(t)}
	\end{pmatrix} \\
	&\quad +\eta \cdot \hat{V}_{dl}\left(\bm{x}^{(t)},z^{(t)}\right) \hat{V}_{dl}\left(\bm{x}^{(t)},z^{(t)}\right)^T \bm{H}\cdot  \Xi\left(\bm{x}^{(t)},z^{(t)}\right),
	\end{split}
	\end{align}
	and 
	$\bm{x}^{(t+1)} \gets \frac{\tilde{\bm{x}}^{(t+1)}}{\norm{\tilde{\bm{x}}^{(t+1)}}_2}$ for $t=0,1,...$, where $\hat{V}_{dl}\left(\bm{x},z\right) = \left[\hat{\bm{v}}_{dl,2}(\bm{x},z),\hat{\bm{v}}_{dl,3}(\bm{x},z) \right]$ has its columns as the last two eigenvectors of the estimated Riemannian Hessian $\mathcal{H} \hat{f}_{dl}(\bm{x},z)$ within the tangent space $T_{(\bm{x},z)}$ of $\mathbb{S}^2\times \mathbb{R}$, $\bm{H} = \Diag\left(\frac{1}{b_1^2},...,\frac{1}{b_1^2}, \frac{1}{b_2^2} \right) \in \mathbb{R}^{4\times 4}$ is a diagonal (bandwidth) matrix, and 
		\begin{equation}
	\Xi\left(\bm{x}^{(t)},z^{(t)}\right) = \begin{pmatrix}
	\frac{\sum\limits_{i=1}^n \bm{X}_i\cdot L'\left(\frac{1-\bm{X}_i^T\bm{x}^{(t)}}{b_1^2} \right)  K\left(\norm{\frac{z^{(t)}-Z_i}{b_2}}_2^2 \right) }{\sum\limits_{i=1}^n L'\left(\frac{1-\bm{X}_i^T\bm{x}^{(t)}}{b_1^2} \right) K\left(\norm{\frac{z^{(t)}-Z_i}{b_2}}_2^2 \right)} -\bm{x}^{(t)}\\ \frac{\sum\limits_{i=1}^n Z_i \cdot L\left(\frac{1-\bm{X}_i^T\bm{x}^{(t)}}{b_1^2} \right)   K'\left(\norm{\frac{z^{(t)}-Z_i}{b_2}}_2^2 \right) }{\sum\limits_{i=1}^n L\left(\frac{1-\bm{X}_i^T\bm{x}^{(t)}}{b_1^2} \right)  K'\left(\norm{\frac{z^{(t)}-Z_i}{b_2}}_2^2 \right)} -z^{(t)}
	\end{pmatrix}, 
	\end{equation}
	is the (directional-linear) mean shift vector at the iterative point $\left(\bm{x}^{(t)},z^{(t)}\right) \in \mathbb{S}^2\times \mathbb{R}$ at step $t$; see Appendix~\ref{App:DirLinSCMS_detail} for more details. We name \eqref{DirLinSCMS} as the \texttt{DirLinSCMS} algorithm.
	
	There are two major differences of the \texttt{DirLinSCMS} iteration in the 3D light cone $\mathbb{S}^2\times \mathbb{R}$ to the standard \texttt{SCMS} algorithm in \autoref{Sec:SCMS_std} and our preceding \texttt{DirSCMS} algorithm in \autoref{Sec:SCMS_sph}. First, we scale the mean shift vector $\Xi\left(\bm{x}^{(t)},z^{(t)}\right)$ by a diagonal (bandwidth) matrix $\bm{H}$ in \eqref{DirLinSCMS} to ensure that the \texttt{DirLinSCMS} iteration follows the correct projected gradient direction of $\hat{f}_{dl}(\bm{x}^{(t)},z^{(t)})$ at step $t$. Second, even under the smooth von Mises and Gaussian kernels, the \texttt{DirLinSCMS} iteration \eqref{DirLinSCMS} still requires a step size parameter to control its convergence. As the step size parameter is notoriously difficult to tune in any application of gradient ascent methods, we provide a theoretically motivated and practically effective guideline to choose the step size parameter $\eta$ in \eqref{DirLinSCMS} as:
	\begin{eqnarray}
	\label{stepsize}
	\eta = \min\left\{b_1b_2, 1 \right\},
	\end{eqnarray}
	where the upper bound $1$ is set to prevent $\eta$ from being too large when the smoothing bandwidth parameters are chosen to be large. In \autoref{Sec:illustris}, we apply our \texttt{DirLinSCMS} algorithm to the friends-of-friends (FoF) halos of the Illustris simulation and study the stability of detected filamentary structures under the effects of peculiar velocities. Analogous to our \texttt{DirSCMS} algorithm on $\mathbb{S}^2$, one can also resort to the bootstrap technique (Appendix~\ref{Sec:UncertMeasure}) to quantify the uncertainties of the estimated density ridge $R(\hat{f}_{dl})$ on $\mathbb{S}^2\times \mathbb{R}$ detected by our \texttt{DirLinSCMS} algorithm.

	\subsection{Cosmic Nodes on the Filaments}
	\label{Sec:cosmic_nodes}
	
    We have discussed so far our methodology to detect cosmic filaments from discrete observations on the celestial sphere $\mathbb{S}^2$ and the 3D light cone $\mathbb{S}^2\times \mathbb{R}$. These one-dimensional structures occupy roughly half of the mass in our universe \citep{cautun2014evolution,Cui2018}. Further, as cosmic time evolves, galaxies within the filaments and in the field gradually fall into galaxy clusters \citep{Navarro1995,Millennium2005,Kuchner2022}, creating the most compact structures that are gravitationally bound in the cosmic web. Here, we describe two potentially overlapping types of cosmic nodes on our density ridge model that are strong candidates for galaxy clusters in an astrophysical sense \citep{Malavasi2022}.
	
	$\bullet$ {\bf Local modes}. One type of regions where galaxy clusters are commonly found are in the local modes (called local peaks or maxima) of the matter/galaxy density function, which signal the locations where galaxies/particles are highly concentrated. 
	They are subsumed in our cosmic filament model based on the density ridge (c.f., Equations~\eqref{ridge}, \eqref{DirRidges}, and \eqref{DL_ridge}), manifesting the property that cosmic filaments bridge the connection between galaxy clusters. We identify the set of local modes on filaments via the mean shift algorithm, which is the prototype of the above \texttt{SCMS} algorithm; see Appendix~\ref{App:mode_seek} for details.
	
	$\bullet$ {\bf Knots}. Besides local modes, another type of indicators on filaments for galaxy clusters are the intersection points of two or more pieces of the filaments \citep{Multiscale2010}, which we call knots. To identify the knots on a given filamentary structure, we implement the metric graph reconstruction algorithm \citep{aanjaneya2011metric,lecci2014statistical}; see Appendix~\ref{App:mode_seek} for details.

\section{Applications}
\label{Sec:Application}
	
This section describes the application of our extended \texttt{SCMS} algorithms on $\mathbb{S}^2$ and $\mathbb{S}^2\times \mathbb{R}$ to the survey and simulation data. 
We first revisit our toy example of a cross-shaped filament at the North pole in \autoref{Sec:SCMS_std} and further compare our proposed \texttt{DirSCMS} algorithm with other \texttt{SCMS} typed filament finders as well as the well-known \texttt{DisPerSE} algorithm in \autoref{Sec:art_cross_fila}. Then, we utilize the \texttt{DirSCMS} algorithm to detect cosmic filaments and nodes on some thin slices of the SDSS-IV galactic data in \autoref{Sec:SDSS_data} and leverage the \texttt{DirLinSCMS} algorithm to identify filamentary structures from dark matter halos of the Illustris simulation in \autoref{Sec:illustris}.

\subsection{Toy example: Cross-Shaped Filament At the North Pole}
\label{Sec:art_cross_fila}
	
We show in \autoref{Sec:SCMS_std} that our proposed \texttt{DirSCMS} algorithm outperforms the standard \texttt{SCMS} method in recovering the true filament structure (\emph{i.e.}, a cross-shaped filament on $\mathbb{S}^2$ with its center at the North pole) given a mock dataset with 2000 noisy observations. Here, we further compare our \texttt{DirSCMS} algorithm with other \texttt{SCMS} typed filament finders, including the Density Ridge Estimation Describing Geospatial Evidence \citep[\texttt{DREDGE};][]{moews2021filaments,Moews2021}  and \texttt{SCMS} algorithm with HEALPix \citep{novel_cat2021}. Additionally, we compare our method with the widely used Discrete Persistent Structure Extractor \citep[\texttt{DisPerSE};][]{sousbie2011I,sousbie2011II}.

	\subsubsection{Method Description}
	
	As mentioned in \autoref{Sec:SCMS_sph}, \texttt{DREDGE}\footnote{The DREDGE code is on \url{https://pypi.org/project/dredge}.} leverages the haversine formula to calculate the geodesic distance between a pair of points and estimate the matter/galaxy density function on $\mathbb{S}^2$. It also inherits the main scheme of the standard \texttt{SCMS} algorithm in its ridge estimation; see \cite{Moews2021}. In our application of \texttt{DREDGE} to the mock dataset under the angular (RA,DEC) coordinate, the smoothing bandwidth parameter is chosen as the average geodesic distance from every observation to its 20 nearest neighbors on $\mathbb{S}^2$.
	
	The \texttt{SCMS} algorithm introduced in \cite{novel_cat2021}, on the other hand, utilizes the HEALPix at resolution $N_{side}=1024$ to compute the estimated galaxy density, its gradient, and Hessian on $\mathbb{S}^2$, whose kernel function is the Gaussian kernel evaluated on the geodesic distance. The choice of the bandwidth parameter does not have a large influence on the \texttt{SCMS} algorithm with HEALPix in this toy example, and we select it to be optional among a reasonable range of candidate values.
	
	The filament detection with \texttt{DisPerSE} consists of three main steps. First, \texttt{DisPerSE} makes use of the Delaunay tessellation to construct the simplicial complex from a set of discrete observations and assigns the density to each simplex according to the inverse volume of its dual Voronoi cell \citep{Weygaert2009cosmic,boots2009spatial}. Then, the discrete gradients, critical simplices, as well as the discrete Morse-Smale complex are derived on top of the discrete density function \citep{scoville2019discrete}. Notice that the filaments by \texttt{DisPerSE} are defined as the gradient flows that directly connect the maxima with saddle points. Second, \texttt{DisPerSE} follows the concept of topological persistence \citep{edelsbrunner2000topological} to purify the discrete Morse-Smale complex. Specifically, it assigns a density ratio to each pair of connected critical simplices and removes those pairs whose ratios are less than some persistence ratio threshold with respect to a random discrete Poisson distribution from the discrete Morse-Smale complex. In the \texttt{DisPerSE} software\footnote{See \url{http://www2.iap.fr/users/sousbie/web/html/indexd41d.html}.}, the persistence ratio threshold is specified in units of ``$\sigma$'' based on the significance level and serves as an important parameter of \texttt{DisPerSE} that controls the smoothness of the initial filamentary structures. Finally, \texttt{DisPerSE} smooths the yielded filaments by iteratively averaging their spatial positions.
	
    Given the versatility of \texttt{DisPerSE}, we apply it to both the 2D angular (RA,DEC) and 3D Cartesian coordinates of the observations in the mock dataset. The persistence ratio threshold for each \texttt{DisPerSE} application is chosen to be optimal among several candidate values, and the final output filaments are smoothed five times. 
	
	\subsubsection{Results} 
	
	\begin{figure*}
		\captionsetup[subfigure]{justification=centering}
		\centering
		\begin{subfigure}[t]{.32\textwidth}
			\centering
			\includegraphics[width=\linewidth]{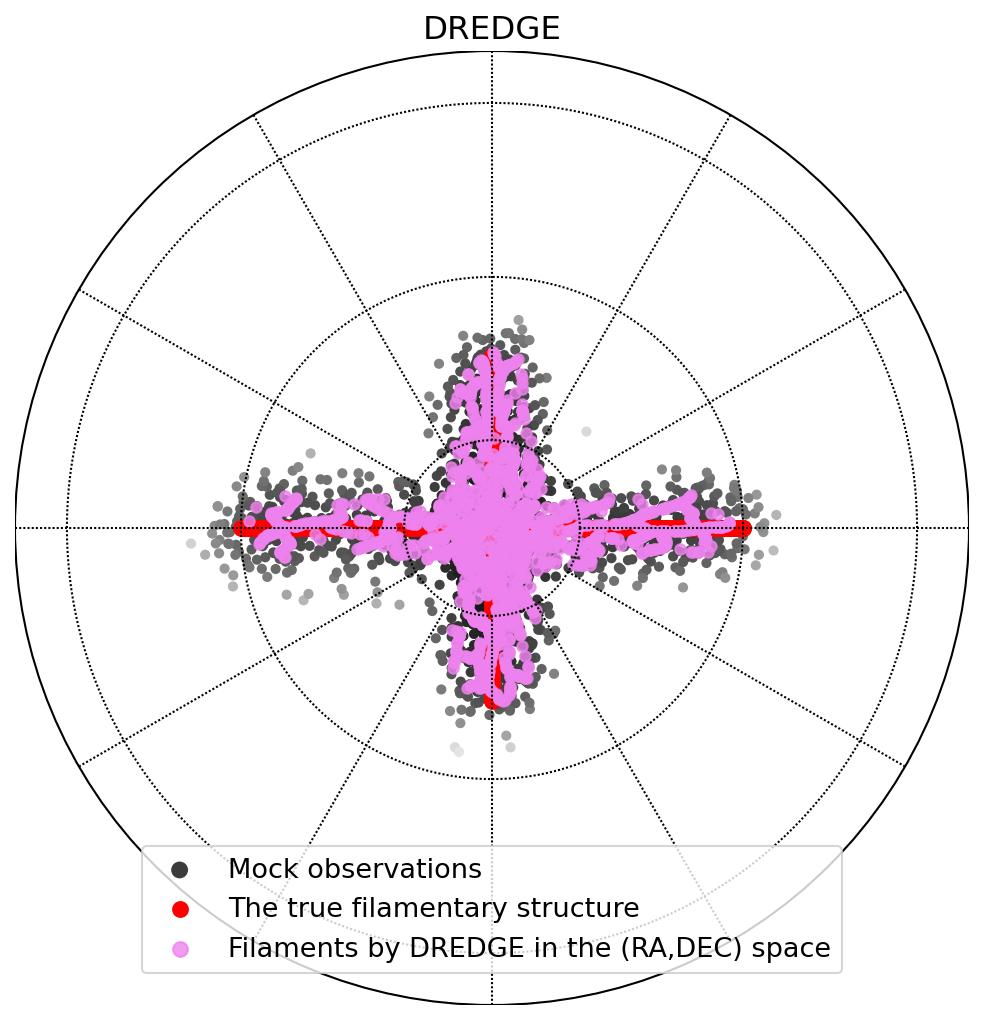}
		\end{subfigure}
		\hfil
		\begin{subfigure}[t]{.32\textwidth}
			\centering
			\includegraphics[width=\linewidth]{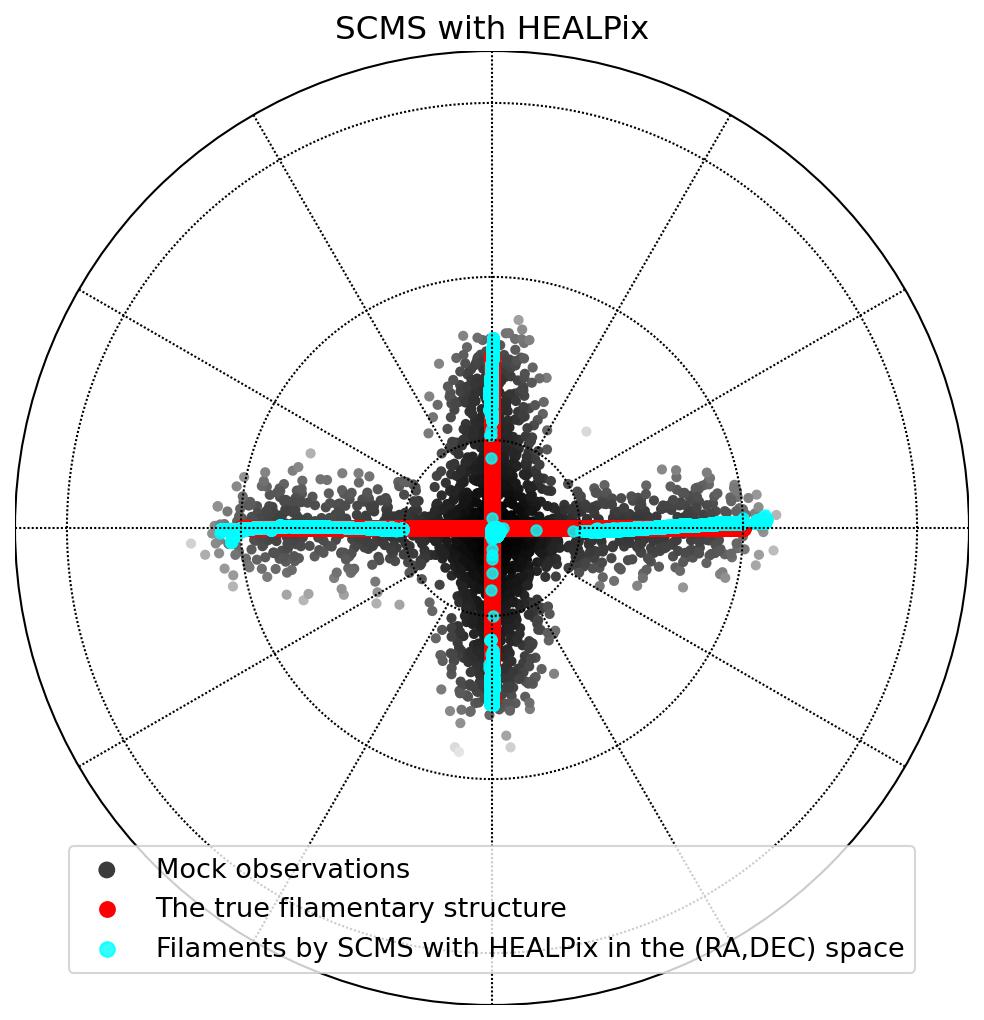}
		\end{subfigure}
		\hfil
		\begin{subfigure}[t]{.32\textwidth}
			\centering
			\includegraphics[width=\linewidth]{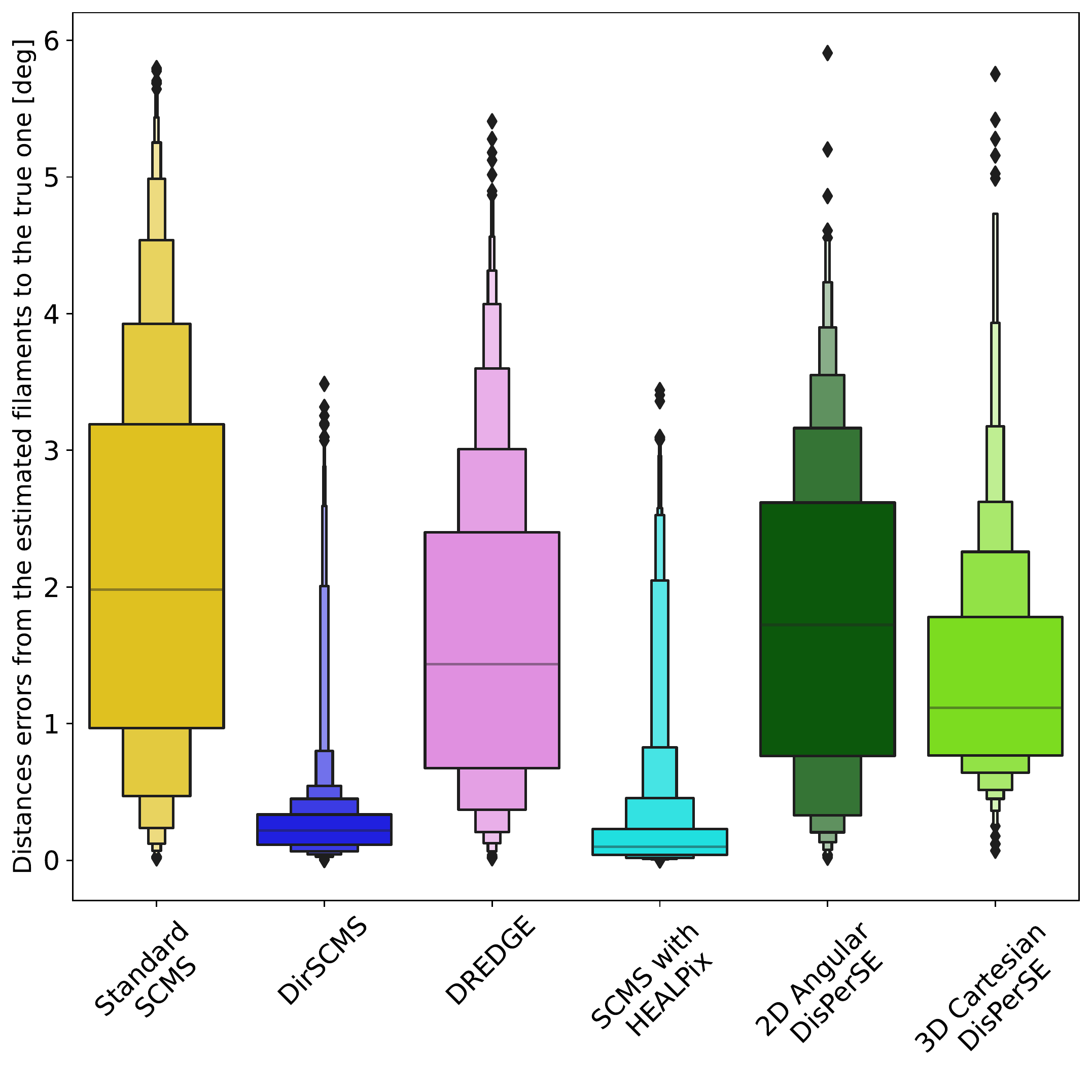}
		\end{subfigure}
		\begin{subfigure}[t]{.32\textwidth}
			\centering
			\includegraphics[width=\linewidth]{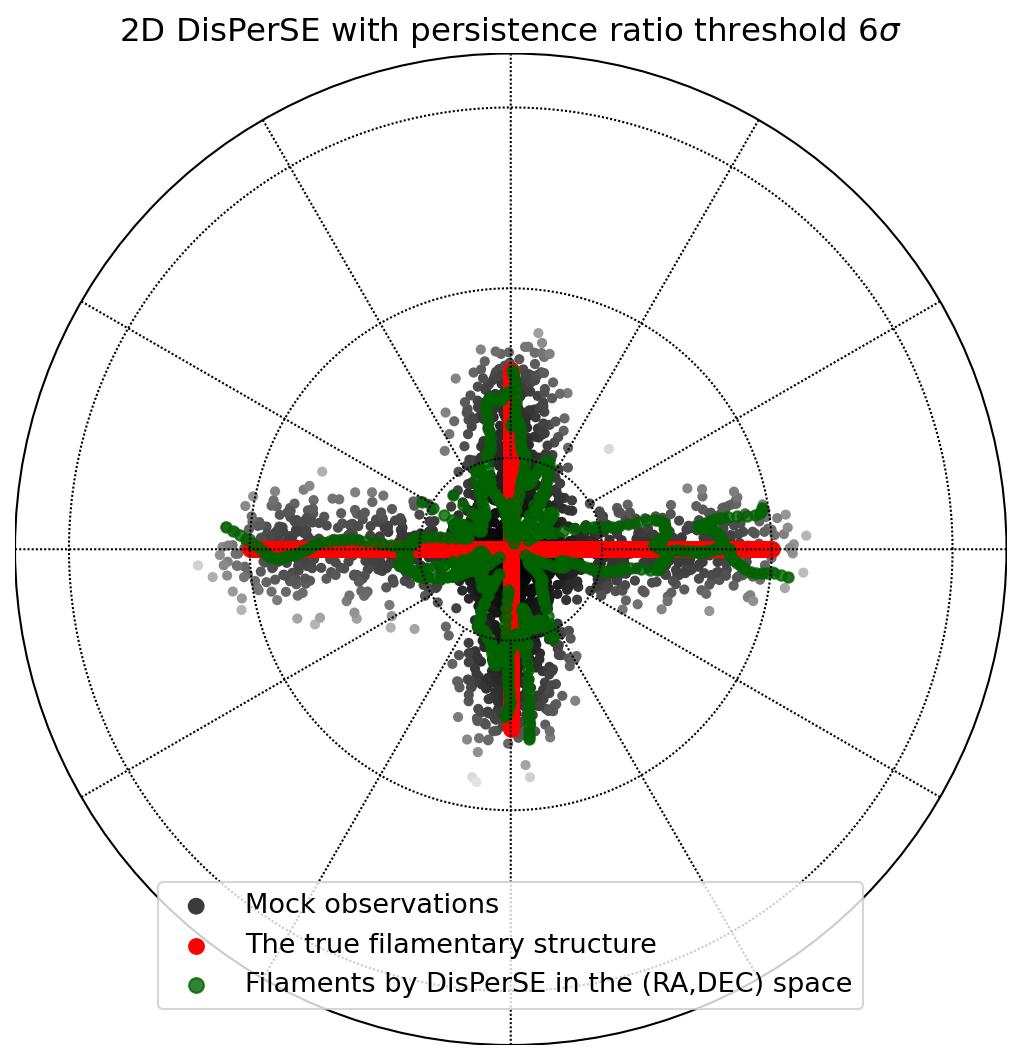}
		\end{subfigure}
		\hfil
		\begin{subfigure}[t]{.32\textwidth}
			\centering
			\includegraphics[width=\linewidth]{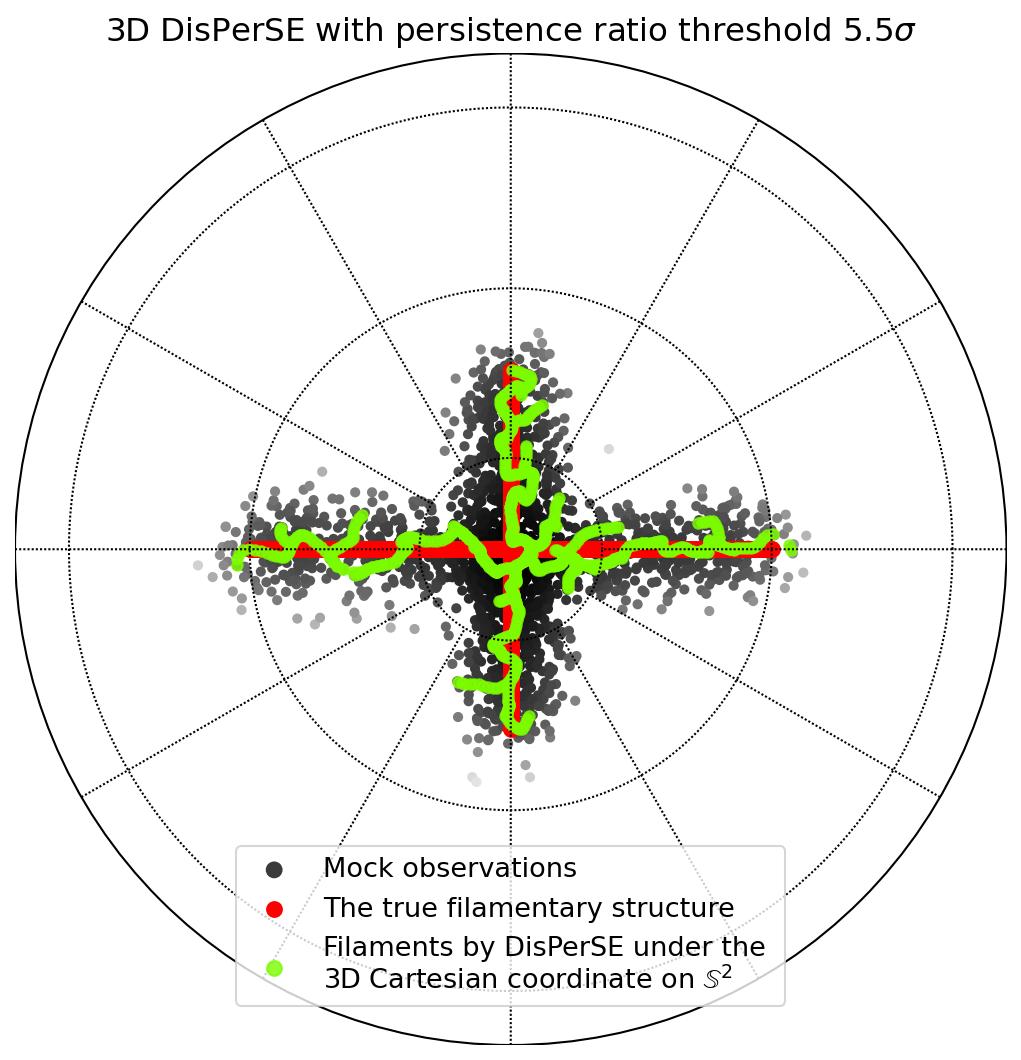}
		\end{subfigure}
		\hfil
		\begin{subfigure}[t]{.32\textwidth}
			\centering
			\includegraphics[width=\linewidth]{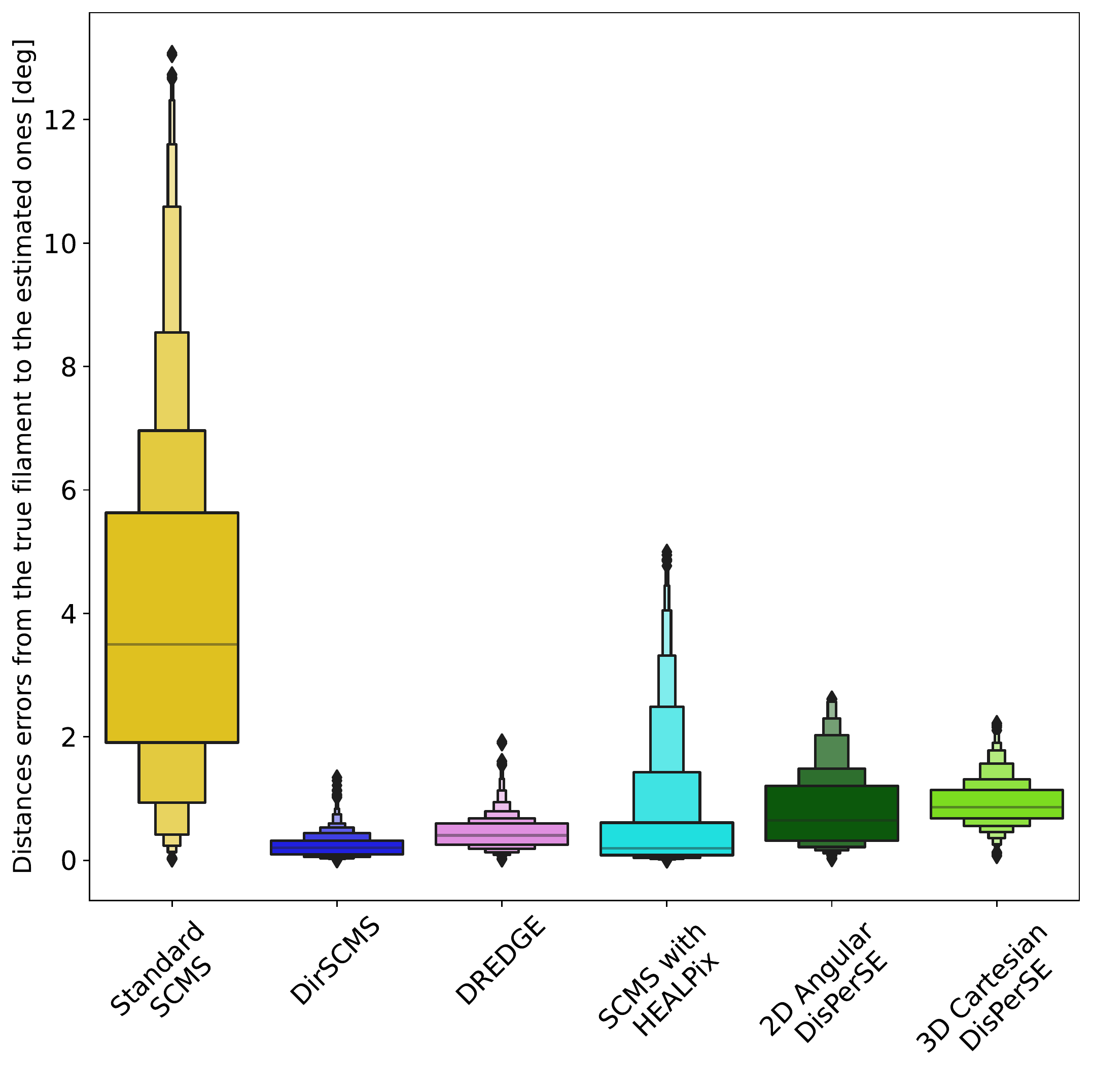}
		\end{subfigure}
		\caption{Comparisons of different filament finding algorithms applied to the mock dataset. \emph{Top Left} and \emph{Top Middle}: The estimated filaments yielded by \texttt{DREDGE} and \texttt{SCMS} with HEALPix together with the true cross-shaped filament and the mock observations around it. \emph{Bottom Left} and \emph{Bottom Middle}: the estimated filaments by \texttt{DisPerSE} under the 2D angular and 3D Cartesian coordinates on $\mathbb{S}^2$, respectively. \emph{Top Right}: The distance error distribution from the estimated filament by each filament finding method to the true cross-shaped filament. \emph{Bottom Right}: The distance error distribution from the true cross-shaped filament to the estimated one by each filament finding method. We present these distance error distributions in terms of the letter-value plots \citep{LetterValue2017}.}
		\label{fig:disperse_cross}
	\end{figure*}
	
	With the mock dataset around the cross-shaped filament at the North pole on $\mathbb{S}^2$ as the inputs, we present the filaments estimated by \texttt{DREDGE}, \texttt{SCMS} algorithm with HEALPix, and \texttt{DisPerSE} in \autoref{fig:disperse_cross}. For better comparison, one can recall the filaments detected by the standard \texttt{SCMS} and our proposed \texttt{DirSCMS} algorithm in \autoref{fig:ridges_Eu_Dir}. All the estimated filaments are post-cleaned by trimming those filamentary points outside the sky footprint of the input observations.
	
	As demonstrated by the top row of \autoref{fig:disperse_cross}, \texttt{DREDGE} fails to converge to the true cross-shaped filament near the North pole on $\mathbb{S}^2$. The \texttt{SCMS} algorithm with HEALPix, on the other hand, precisely approximates the true cross-shaped filament in the regions with relatively low declination but fails to recover the filament with declination greater than $80^{\circ}$. This inferior performance around the North pole may be due to its usage of the angular coordinate system, which is not completely isotropic on $\mathbb{S}^2$, and the numerical instabilities in computing the derivatives of the estimated density with the HEALPix scheme on $\mathbb{S}^2$. As for the bottom row of \autoref{fig:disperse_cross}, \texttt{DisPerSE} produces some estimated filaments that recover parts of the true filamentary structure. However, the filaments by \texttt{DisPerSE} are irregular and contain many spurious components compared with the filament yielded by our \texttt{DirSCMS} algorithm. 
	
	To conclude this artificial cross-shaped filament example, we compute the distance error distribution $\big\{d_g(\bm{x}, R) \text{ for all } \bm{x}\in \hat{R} \big\}$ from each estimated filament $\hat{R}$ yielded by the standard \texttt{SCMS}, \texttt{DREDGE}, \texttt{SCMS} with HEALPix, \texttt{DisPerSE}, and our \texttt{DirSCMS} algorithms to the true cross-shaped filament $R$, where $d_g(\bm{x},R) = \min\big\{d_g(\bm{x},\bm{y}): \bm{y}\in R \big\}$ with $d_g(\cdot,\cdot)$ defined in \eqref{geo_dist}. This metric reveals how close the filamentary points in $\hat{R}$ to the true structure $R$; see the top right panel of \autoref{fig:disperse_cross}. On the other hand, we also calculate the distance error distribution $\big\{d_g(\bm{y}, \hat{R}) \text{ for all } \bm{y}\in R \big\}$ from the true cross-shaped filament $R$ to each estimated filament $\hat{R}$, which measures how well $\hat{R}$ approximates the entire true structure $R$; see the bottm right panel of \autoref{fig:disperse_cross}. Our proposed \texttt{DirSCMS} algorithm exhibits relatively low distance errors in terms of the above two metrics, which can be further demonstrated by the Hausdorff distance
	$$d_H\left(R,\hat{R} \right) = \max\left\{\sup_{\bm{x}\in \hat{R}} d_g(\bm{x},R), \sup_{\bm{y}\in R} d_g(\bm{y},\hat{R}) \right\}$$
	in Table~\ref{table:Haus_dist}. Thus, our \texttt{DirSCMS} algorithm outperforms all the competitors in recovering the true filament under the spherical geometry.
	
	\begin{table}
		\centering 
		\caption{Hausdorff distance [deg] between the estimated filament $\hat{R}$ by each filament finding method and the true cross-shaped filament $R$.} 
		\label{table:Haus_dist}
		\begin{tabular}{C{0.09\linewidth}C{0.09\linewidth}C{0.07\linewidth}C{0.14\linewidth}C{0.155\linewidth}C{0.175\linewidth}}
			\hline
			Standard \texttt{SCMS} & \multirow{2}{*}{\texttt{DirSCMS}} & \multirow{2}{*}{\texttt{DREDGE}} & \texttt{SCMS} with HEALPix & 2D Angular \texttt{DisPerSE} & 3D Cartesian \texttt{DisPerSE}\\
			\hline
		13.079 & \bf 3.486 & 5.408 & 4.997 & 5.907 & 5.754 \\
			\hline
		\end{tabular}
	\end{table}
	
	\subsection{Sloan Digital Sky Survey}
	\label{Sec:SDSS_data}
	
    We now apply our method to the galaxy sample from Data Release 16 of the Sloan Digital Sky Survey, Fourth Phase (SDSS-IV; \citealt{SDSS-IV2020}). We aim at comparing our proposed \texttt{DirSCMS} algorithm with the standard \texttt{SCMS} and \texttt{DisPerSE} methods in recovering the cosmic filaments in the region of high declination. Given that the SDSS-IV galaxies mainly come from regions of relatively low declination, we create a mock galactic data sample by rotating the galaxies on the celestial sphere $\mathbb{S}^2$.
	
	\subsubsection{Data Description}
	
	The SDSS-IV consists of three main surveys in this release: the Extended Baryon Oscillation Spectroscopic Survey (eBOSS; \citealt{eBOSS2016}), Mapping Nearby Galaxies at APO (MaNGA; \citealt{MaNGA2015}), and the APO Galactic Evolution Experiment 2 (APOGEE-2; \citealt{APOGEE2017}), which together covers around 35.28\% of the sky (14,555 $\text{deg}^2$). In particular, eBOSS extends the galaxy measurement of its predecessor BOSS, the eBOSS, from the redshift range $z< 0.6$ to $z < 1$. Moreover, compared with the previous SDSS data release, eBOSS provides 860,935 new optical spectra of galaxies and quasars in this release and reprocesses its spectra using the latest version of the data reduction pipelines \citep{Bolton2012,Dawson2013}. Our following analysis relies primarily on all the spectra classified as galaxies by SDSS-IV. Among them, we focus on those galaxies with definite positive redshift values.
	
	As revealed by \autoref{fig:disperse_cross}, the standard \texttt{SCMS} and \texttt{DisPerSE} algorithms can produce some highly biased filaments under the (RA,DEC) coordinate system of the survey data, especially in the regions of high declination. While the SDSS-IV generally observes astronomical objects in the regions of relatively low declination, some future wide sky surveys, such as EUCLID \citep{Euclid2021} and LSST \citep{LSST2022}, are planning to observe the sky footprints with higher declination values. Therefore, we intend to study how the standard \texttt{SCMS}, \texttt{DisPerSE}, and our proposed \texttt{DirSCMS} algorithms behave if our observational data lie mainly on the region of high declination. To this end, we consider two thin slices $0.05\leq z< 0.055$ at low redshift and $0.46 \leq z < 0.465$ at high redshift and extract the above SDSS-IV galaxies within these two redshift slices that locate in the North Galactic Cap ($100^{\circ}< \text{RA}<270^{\circ}, -5^{\circ}< \text{DEC}< 70^{\circ}$). 
	The total number of resulting galaxies within the selected celestial region is 22,709 for the low redshift slice $0.05\leq z< 0.055$ and 14,486 for the high redshift slice $0.46\leq z< 0.465$. Then, we compute the spherical mean $\frac{\sum_{i=1}^n \bm{X}_i}{\norm{\sum_{i=1}^n \bm{X}_i}_2}$ of the galaxies based on their Cartesian coordinates $\bm{X}_1,...,\bm{X}_n \in \mathbb{S}^2$ in each redshift slice and rotate the galaxies within each redshift slice such that their new spherical mean is $\bm{\mu}_0=(0,0,1)\in \mathbb{S}^2$. Such a rotation can be easily done by multiplying the rotation matrix \eqref{Rodrigues_rot} to the Cartesian coordinate of each galaxy. Hence, we obtain two mock galaxy samples for the low and high redshift slices on $\mathbb{S}^2$ whose center is at the North pole.
	
	\subsubsection{Results}
	
	\begin{figure*}
		\captionsetup[subfigure]{justification=centering}
		\centering
		\begin{subfigure}[t]{.32\textwidth}
			\centering
			\includegraphics[width=\linewidth]{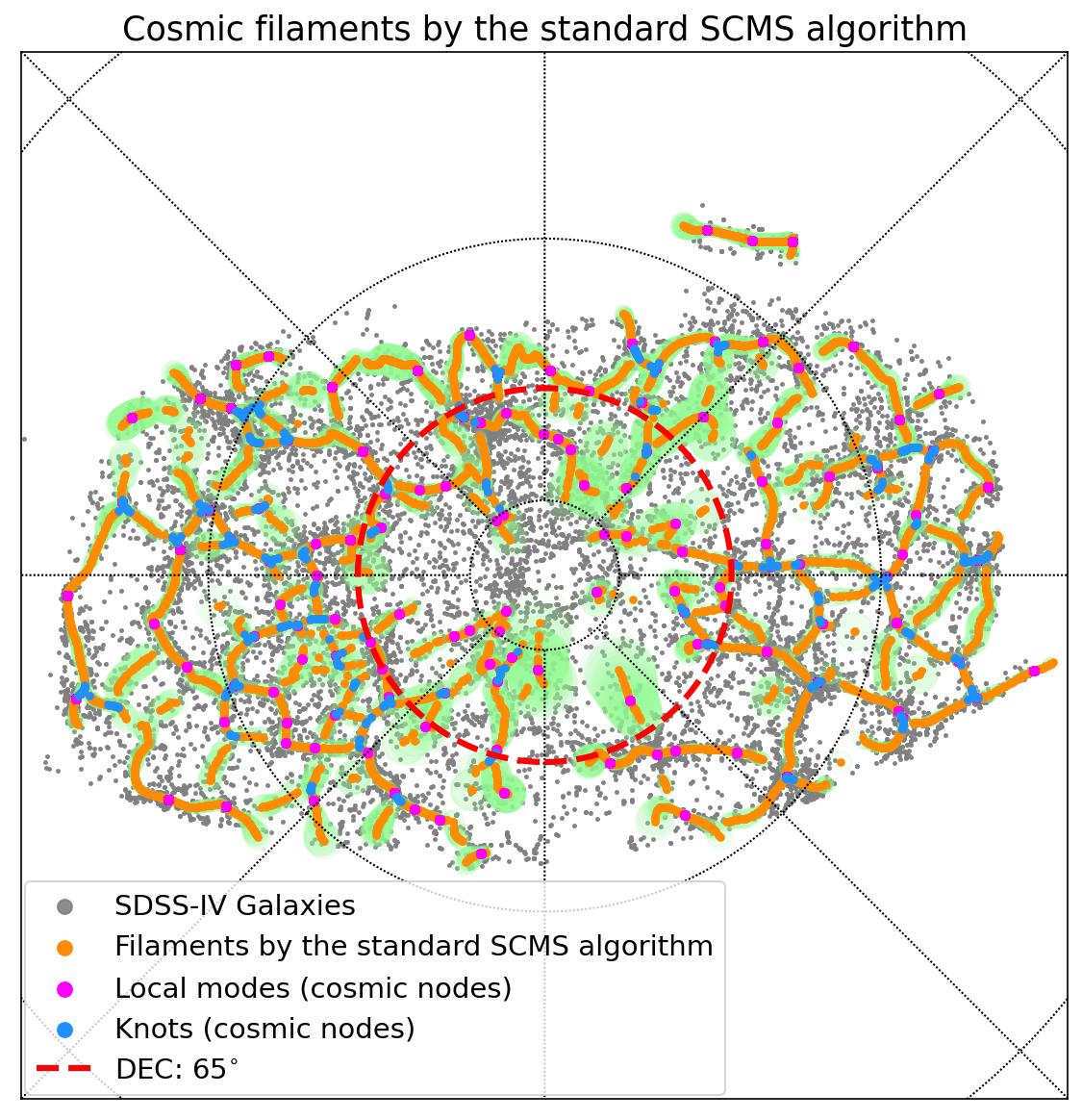}
		\end{subfigure}
		\hfil
		\begin{subfigure}[t]{.32\textwidth}
			\centering
			\includegraphics[width=\linewidth]{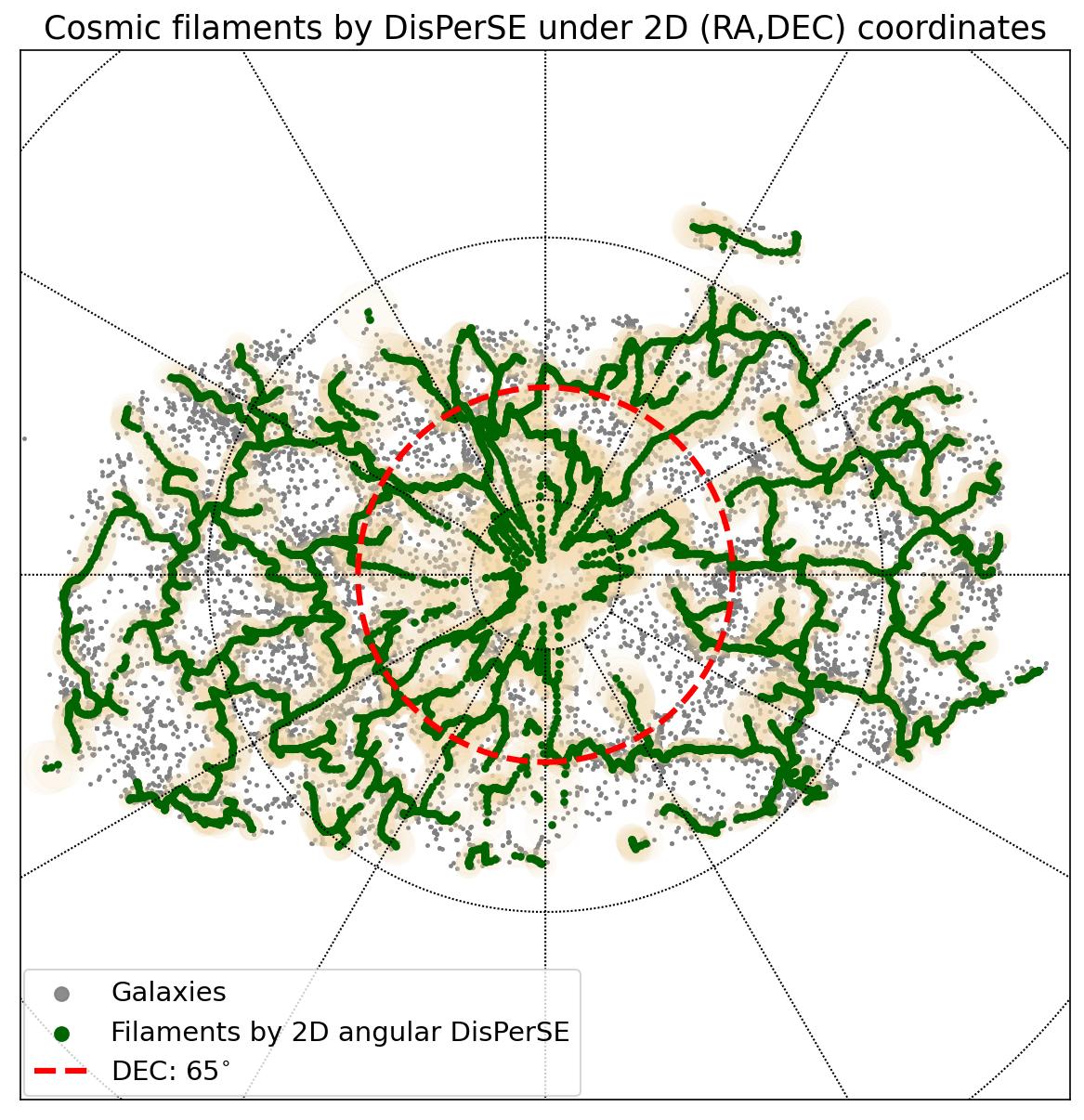}
		\end{subfigure}
		\hfil
		\begin{subfigure}[t]{.32\textwidth}
			\centering
			\includegraphics[width=\linewidth]{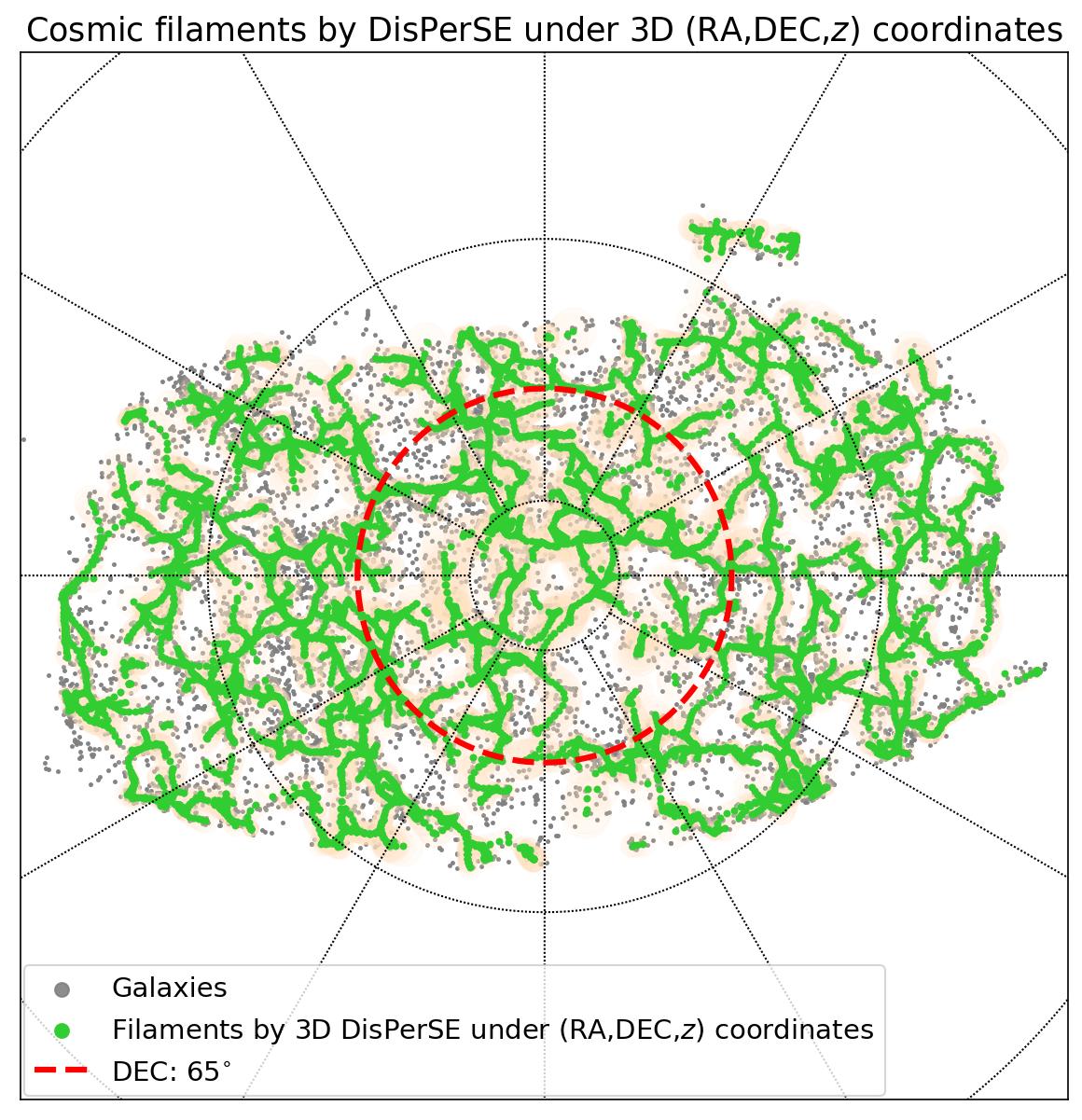}
		\end{subfigure}
		\begin{subfigure}[t]{.32\textwidth}
			\centering
			\includegraphics[width=\linewidth]{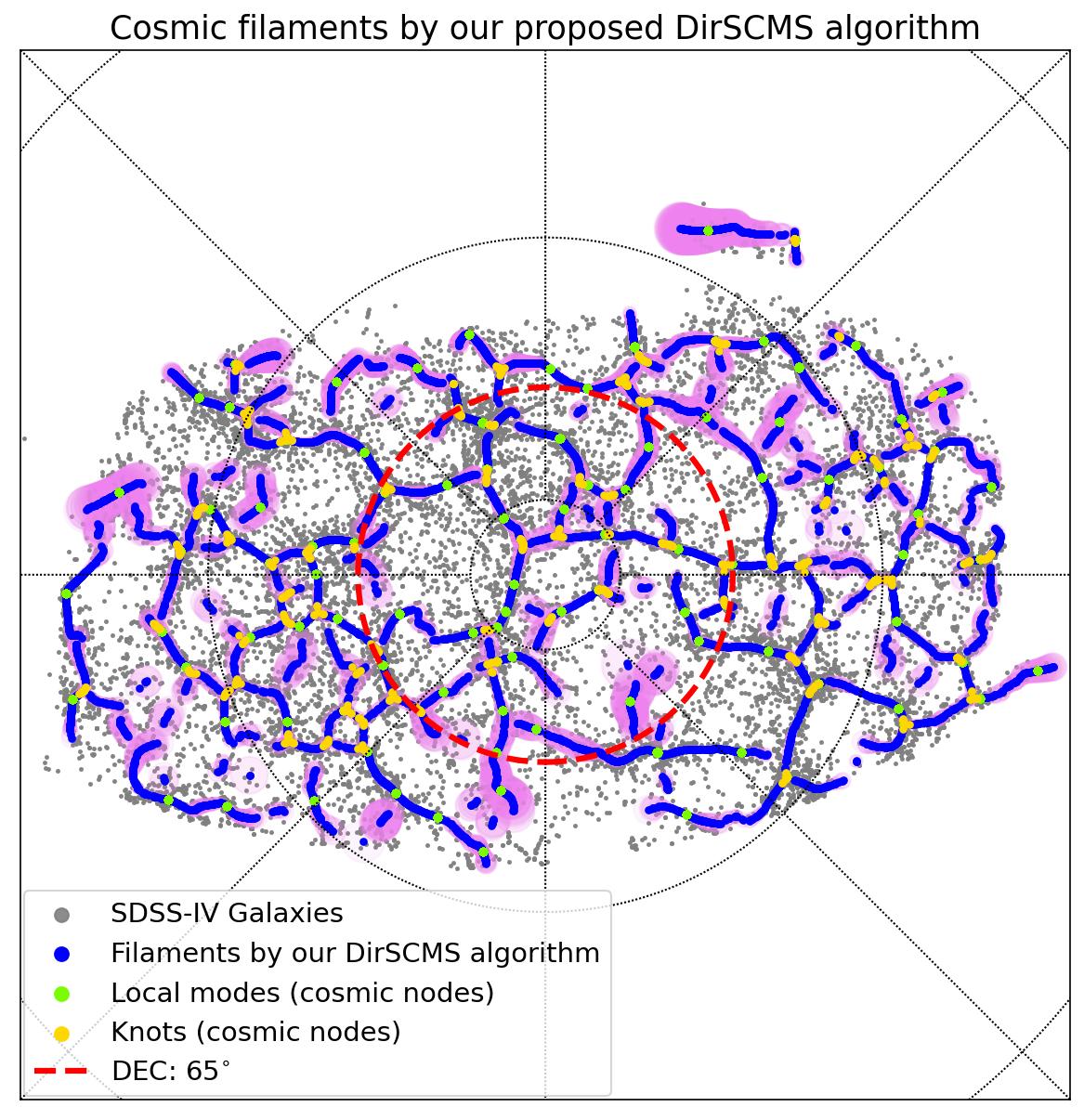}
		\end{subfigure}
		\hfil
		\begin{subfigure}[t]{.32\textwidth}
			\centering
			\includegraphics[width=\linewidth]{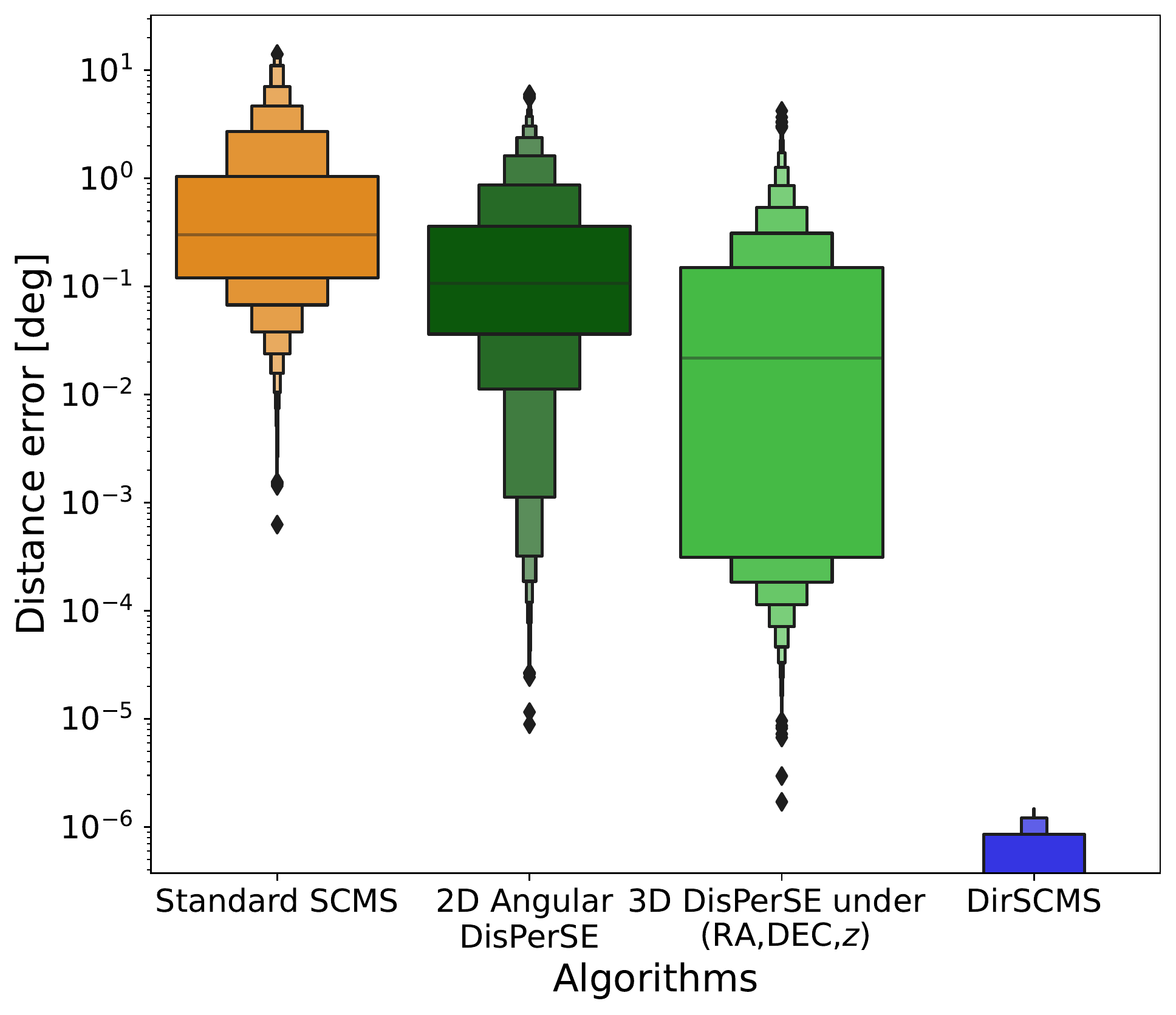}
		\end{subfigure}
		\hfil
		\begin{subfigure}[t]{.32\textwidth}
			\centering
			\includegraphics[width=\linewidth]{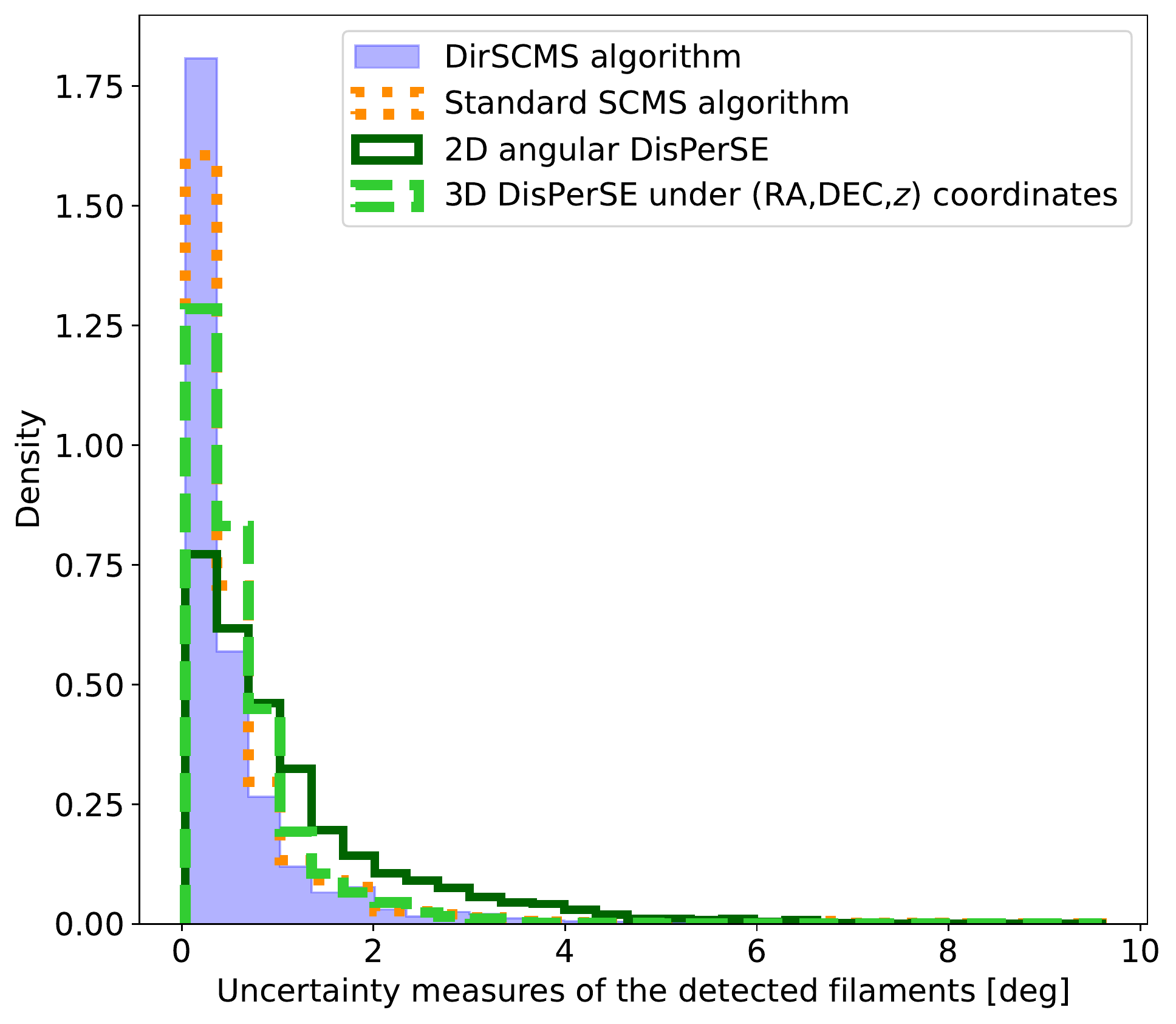}
		\end{subfigure}
		\caption{Comparisons of cosmic filaments detected by the standard \texttt{SCMS}, \texttt{DisPerSE}, and our proposed \texttt{DirSCMS} algorithms on the mock galaxy sample in the low redshift slice $0.05\leq z < 0.055$. \emph{Top Row} and \emph{Bottom Left}: the cosmic filaments (and associated cosmic nodes) recovered by the standard \texttt{SCMS}, \texttt{DisPerSE} under the 2D angular (RA,DEC) and 3D (RA,DEC,$z$) coordinate systems, and our \texttt{DirSCMS} algorithms, respectively, where the light colored areas around the filaments indicate the magnitude of uncertainty measure for each filamentary point. In the \emph{Upper Right} panel, we project the filaments in the 3D (RA,DEC,$z$) space onto the 2D angular (RA,DEC) space for visualization. \emph{Bottom Middle}: The letter-value plots of distance errors from the filaments in the high-declination region to the ones in the original low-declination region on $\mathbb{S}^2$ estimated by the standard \texttt{SCMS}, \texttt{DisPerSE}, and our \texttt{DirSCMS} algorithms. \emph{Bottom Right}: The histograms of uncertainty measures for the filaments recovered by all these algorithms.}
		\label{fig:fila_high_dec}
	\end{figure*}
	
	We first apply the standard \texttt{SCMS}, \texttt{DisPerSE}, and our proposed \texttt{DirSCMS} algorithms to the mock galaxy sample in the low redshift slice $0.05\leq z < 0.055$ in order to detect the cosmic filaments (as well as the associated cosmic nodes). The bandwidth parameter for the standard \texttt{SCMS} algorithm is taken according to Equation (A1) in Appendix A of \cite{Chen2015methods} as: 
	\begin{eqnarray}
	\label{bw_Eu}
	b_{\text{Euc}} = A_0\left(\frac{1}{d+2} \right)^{\frac{1}{d+4}} n^{-\frac{1}{d+4}} \sigma_{\min}
	\end{eqnarray} 
	with dimension $d=2$ and scale factor $A_0=0.6$ in this case, where $\sigma_{\min}$ is minimal standard deviation across all the coordinates; see also Proposition 1 in \cite{chacon2011asymptotics}. During the application of \texttt{DisPerSE}, we use both the 2D angular (RA,DEC) coordinates and 3D (RA,DEC,$z$) coordinates of the mock galaxies as inputs and set the persistence threshold ratio as $3\sigma$. In addition, when utilizing \texttt{DisPerSE} to detect cosmic filaments under the 3D (RA,DEC,$z$) space, we transform the coordinates of input galaxies to their Cartesian counterparts via the WMAP-9 cosmology. The output filaments by \texttt{DisPerSE} are iteratively smoothed for 5 times. As for the bandwidth parameter for our \texttt{DirSCMS} algorithm, we slightly modify the rule of thumb on $\mathbb{S}^2$ in Proposition 2 of \cite{garcia2013exact} as:
	\begin{eqnarray}
	\label{bw_Dir}
	b_{\text{Dir}} = B_0 \left[\frac{8\sinh^2(\hat{\kappa})}{\hat{\kappa}\left[(1+4\hat{\kappa}^2) \sinh(2\hat{\kappa}) -2\hat{\kappa} \cosh(2\hat{\kappa})  \right]n} \right]^{\frac{1}{6}}
	\end{eqnarray}
	with $\hat{\kappa} = \frac{\bar{R}(3-\bar{R})}{1-\bar{R}^2}$, where $\bar{R}=\frac{\norm{\bm{X}_i}_2}{n}$ given the Cartesian coordinates of galaxies $\bm{X}_1,...,\bm{X}_n$ on $\mathbb{S}^2$ and we take $B_0=0.25$ to avoid the over-smoothing effect \citep{sheather2004density}. Finally, to reduce the production of spurious filaments, we remove 20\% of galaxies with low estimated density values before the iterations of the standard \texttt{SCMS} and our \texttt{DirSCMS} algorithms; see Step 2 of Algorithm~\ref{Algo:Dir_SCMS} in Appendix~\ref{App:DirSCMS_detail}.
	
	\begin{figure*}
		\captionsetup[subfigure]{justification=centering}
		\centering
		\begin{subfigure}[t]{.32\textwidth}
			\centering
			\includegraphics[width=\linewidth]{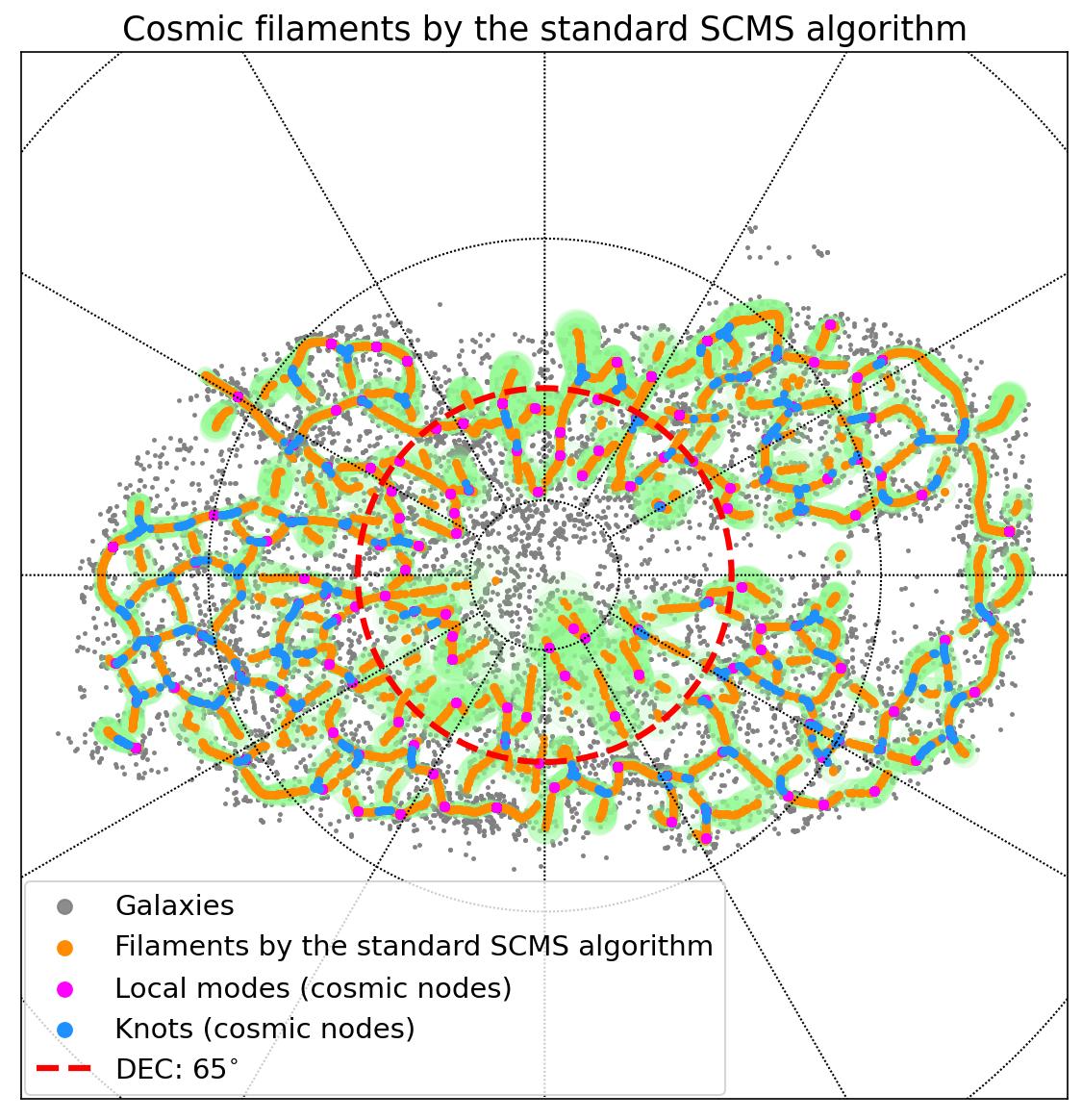}
		\end{subfigure}
		\hfil
		\begin{subfigure}[t]{.32\textwidth}
			\centering
			\includegraphics[width=\linewidth]{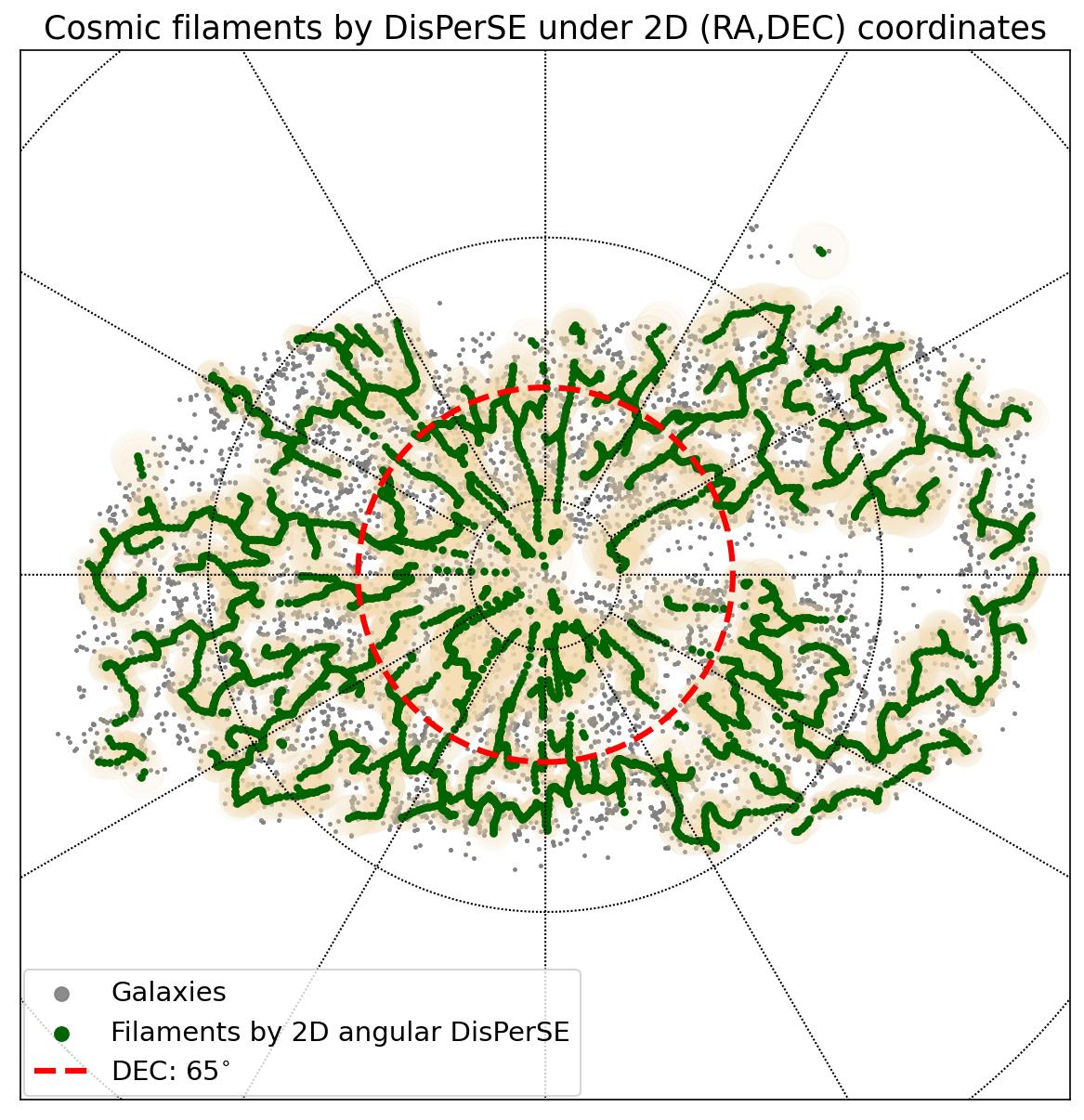}
		\end{subfigure}
		\hfil
		\begin{subfigure}[t]{.32\textwidth}
			\centering
			\includegraphics[width=\linewidth]{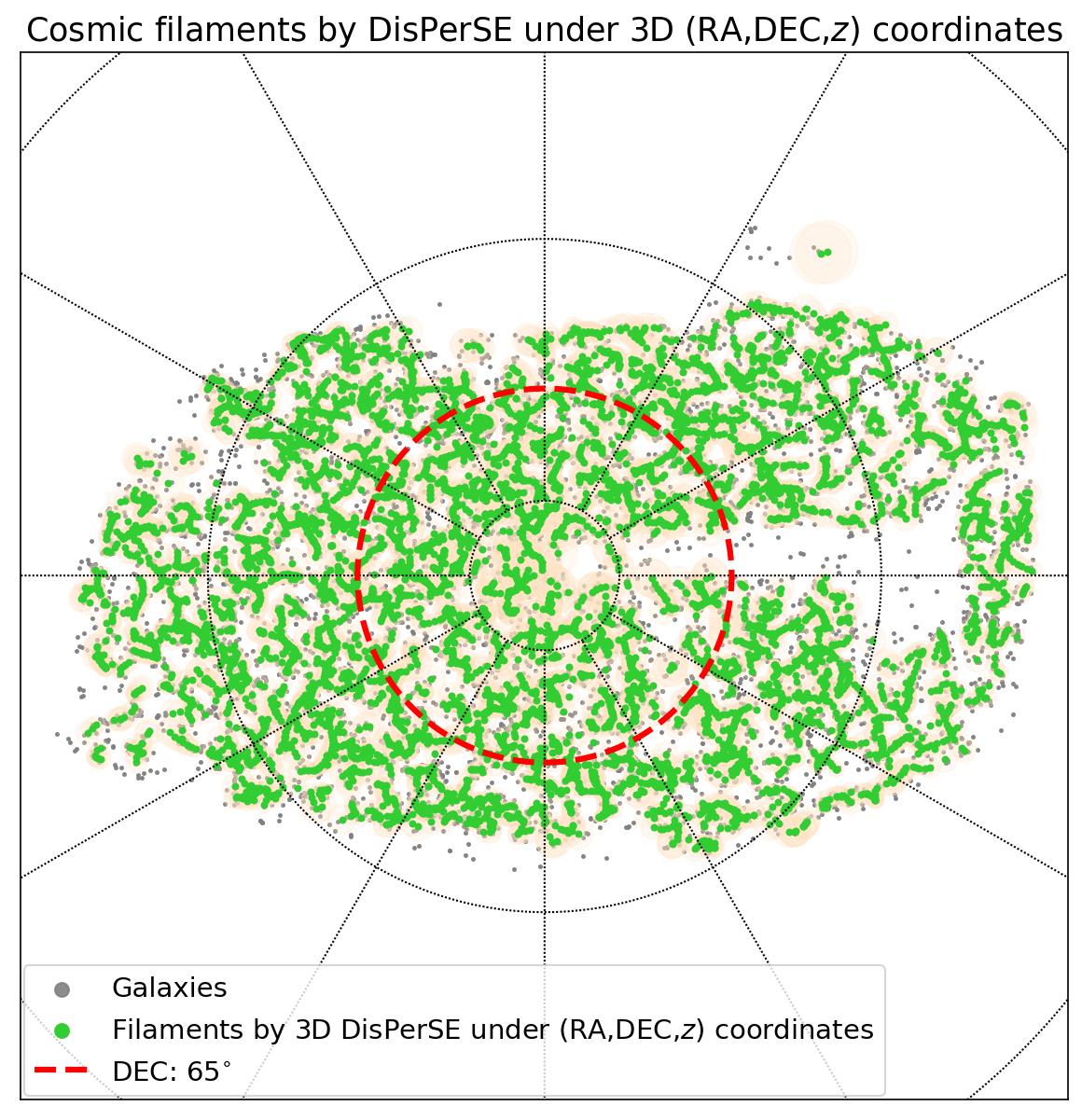}
		\end{subfigure}
		\begin{subfigure}[t]{.32\textwidth}
			\centering
			\includegraphics[width=\linewidth]{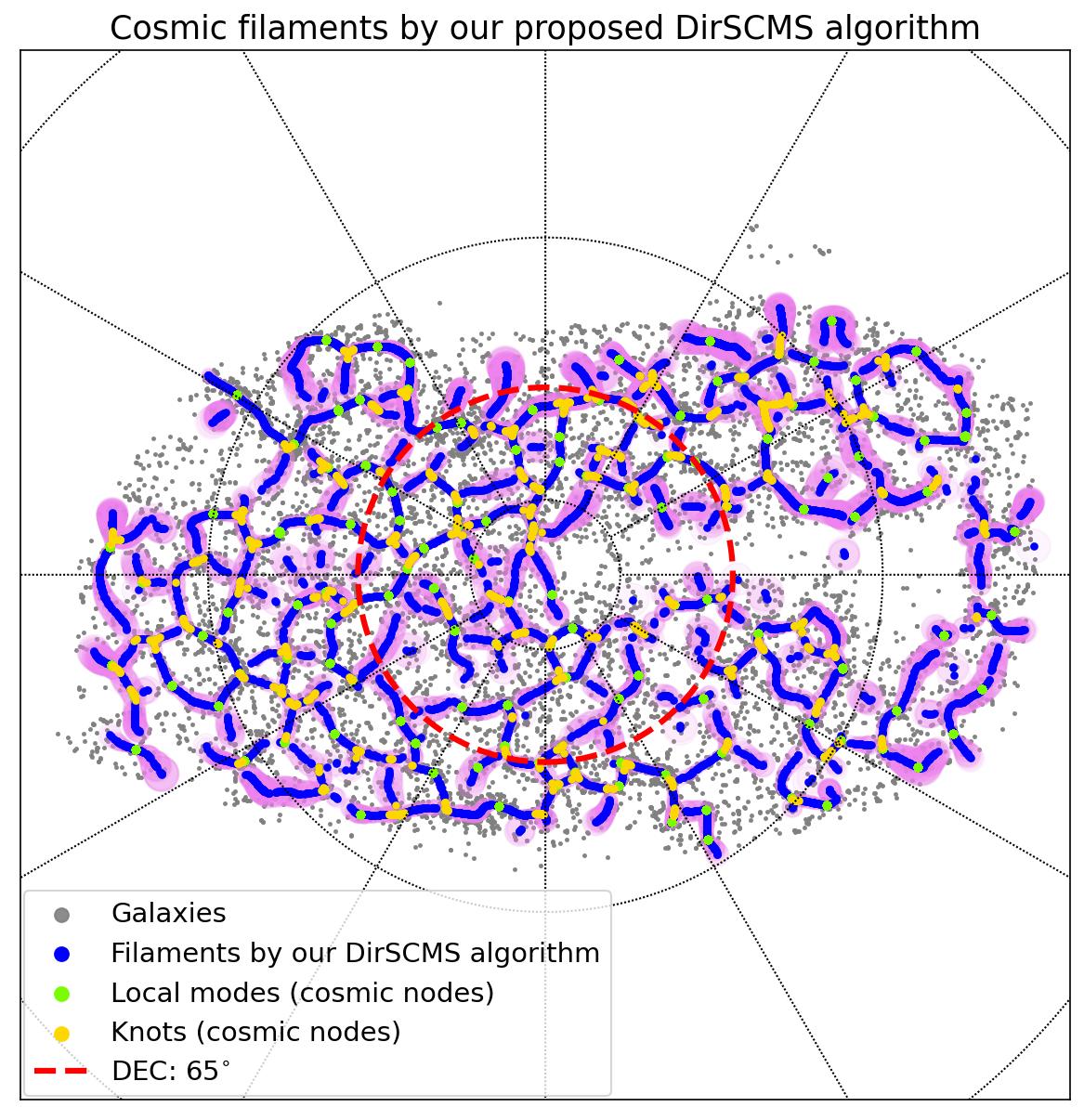}
		\end{subfigure}
		\hfil
		\begin{subfigure}[t]{.32\textwidth}
			\centering
			\includegraphics[width=\linewidth]{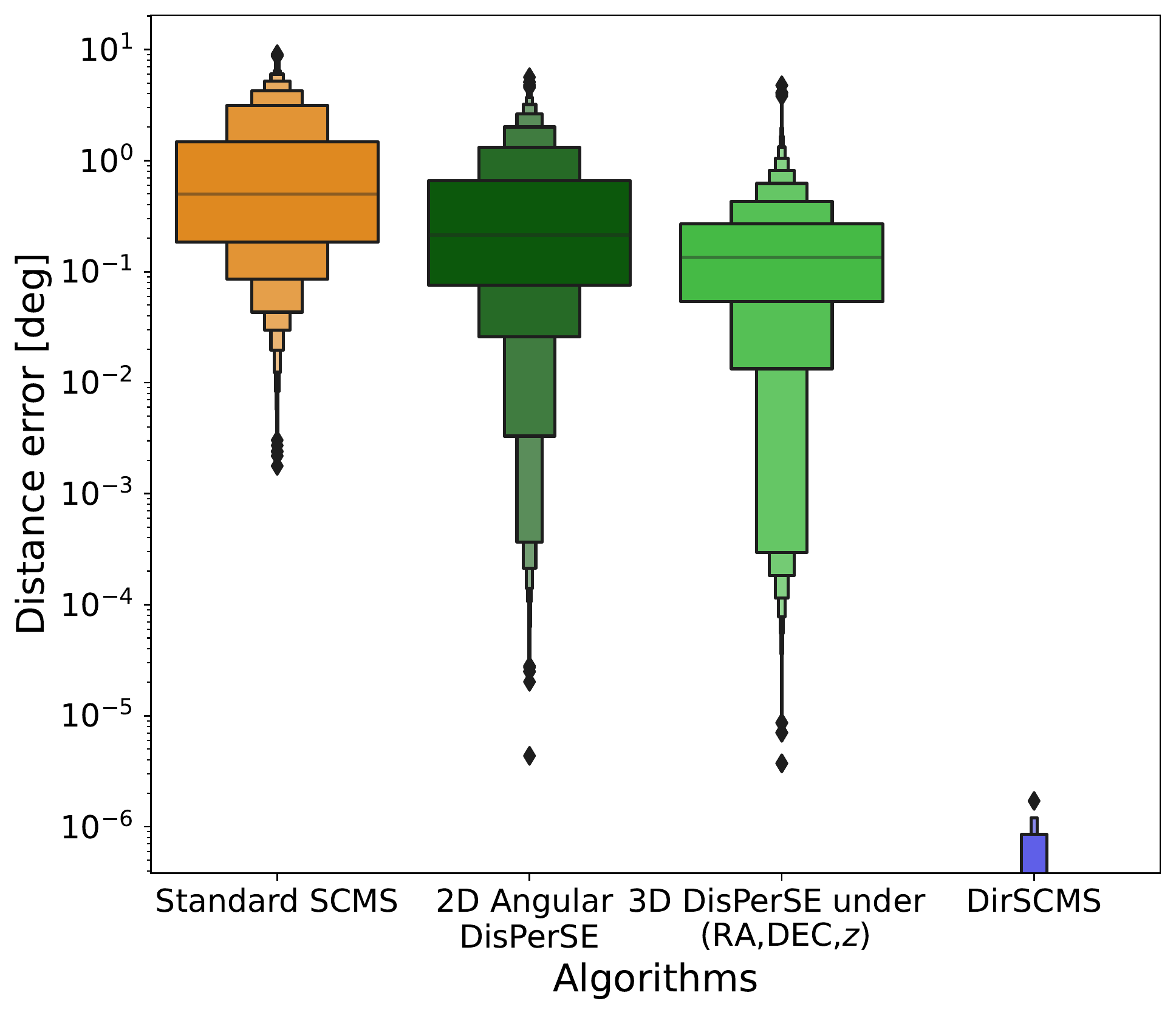}
		\end{subfigure}
		\hfil
		\begin{subfigure}[t]{.32\textwidth}
			\centering
			\includegraphics[width=\linewidth]{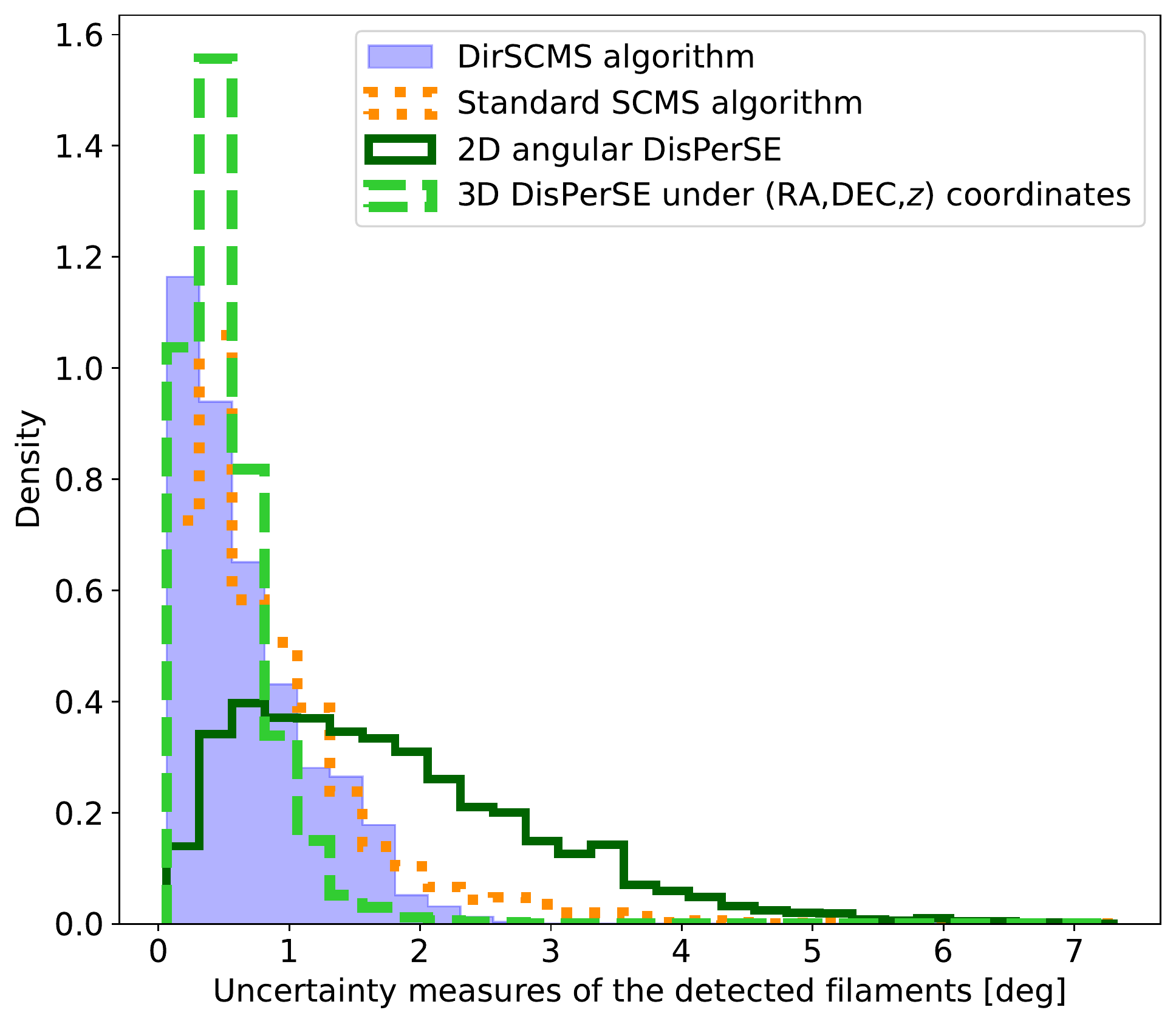}
		\end{subfigure}
		\caption{Comparisons of cosmic filaments detected by the standard \texttt{SCMS}, \texttt{DisPerSE}, and our proposed \texttt{DirSCMS} algorithms on the mock galaxy sample in the high redshift slice $0.46\leq z < 0.465$. The detailed description for each panel can be found in \autoref{fig:fila_high_dec}.}
		\label{fig:fila_high_dec_high_rs}
	\end{figure*}

	As shown by \autoref{fig:fila_high_dec}, the standard \texttt{SCMS} algorithm misses a lot of potential filamentary structures within the regions with $\text{DEC} > 65^{\circ}$. \texttt{DisPerSE} under both the 2D angular (RA,DEC) and 3D (RA,DEC,$z$) coordinates, though capturing parts of sensible filaments within the high-declination regions, produces many spurious filamentary structures. Moreover, the filamentary points detected by the standard \texttt{SCMS} and \texttt{DisPerSE} algorithms have considerably high uncertainties, especially when the filamentary points lie in the region of high declination. On the contrary, our proposed \texttt{DirSCMS} algorithm is able to recover the filaments with high confidence whatever declination the galaxies are located at. To further demonstrate the isotropy and adaptivity of our \texttt{DirSCMS} algorithm on $\mathbb{S}^2$, we quantitatively analyze how the filamentary structures recovered by the above filament finding algorithms would change when the declination values of the galaxies vary. To this end, we estimate the two sets of filaments with each filament finding algorithm based on the mock galaxy sample before rotation (\emph{i.e.}, in the original SDSS observational region of low declination) and after rotation (\emph{i.e.}, in the region of high declination and centering at the North pole), respectively. Then, we compute the distance error from each point of the filamentary set in the high-declination region to the one in the original low-declination region on $\mathbb{S}^2$, conditioning that the spherical means of these two sets of filaments are the same. The distance metric is based on the geodesic distance on $\mathbb{S}^2$; see \eqref{geo_dist}. These distance error distributions are depicted in the bottom middle panel of \autoref{fig:fila_high_dec}. The filamentary structures in low and high-declination regions by the standard \texttt{SCMS} and \texttt{DisPerSE} algorithms exhibit large discrepancies in terms of their distance error distributions, suggesting that these filament finding algorithm is highly unstable with respect to any change of the declination values of the input galaxies on $\mathbb{S}^2$. In stark contrast to them, the filaments estimated by our \texttt{DirSCMS} algorithm in low and high-declination regions are identical (\emph{i.e.}, having zero distance errors for all the filamentary points) whenever the bandwidth parameter is held constant, proving that our \texttt{DirSCMS} algorithm is invariant to any data rotation on $\mathbb{S}^2$ and thus adaptive to the spherical geometry. The results for the mock galaxy sample in the high redshift slice $0.46\leq z < 0.465$ are similar and shown in \autoref{fig:fila_high_dec_high_rs}.

	\subsection{Illustris Simulation}
	\label{Sec:illustris}
	
	Our second astronomical application targets at demonstrating the effectiveness and stability of our \texttt{DirLinSCMS} algorithm (\autoref{Sec:SCMS_cone}) when it is applied to astronomical data in the 3D (RA,DEC,$z$) space. The analysis is performed on the friends-of-friends (FoF) halo of the Illustris simulation project \citep{Illustris-I2014,Illustris-II2014,Illustris2014properties,Illustris2015time}.
	
	\subsubsection{Data Description}
	
	The Illustris presents a collection of cosmological hydrodynamical simulations with delicate physical models targeting at the formation and evolution of galaxies across cosmic time \citep{Illustris2015public}. Each snapshot of the Illustris simulation is self-consistently conducted in a 3D cube with side length $75h^{-1}\Mpc \simeq 106.5 \Mpc$ under the WMAP-9 cosmology and evolves from redshift $z=127$ to the present day $z=0$. Within these periodic simulation boxes, there are five types of resolution elements, consisting of dark matter particles, gas cells, gas tracers, stellar and stellar wind particles, as well as black hole sinks. The final public data release of the Illustris simulation includes six primary realizations with different resolution levels or physics implementations of the Illustris volume, each of which has 136 snapshots at different available redshifts from $z=46.77$ to $z=0$. Every snapshot also contains a group catalog with FoF halos and SUBFIND \citep{Springel2001subfindorg,Onions2012subhalo} subhalos.
	
	Given that dark matter halos incubate galaxies and serve as fundamental nonlinear units of cosmic structures \citep{Frenk2012,Wechsler2018}, our following filament detection tasks are based on the FoF halos in the snapshot at redshift $z=0$ from the Illustris-3 simulation\footnote{The data can be downloaded at \url{https://www.illustris-project.org/data/downloads/Illustris-3/}.}. The standard FoF algorithm \citep{Turner1976,Zel1982} with linking length $0.2$ is performed on the dark matter particles so as to identify 131,727 halos within the associated simulation box at redshift $z=0$.
	
	To study the stability of our proposed \texttt{DirLinSCMS} algorithm in detecting filamentary structures from Illustris halos against the redshift distortions due to the effects of peculiar velocities, we mimic the data coordinates that are commonly encounter in astronomical surveys. Specifically, we convert the 3D coordinates of the above Illustris halo data in the periodic box to their (RA,DEC,$z$) coordinates by assuming an observer placed at $(0,0,0)$. This can be achieved, for instance, by modifying the ``\textsf{ra\_dec\_z}'' function in the ``\textsf{mock\_observables}'' module of Halotools (v0.7); see \cite{Halotools2017} for details. After the change of coordinates, each halo has two different redshift values, one in the observed redshift space (with the effects of peculiar velocities) and the other in the true/cosmological redshift space (without the effects of peculiar velocities). These two redshift values, together with the (RA,DEC) coordinate, encodes two distinct positions in the 3D light cone $\mathbb{S}^2 \times \mathbb{R}$ for each halo. 

	\subsubsection{Results}
	
	\begin{figure*}
		\captionsetup[subfigure]{justification=centering}
		\centering
		\begin{subfigure}[t]{.32\textwidth}
			\centering
			\includegraphics[width=\linewidth]{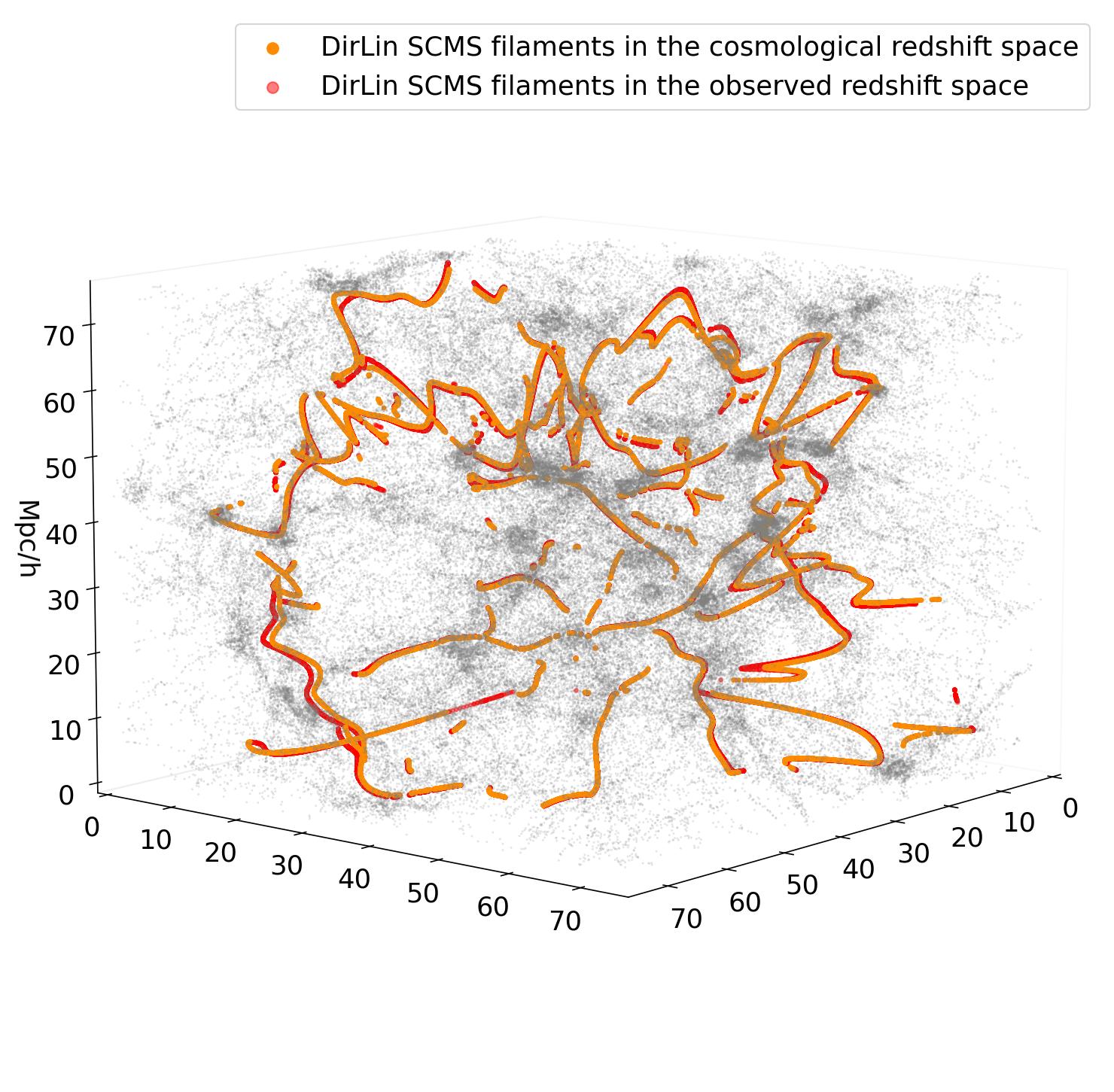}
		\end{subfigure}
		\hfil
		\begin{subfigure}[t]{.32\textwidth}
			\centering
			\includegraphics[width=\linewidth]{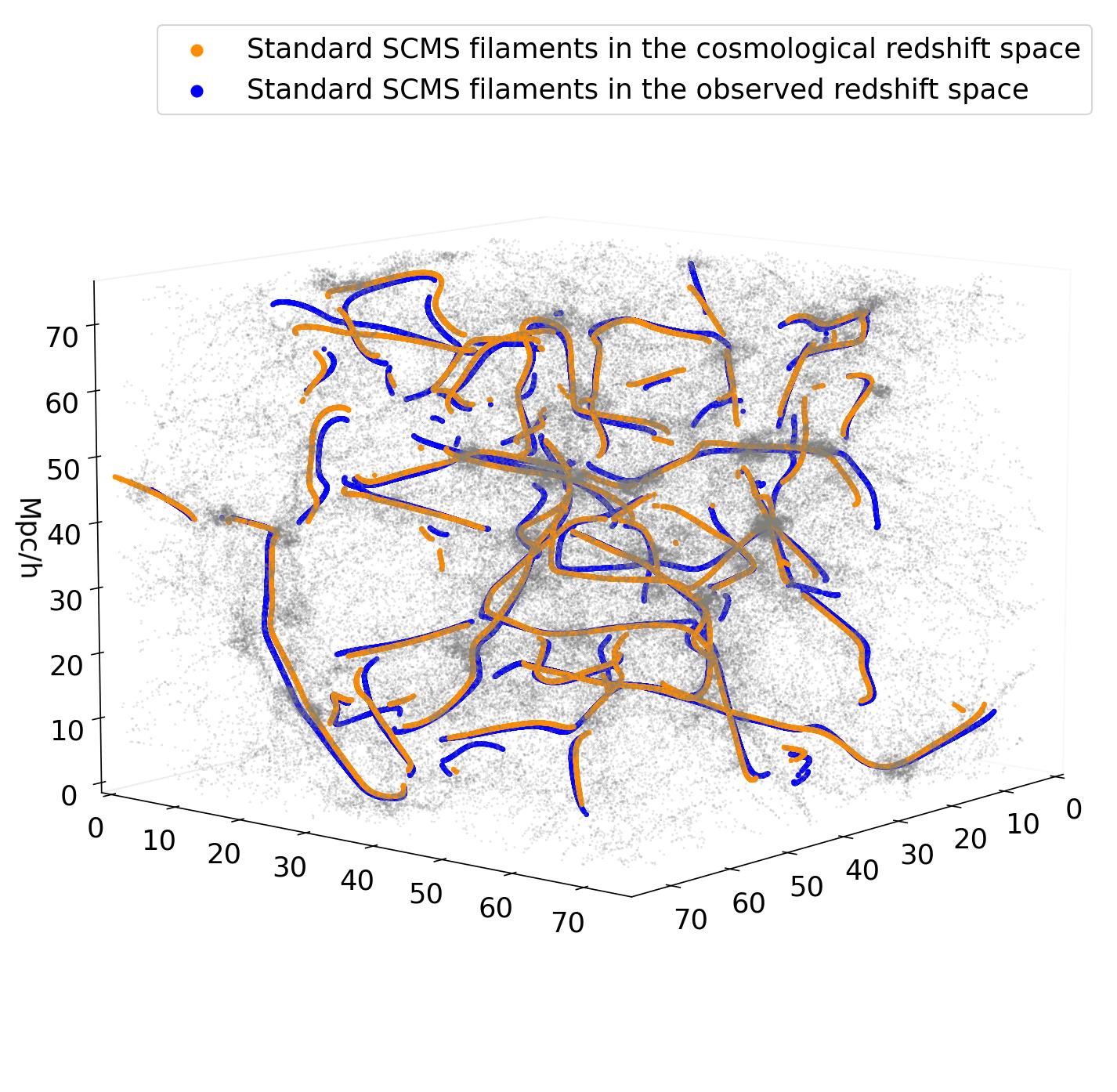}
		\end{subfigure}
		\hfil
		\begin{subfigure}[t]{.32\textwidth}
			\centering
			\includegraphics[width=\linewidth]{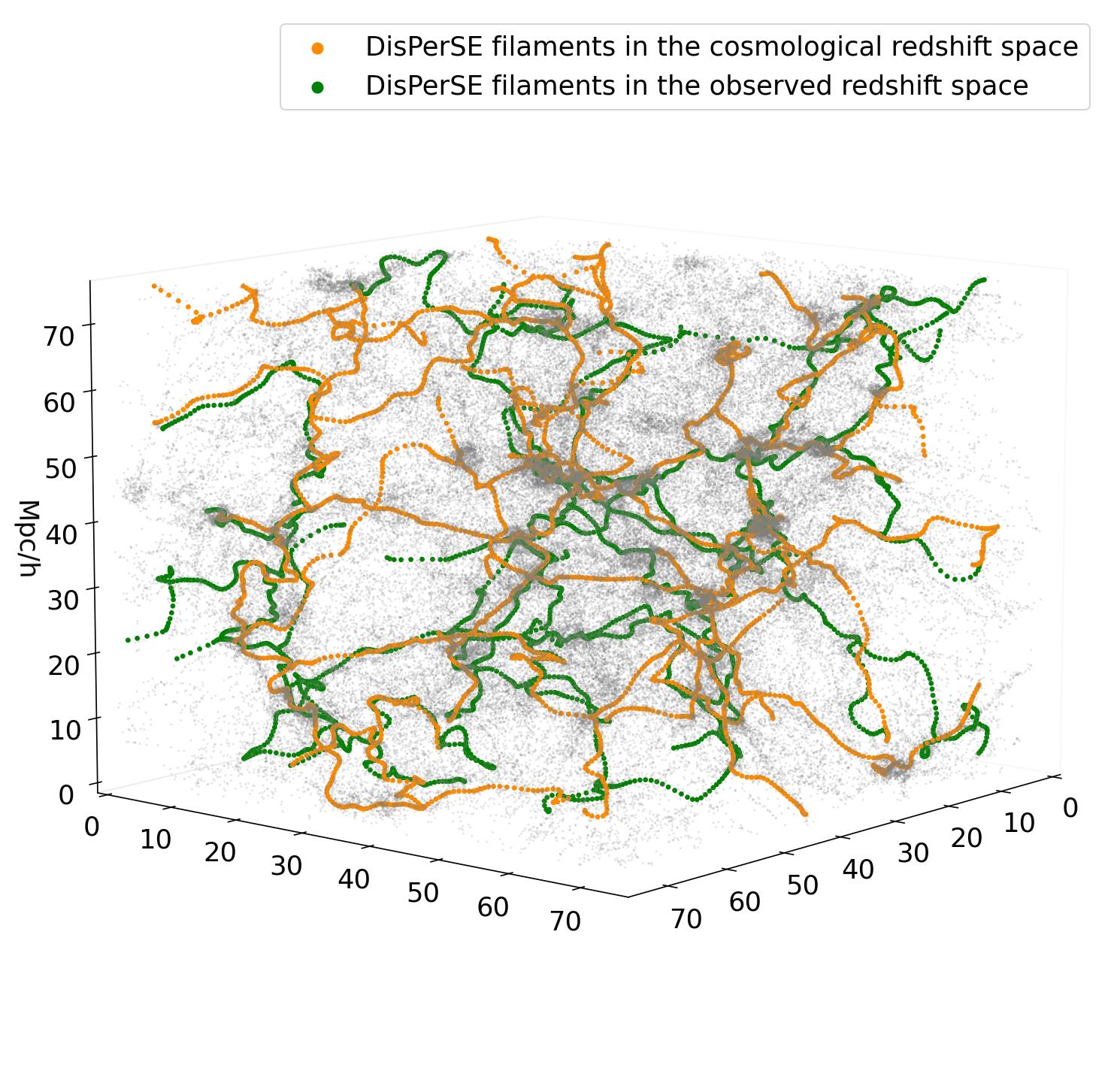}
		\end{subfigure}
		\begin{subfigure}[t]{.99\textwidth}
			\centering
			\includegraphics[width=\linewidth]{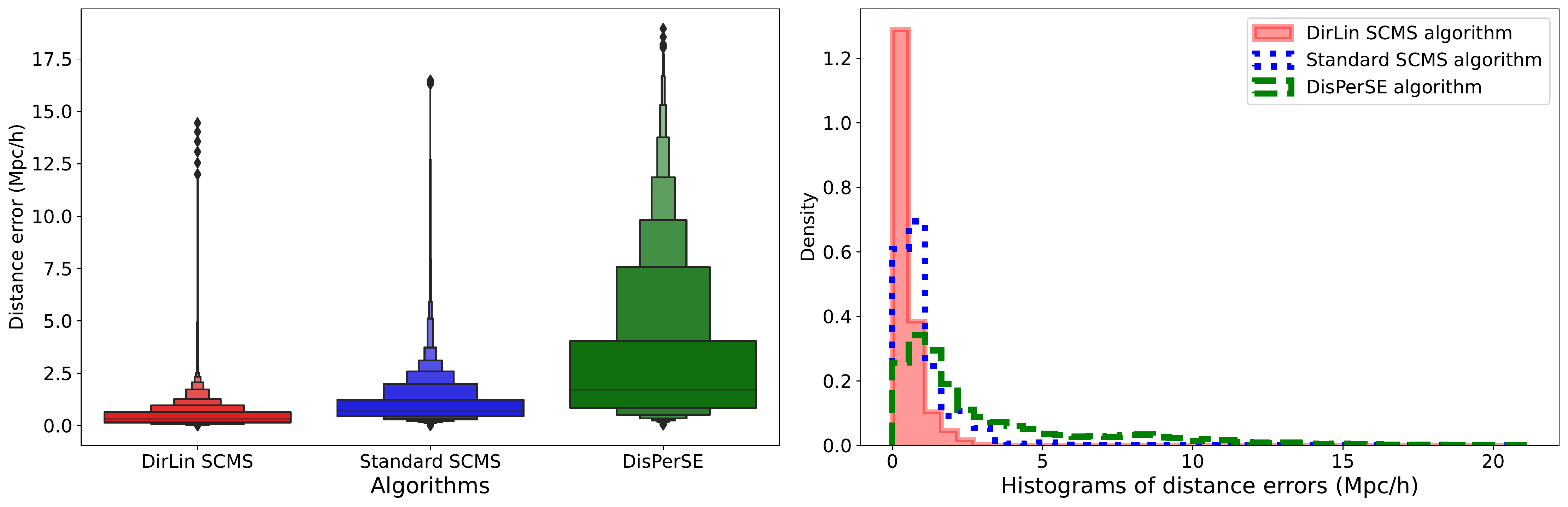}
		\end{subfigure}
		\caption{Comparisons of filamentary structures in the observed and cosmological redshift spaces detected by our \texttt{DirLinSCMS}, standard \texttt{SCMS}, and \texttt{DisPerSE} algorithms on the FoF halos of Illustris-3 at redshift $z=0$. \emph{Top Row}: the filaments detected by our \texttt{DirLinSCMS}, standard \texttt{SCMS}, and \texttt{DisPerSE} algorithms under the observed and true/cosmological redshift values, respectively. \emph{Bottom Row}: The comparative letter-value plots and histograms of distance errors from the filament detected by each method in the observed redshift space to the one in the cosmological redshift space.}
		\label{fig:Illustris_DirLin_Eu_comp}
	\end{figure*}
	
	We apply our \texttt{DirLinSCMS} algorithm to the (RA,DEC,$z$) coordinates of the halo data under the observed and cosmological redshifts respectively, yielding two set of filamentary structures; see the left panel of \autoref{fig:Illustris_DirLin_Eu_comp}. The bandwidth parameters $b_1,b_2$ for the directional and linear data parts are chosen via \eqref{bw_Dir} and \eqref{bw_Eu} with scale factors $B_0=0.5$ and $A_0=40$ in the observed redshift space. When detecting the filaments in the true/cosmological redshift space, we apply the same set of bandwidth parameters as the one in the observed redshift space. It is worth mentioning that the resulting filamentary structures from our \texttt{DirLinSCMS} algorithm are sensitive to the choices of its bandwidth parameters and a useful guideline to select $b_1,b_2$ is to ensure that their values are balanced and of roughly the same scale under the input data. The stability of our \texttt{DirLinSCMS} algorithm against the redshift distortions, however, is relatively robust to different choices of the bandwidth parameters. More discussions about the bandwidth selection for our \texttt{DirLinSCMS} algorithm and its stability results can be found in Appendix~\ref{App:DirLinSCMS_detail}. For comparison, we also implement the standard \texttt{SCMS} and \texttt{DisPerSE} algorithms on the Cartesian coordinates of the halo data under the WMAP-9 cosmology in both the observed and cosmological redshift spaces. The bandwidth parameter of the standard \texttt{SCMS} algorithm is selected as \eqref{bw_Eu} with $A_0=1.5$ and $d=3$. The persistence ratio threshold is set to be $9\sigma$ during the application of \texttt{DisPerSE}, and the output filaments are smoothed 5 times. All the final filamentary structures produced by each algorithm are presented under their 3D coordinates inside the $(75\Mpc/h)^3$ simulation box; see the top row of \autoref{fig:Illustris_DirLin_Eu_comp}.
	
	It is noticeable from the 3D visualizations of filaments in \autoref{fig:Illustris_DirLin_Eu_comp} that the two sets of filaments in the observed and cosmological redshift shapes are more aligned and consistent with each other under our \texttt{DirLinSCMS} algorithm than the standard \texttt{SCMS} and \texttt{DisPerSE} method. We compute the distance errors from the filaments in the observed redshift space to the ones in the cosmological redshift space based on these three filament finding algorithms respectively, and plot their distributions in terms of the comparative letter-value plots and histograms at the bottom row of \autoref{fig:Illustris_DirLin_Eu_comp}. Both the two-sample Welch's t-test and Kolmogorov-Smirnov test between the distance errors by our \texttt{DirLinSCMS} algorithm against the other two methods yield essentially zero p-values. It validates the visual finding that our \texttt{DirLinSCMS} algorithm is more robust to the redshift distortions introduced by the peculiar velocities of the FoF halos.

	\section{Conclusion and Discussion}
	\label{sec:conclusion}
	
	In this paper, we extend the well-known \texttt{SCMS} algorithm \citep{ozertem2011locally,NonparRidges2014,Chen2015methods} to recover the cosmic filaments from discrete observations (e.g., galaxies or dark matter halos) on the 2D (RA,DEC) celestial sphere $\mathbb{S}^2$ and 3D (RA,DEC,$z$) light cone $\mathbb{S}^2\times \mathbb{R}$, the two nonlinear data space that are commonly encountered in astronomical surveys. Similar to the standard \texttt{SCMS} method in the flat Euclidean space $\mathbb{R}^d$, our extended \texttt{SCMS} algorithms in \texttt{SCONCE}, \emph{i.e.}, \texttt{DirSCMS} on $\mathbb{S}^2$ and \texttt{DirLinSCMS} on $\mathbb{S}^2\times \mathbb{R}$, model the cosmic filaments as one-dimensional density ridges that trace over the regions where the observations are highly concentrated on $\mathbb{S}^2$ or $\mathbb{S}^2\times \mathbb{R}$. As demonstrated through the cross-shaped filament example in \autoref{fig:ridges_Eu_Dir} and \autoref{fig:disperse_cross} and the application to SDSS-IV data, our \texttt{DirSCMS} algorithm is invariant to any rotation of the observations on $\mathbb{S}^2$ and more accurate than the other \texttt{SCMS} typed and \texttt{DisPerSE} methods in approximating the true filamentary structure at the region of high declination. The same advantage also applies to its 3D \texttt{DirLinSCMS} extension. It makes our proposed methods not only useful for the current SDSS data but also capable of analyzing the data products from other upcoming wide sky surveys, such as EUCLID \citep{Euclid2021} and LSST \citep{LSST2019,LSST2022}, which cover the sky footprint with high declination values. Furthermore, our applications to the FoF halos at $z=0$ of the Illustris-3 simulation reveal that our \texttt{DirLinSCMS} algorithm is more robust to the redshift distortions brought by the effects of peculiar velocities.
	
	We summarize the strengths and weaknesses of our extended \texttt{SCMS} methods in \autoref{table:pros_cons}, some of which are inherited from the standard \texttt{SCMS} method \citep{Chen2015methods,novel_cat2021}. 
	
	\begin{table*}
		\centering
		\caption{Pros and cons of our proposed \texttt{DirSCMS} and \texttt{DirLinSCMS} algorithms.} 
		\label{table:pros_cons}
	\begin{tabular}{ C{0.12\linewidth} C{0.4\linewidth} C{0.4\linewidth} }
		& \cellcolor[HTML]{33FF8A} {\bf Strengths} & \cellcolor[HTML]{FFA833} {\bf Weaknesses} \\
		\toprule
		& $+$ It is more accurate in recovering the filaments on the celestial sphere $\mathbb{S}^2$ than the standard \texttt{SCMS} and \texttt{DisPerSE} algorithms, especially in the regions of high declination.  &  \\
		\multirow{5}{*}{\texttt{DirSCMS}} & $+$ It controls the false detection of light-of-sight filaments caused by converting the redshifts to their comoving distances or the finger-of-god effects \citep{Jackson1972}. & $-$ It only recovers the filaments on $\mathbb{S}^2$ (or within some thin redshift slices) and will inevitably miss those light-of-sight structures. \\
		& $+$ It has a single tuning bandwidth parameter, and the yielded filaments are stable under different choices of the parameter. & \\
		& $+$ It is easy to implement and fast to execute with parallel programming under the Ray environment \citep{moritz2018ray} provided by our \texttt{python} package \texttt{SCONCE-SCMS}. & \\
		\hline
	   \multirow{5}{*}{\texttt{DirLinSCMS}} &  $+$ It recovers the filaments in the 3D redshift space that are both parallel and perpendicular to the light-of-sight direction.  & $-$ There are two bandwidth parameters and a step size parameter that need tuning, and the yielded filaments are sensitive to the choices of the bandwidth parameters. \\
	     &  $+$ The output filaments are relatively robust to the redshift distortions due to the peculiar velocities.  &  $-$ Depending on the step size parameter, it is relatively slower to converge when compared with the \texttt{DirSCMS} algorithm. \\
	    \hline
	   & $+$ They are adaptive to (either the spherical or conic) survey geometries, rotationally invariant, and independent of the reference coordinate system. & \multirow{2}{\linewidth}{\centering $-$ They determine the filaments (\emph{i.e.}, estimated density ridges) as a set of discrete realizations; one needs to separate the filaments into individual components and compute their lengths and orientations using other methods.} \\ 
	  \multirow{2}{\linewidth}{\centering Shared by these two algorithms} & $+$ The yielded filaments are independent of any cosmology and can be used to design the cosmological probes. &  \\
	    & $+$ They provide the uncertainty measure for each filamentary point by bootstrap techniques; see Appendix~\ref{Sec:UncertMeasure}. & $-$ They only specify the central spines of the filaments but provide no information about their physical widths. \\ \bottomrule
		\hline
	\end{tabular}
    \end{table*}
	
	We conclude the paper by stating that our proposed \texttt{DirSCMS} algorithm and its estimated cosmic filaments on $\mathbb{S}^2$ provide a feasible way to identify cosmic voids in both the spectroscopic and photometric redshift surveys. As in \cite{Clampitt2015}, one can project the galaxies within a sufficiently large redshift slice onto the 2D celestial sphere and identify the low-density regions surrounded by the estimated filaments on $\mathbb{S}^2$ from our \texttt{DirSCMS} algorithm as cosmic voids. This approach is able to seek out cosmic voids with high confidence as long as the light-of-sight width of the redshift slice is at least twice of the redshift resolution \citep{voidsDES2017}. In addition, the cosmic filament model based on density ridges on $\mathbb{S}^2\times \mathbb{R}$ and the \texttt{DirLinSCMS} algorithm can be generalized to detect the two-dimensional cosmic sheets/walls with little effort (see Figure 4 in \citealt{DirLinProd2021}), though the convergence of our \texttt{DirLinSCMS} algorithm could be slower and the yielded two-dimensional structures would be less salient and robust.

	\section*{Acknowledgement}
	
	We thank Javier Carr\'on Duque for implementing the \texttt{SCMS} algorithm with HEALPix on our cross-shaped filament example and Martin A. Fernandez for the helpful discussion about the \texttt{DisPerSE} software.
	
	Funding for the Sloan Digital Sky 
	Survey IV has been provided by the 
	Alfred P. Sloan Foundation, the U.S. 
	Department of Energy Office of 
	Science, and the Participating 
	Institutions. 
	
	SDSS-IV acknowledges support and 
	resources from the Center for High 
	Performance Computing  at the 
	University of Utah. The SDSS 
	website is \url{www.sdss.org}.
	
	SDSS-IV is managed by the 
	Astrophysical Research Consortium 
	for the Participating Institutions 
	of the SDSS Collaboration including 
	the Brazilian Participation Group, 
	the Carnegie Institution for Science, 
	Carnegie Mellon University, Center for 
	Astrophysics | Harvard \& 
	Smithsonian, the Chilean Participation 
	Group, the French Participation Group, 
	Instituto de Astrof\'isica de 
	Canarias, The Johns Hopkins 
	University, Kavli Institute for the 
	Physics and Mathematics of the 
	Universe (IPMU) / University of 
	Tokyo, the Korean Participation Group, 
	Lawrence Berkeley National Laboratory, 
	Leibniz Institut f\"ur Astrophysik 
	Potsdam (AIP),  Max-Planck-Institut 
	f\"ur Astronomie (MPIA Heidelberg), 
	Max-Planck-Institut f\"ur 
	Astrophysik (MPA Garching), 
	Max-Planck-Institut f\"ur 
	Extraterrestrische Physik (MPE), 
	National Astronomical Observatories of 
	China, New Mexico State University, 
	New York University, University of 
	Notre Dame, Observat\'orio 
	Nacional / MCTI, The Ohio State 
	University, Pennsylvania State 
	University, Shanghai 
	Astronomical Observatory, United 
	Kingdom Participation Group, 
	Universidad Nacional Aut\'onoma 
	de M\'exico, University of Arizona, 
	University of Colorado Boulder, 
	University of Oxford, University of 
	Portsmouth, University of Utah, 
	University of Virginia, University 
	of Washington, University of 
	Wisconsin, Vanderbilt University, 
	and Yale University.
	
	\section*{Data Availability}
	
	The data and code underlying this paper are available at \url{https://github.com/zhangyk8/sconce-scms/tree/main/examples/Theory_Method_Code}, where the SDSS-IV and Illustris-3 simulation data are obtained from their official websites and post-processed according to our descriptions in the paper. 
	

	
	
	\bibliographystyle{mnras}
	\bibliography{DirLinSCMS_met.bib} 

	
	
	
	\appendix
	
	\section{Riemannian Gradient and Hessian}
	\label{App:Riem_terms}
	
	Given a smooth function $f$ on a Riemannian manifold $\mathcal{M}$, where $\mathcal{M}$ is the celestial sphere $\mathbb{S}^2$ or 3D light cone $\mathbb{S}^2 \times \mathbb{R}$ in our cosmic filament detection scenario, its \emph{Riemannian gradient} $\grad f(\bm{x})$ is defined as a unique vector in the tangent space $T_{\bm{x}}$ such that 
	\begin{eqnarray}
	\label{Riem_grad}
	d f_{\bm{x}}(\bm{v}) = \langle \grad f(\bm{x}), \bm{v} \rangle_{\bm{x}}
	\end{eqnarray} 
	for any $\bm{v}\in T_{\bm{x}}$, where $d f_{\bm{x}}(\bm{v})$ is the directional derivative (or more technically, the differential) of $f$ at $\bm{x}\in \mathcal{M}$ along the tangent direction $\bm{v}$ and $\langle\cdot, \cdot \rangle_{\bm{x}}$ is the Riemannian metric. 
	
	The \emph{Riemannian Hessian} $\mathcal{H} f(\bm{x})$ at any point $\bm{x}\in \mathcal{M}$ is the linear map from the tangent space $T_{\bm{x}}$ to itself as:
	\begin{eqnarray}
	\label{Riem_Hess}
	\mathcal{H} f(\bm{x})[\bm{u}] = \bar{\nabla}_{\bm{u}} \grad f(\bm{x})
	\end{eqnarray}
	for any tangent vector $\bm{u}\in T_{\bm{x}}$, where $\bar{\nabla}_{\bm{u}}$ is the Riemannian connection on $\mathcal{M}$. More details about the definitions of Riemannian gradient and Hessian on a general manifold can be found in, e.g., \cite{absil2009optimization} or \cite{boumal2020introduction}.
	
	\subsection{On the Celestial sphere $\mathbb{S}^2$}
	
	While it is mathematically involved to calculate the Riemannian gradient and Hessian directly from the definitions \eqref{Riem_grad} and \eqref{Riem_Hess}, we can view the celestial sphere $\mathbb{S}^2$ as a 2D (Riemannian) submanifold in the ambient Euclidean space $\mathbb{R}^3$. After a proper extension of $f$ from $\mathbb{S}^2$ to its neighborhood in $\mathbb{R}^3$, the Riemannian gradient and Hessian of $f$ on $\mathbb{S}^2$ are related to the regular gradient $\nabla f(\bm{x}) \in \mathbb{R}^3$ and Hessian $\nabla\nabla f(\bm{x}) \in \mathbb{R}^{3\times 3}$ in the ambient Euclidean space $\mathbb{R}^3$ as \citep{absil2009optimization,absil2013extrinsic,DirMS2020}:
	\begin{align}
	\label{Riem_grad_Hess_sph}
	\begin{split}
	&\grad f(\bm{x}) = \left(\bm{I}_3 -\bm{x}\bm{x}^T \right) \nabla f(\bm{x}),\\
	&\mathcal{H} f(\bm{x})= \left(\bm{I}_3 -\bm{x}\bm{x}^T \right) \Big[\nabla\nabla f(\bm{x}) -\bm{x}^T \nabla f(\bm{x})\cdot \bm{I}_3 \Big] \left(\bm{I}_3 -\bm{x}\bm{x}^T \right),
	\end{split}
	\end{align}
	where $\bm{I}_3\in \mathbb{R}^{3\times 3}$ is the identity matrix. Since the eigenvectors $\bm{v}_i(\bm{x}), i=1,2$ of $\mathcal{H}f(\bm{x})$ within the tangent space $T_{\bm{x}}$ are orthogonal to $\bm{x}$ by definition, it follows from \eqref{Riem_grad_Hess_sph} that 
	$$\bm{v}_i(\bm{x})^T \grad f(\bm{x}) = \bm{v}_i(\bm{x})^T \left(\bm{I}_3 -\bm{x}\bm{x}^T \right) \nabla f(\bm{x}) = \bm{v}_i(\bm{x})^T \nabla f(\bm{x})$$
	for $i=1,2$, and we obtain an equivalent definition of the (directional) density ridge of $f$ to \eqref{DirRidges} as:
	\begin{eqnarray}
	\label{DirRidges_eq}
	R(f) = \left\{\bm{x}\in \mathbb{S}^2: \bm{v}_2(\bm{x})^T \nabla f(\bm{x}) = \bm{0}, \lambda_2(\bm{x}) < 0 \right\}.
	\end{eqnarray}
	The definition \eqref{DirRidges} or \eqref{DirRidges_eq} of density ridge $R(f)$ can also be arguably generalized to the sphere with arbitrary dimension and even any smooth density function on a general manifold; see Section 4.1 in \cite{DirSCMS2021}.
	
	\subsection{On the 3D Light Cone $\mathbb{S}^2\times \mathbb{R}$}
	
	Similarly, we regard the 3D light cone $\mathbb{S}^2 \times \mathbb{R}$ as a submanifold of its ambient Euclidean space $\mathbb{R}^4$ and compute the Riemannian gradient and Hessian of $f_{dl}$ (after a proper extension) according to the regular gradient $\nabla f_{dl}(\bm{x},z)\in \mathbb{R}^4$ and $\nabla\nabla f_{dl}(\bm{x},z) \in \mathbb{R}^{4\times 4}$ as:
	\begin{align}
	\label{Riem_grad_hess_cone}
	\begin{split}
	\grad f_{dl}(\bm{x},z) &= \mathcal{P}_{\bm{x}} \nabla f_{dl}(\bm{x},z),\\
	\mathcal{H} f_{dl}(\bm{x},z) &= \mathcal{P}_{\bm{x}} \left[\nabla\nabla f_{dl}(\bm{x},z) - \begin{pmatrix}
	\bm{x}^T \nabla_{\bm{x}} f_{dl}(\bm{x},z) & \bm{0}\\
	\bm{0}^T & 0
	\end{pmatrix} \right] \mathcal{P}_{\bm{x}},
	\end{split}
	\end{align}
	where $\mathcal{P}_{\bm{x}} = \begin{pmatrix}
	\bm{I}_3 -\bm{x}\bm{x}^T & \bm{0}\\
	\bm{0}^T & 1
	\end{pmatrix} \in \mathbb{R}^{4\times 4}$
	is the projection matrix, $\bm{0}=(0,0,0)^T \in \mathbb{R}^3$ is a zero vector, and $\nabla f_{dl}(\bm{x},z)=\begin{pmatrix}
	\nabla_{\bm{x}} f_{dl}(\bm{x})\\
	\nabla_z f_{dl}(\bm{x})
	\end{pmatrix} \in \mathbb{R}^4$. Likewise, given that the columns of $V_{dl}(\bm{x},z)=\left[\bm{v}_{dl,2}(\bm{x},z), \bm{v}_{dl,3}(\bm{x},z) \right] \in \mathbb{R}^{4\times 2}$ derived from the eigenvectors of $\mathcal{H} f_{dl}(\bm{x},z)$ within the tangent space $T_{(\bm{x},z)}$ of $\mathbb{S}^2\times \mathbb{R}$ are orthogonal to $(\bm{x},z)$ by definition, we have that
	$$V_{dl}(\bm{x},z)^T \grad f_{dl}(\bm{x},z)  = V_{dl}(\bm{x},z)^T \nabla f_{dl}(\bm{x},z)$$
	and an equivalent form for the density ridge \eqref{DL_ridge} on $\mathbb{S}^2\times \mathbb{R}$ as:
	\begin{align}
	\label{DL_ridge_eq}
	\begin{split}
	R(f_{dl}) = \Big\{(\bm{x},z) \in \mathbb{S}^2\times \mathbb{R}: & V_{dl}(\bm{x},z)^T \nabla f_{dl}(\bm{x},z) = \bm{0},\\ 
	&\lambda_{dl,2}(\bm{x},z)<0\Big\}.
	\end{split}
	\end{align}

	\section{Detailed Descriptions of Our Proposed Algorithms in {\tt SCONCE}}
	
	In \autoref{Sec:SCMS_sph} and \autoref{Sec:SCMS_cone}, we have introduced our extended \texttt{SCMS} algorithms on $\mathbb{S}^2$ and $\mathbb{S}^2\times \mathbb{R}$, especially their key iterative formulae \eqref{DirSCMS_iter} and \eqref{DirLinSCMS}. Here, we provide additional details on their implementations in practice, including the computations of estimated Riemannian Hessian matrices and stopping criteria. Additionally, we discuss how to incorporate stellar properties into the filament detection process with our \texttt{DirSCMS} algorithm in Appendix~\ref{Sec:GenDirEst} and delineate the algorithmic procedures for seeking out the cosmic nodes in Appendix~\ref{App:mode_seek}.
	
	\subsection{{\tt DirSCMS} Algorithm on $\mathbb{S}^2$}
	\label{App:DirSCMS_detail}
	
	\begin{algorithm*}[!ht]
		\caption{Cosmic Filament Detection on the Celestial Sphere $\mathbb{S}^2$ via the \texttt{DirSCMS} Algorithm}
		\label{Algo:Dir_SCMS}
		\begin{algorithmic}
			\State \textbf{Input}: 
			\begin{itemize}
				\item The data sample with the Cartesian coordinate $\mathcal{D}=\left\{\bm{X}_1,...,\bm{X}_n \right\}$ on $\mathbb{S}^2$.
				\item The smoothing bandwidth $b>0$, threshold value $\tau>0$, precision level $\epsilon >0$ for stopping the algorithm, and a suitable mesh $\mathcal{M}_D \subset \mathbb{S}^2$ of initial points covering the region of interest. 
			\end{itemize} 
			\State \textbf{Step 1}: Compute the estimated density function $\hat{f}(\bm{x})=\frac{C_L(b)}{n} \sum\limits_{i=1}^n L\left(\frac{1-\bm{x}^T \bm{X}_i}{b^2} \right)$ or more generally, via \eqref{GenDirKDE}, on the data sample $\mathcal{D}$.
			\State \textbf{Step 2}: Remove $\bm{X}'\in \mathcal{D}$ whenever $\hat{f}(\bm{X}') < \tau$. Denote the remaining data sample by $\tilde{\mathcal{D}}$.
			\State \textbf{Step 3}: For each $\bm{x}^{(0)}\in \mathcal{M}_D$, iterate the following procedure until convergence:
			\While {$\norm{\hat{\bm{v}}_2(\bm{x}^{(t)})^T \nabla \hat{f}(\bm{x}^{(t)})}_2 > \epsilon$}: (At iteration step $t$)
			\State \textbf{Step 3-1}: Compute the estimated Riemannian Hessian matrix $\mathcal{H} \hat{f}(\bm{x}^{(t)})$ as in \eqref{Riem_Hess_est} or more generally, \eqref{Riem_Hess_wt_est}.
			\State \textbf{Step 3-2}: Compute $\hat{\bm{v}}_2(\bm{x}^{(t)})$ as the last eigenvector of $\mathcal{H} \hat{f}(\bm{x}^{(t)})$ inside the tangent space $T_{\bm{x}^{(t)}}$ of $\mathbb{S}^2$ via the spectral decomposition.
			\State \textbf{Step 3-3}: Update $\tilde{\bm{x}}^{(t+1)} \gets \bm{x}^{(t)} - \hat{\bm{v}}_2(\bm{x}^{(t)}) \hat{\bm{v}}_2(\bm{x}^{(t)})^T  \left[\frac{\sum_{i=1}^n \bm{X}_i L'\left(\frac{1-\bm{X}_i^T \bm{x}^{(t)}}{b^2} \right)}{\norm{\sum_{i=1}^n \bm{X}_i L'\left(\frac{1-\bm{X}_i^T \bm{x}^{(t)}}{b^2} \right)}}_2 \right]$ or more generally, via \eqref{DirSCMS_wt}.
			\State \textbf{Step 3-4}: Standardize $\tilde{\bm{x}}^{(t+1)}$ as $\bm{x}^{(t+1)} \gets \frac{\tilde{\bm{x}}^{(t+1)}}{\norm{\tilde{\bm{x}}^{(t+1)}}_2}$.
			\EndWhile
			\State \textbf{Output}: The estimated filament set (or density ridge) $R(\hat{f})$ on $\mathbb{S}^2$ represented by the collection of converged points.
		\end{algorithmic}
	\end{algorithm*}
	
	The implementation of our \texttt{DirSCMS} algorithm requires the calculations of the estimated Riemannian Hessian $\mathcal{H} \hat{f}(\bm{x})$ from the directional KDE \eqref{DirKDE} and its eigenvectors. Based on \eqref{Riem_grad_Hess_sph}, the estimated Riemannian Hessian $\mathcal{H} \hat{f}(\bm{x})$ has its formula as:
	\begin{align}
	\label{Riem_Hess_est}
	\begin{split}
	\mathcal{H} \hat{f}(\bm{x}) 
	&= \frac{C_L(b)}{nb^2}\left(\bm{I}_3 -\bm{x}\bm{x}^T \right) \Bigg[\frac{1}{b^2}\sum_{i=1}^n \bm{X}_i\bm{X}_i^T\cdot L''\left(\frac{1-\bm{X}_i^T \bm{x}}{b^2} \right) \\
	& \quad + \sum_{i=1}^n \bm{X}_i^T \bm{x} \cdot \bm{I}_3\cdot L'\left(\frac{1-\bm{X}_i^T \bm{x}}{b^2} \right) \Bigg] \left(\bm{I}_3 -\bm{x}\bm{x}^T \right),
	\end{split}
	\end{align}
	and we compute its last eigenvector $\hat{\bm{v}}_2(\bm{x})$ within the tangent space $T_{\bm{x}}$ of $\mathbb{S}^2$ via the spectral decomposition \citep{HJ2012}. We run the \texttt{DirSCMS} iterative formula \eqref{DirSCMS_iter} in \autoref{Sec:DirSCMS} or its general form \eqref{DirSCMS_wt} in Appendix~\ref{Sec:GenDirEst} until
	$$\norm{\hat{\bm{v}}_2(\bm{x})^T \grad \hat{f}(\bm{x})}_2 = \norm{\hat{\bm{v}}_2(\bm{x})^T \nabla \hat{f}(\bm{x})}_2 \leq \epsilon,$$
	where $\epsilon >0$ is a precision level for stopping the algorithm. Notice also that we leverage the normalized gradient 
	$$\frac{\nabla \hat{f}(\bm{x}^{(t)})}{\norm{\nabla \hat{f}(\bm{x}^{(t)})}_2}=-\frac{\sum_{i=1}^n \bm{X}_i L'\left(\frac{1-\bm{X}_i^T \bm{x}^{(t)}}{b^2} \right)}{\norm{\sum_{i=1}^n \bm{X}_i L'\left(\frac{1-\bm{X}_i^T \bm{x}^{(t)}}{b^2} \right)}}_2$$ 
	in our design of the \texttt{DirSCMS} iteration \eqref{DirSCMS_iter} instead of the standard gradient $\nabla \hat{f}(\bm{x}^{(t)})$ in pursuit of a faster convergence among low-density regions \citep{DirSCMS2021}. 
	
	We summarize the full filament detection procedures on $\mathbb{S}^2$ in Algorithm~\ref{Algo:Dir_SCMS}. Given that the observational data always contain a lot of noise, we incorporate an extra denoising step before applying the \texttt{DirSCMS} iteration \eqref{DirSCMS_iter} in our filament detection procedures on $\mathbb{S}^2$; see Step 2 in Algorithm~\ref{Algo:Dir_SCMS}. One can follow the discussion in Appendix A of \cite{Chen2015methods} to select the denoising threshold $\tau$. In our applications of the \texttt{DirSCMS} algorithm, we use the von Mises kernel, leading to the equation $L''(r)=-L'(r)=L(r)=e^{-r}$.
	
	\subsection{Weighted {\tt DirSCMS} Algorithm on $\mathbb{S}^2$}
	\label{Sec:GenDirEst}
	
	In statistical parlance, an astronomical survey can be understood as a marked point process, where the marks are represented by stellar properties of the observed galaxies or simulated dark matter halos \citep{SDSS-IV2020,Vogelsberger2020Cosmological}. These extra properties of the astronomical objects are not only useful in analyzing its intrinsic evolution but also facilitate the estimation of the underlying matter distribution and its filamentary structures. For instance, \cite{Outskirts2020} has demonstrated by mock galaxy samples that the filaments can be better recovered by the mass-weighted galaxies. Here, we describe how to generalize our proposed \texttt{DirSCMS} algorithm to incorporate a given property of interest into the filament estimation on $\mathbb{S}^2$.

	\subsubsection{Methodology}

	Assume that the data are $(\bm{X}_1,Y_1),...,(\bm{X}_n,Y_n)$ sampled from a (directional-linear) density function $f_{dl}(\bm{x},y)$ on $\mathbb{S}^2 \times \mathbb{R}$, where $\bm{X}_i, i=1,...,n$ are Cartesian coordinates of their locations on $\mathbb{S}^2$ and $Y_i,i=1,...,n$ are their associated stellar properties of interest. Instead of directly estimating the density function $f_{dl}$, which has been discussed in \autoref{Sec:SCMS_cone}, we utilize $Y_i,i=1,...,n$ as the weights of the directional KDE \eqref{DirKDE} to construct an estimator as:
	\begin{eqnarray}
	\label{GenDirKDE}
	\hat{f}_g(\bm{x}) = C_L(b) \sum_{i=1}^n Y_i \cdot L\left(\frac{1-\bm{x}^T \bm{X}_i}{b^2} \right),
	\end{eqnarray}
	where $L(\cdot)$ and $b$ are the directional kernel function and smoothing bandwidth parameter, respectively. This weighted estimator better approximates the matter density function (up to a scaled factor) by, e.g., placing more weights on the heavier objects. The intended estimand of \eqref{GenDirKDE}, however, is no longer a probability density function but a \emph{generalized density function} on $\mathbb{S}^2$ as \citep{chen2014Generalized}:
	\begin{eqnarray}
	\label{GenDen}
	f_g(\bm{x}) = \int_{\mathbb{R}} y\cdot f_{dl}(\bm{x},y) dy = f(\bm{x}) \cdot \mathbb{E}\left(Y_i|\bm{X}_i=\bm{x}\right),
	\end{eqnarray}
	where $f(\bm{x})$ is the marginal density function of $\left\{\bm{X}_i\right\}_{i=1}^n$ on $\mathbb{S}^2$ and $\mathbb{E}\left(Y_i|\bm{X}_i=\bm{x}\right)$ is the conditional expectation of $Y_i$ given $\bm{X}_i=\bm{x}$. This generalized density function $f_g$ is also known as the intensity function in the domain of spatial (marked) point process \citep{PoissonProcesses1993}. The density ridge or cosmic filament model based on $f_g$ is defined similarly as:
	\begin{eqnarray}
	\label{GenDirRidges}
	R(f_g) = \left\{\bm{x}\in \mathbb{S}^2: \bm{v}_{g,2}(\bm{x})^T \grad f_g(\bm{x}) = \bm{0}, \lambda_{g,2}(\bm{x}) < 0 \right\}
	\end{eqnarray}
	and can be also estimated by the plug-in estimator $R(\hat{f}_g)$, where the Riemannian Hessian $\mathcal{H} f_g(\bm{x})$ has its eigen-system as $\left\{\left(\bm{v}_{g,i}(\bm{x}), \lambda_{g,i}(\bm{x}) \right) \text{ for } i=1,2 \text{ with } \lambda_{g,1}(\bm{x}) \geq \lambda_{g,2}(\bm{x}) \right\}$ in the tangent space $T_{\bm{x}}$ at $\bm{x}\in \mathbb{S}^2$. Identifying the (estimated) density ridges of the generalized density function is analogous to our \texttt{DirSCMS} algorithm in \autoref{Sec:DirSCMS} but with a more general \texttt{SCMS} iteration as:
	\begin{align}
	\label{DirSCMS_wt}
	\begin{split}
	\tilde{\bm{x}}^{(t+1)} &\gets \bm{x}^{(t)}\\
	& \quad - \hat{\bm{v}}_{g,2}(\bm{x}^{(t)}) \hat{\bm{v}}_{g,2}(\bm{x}^{(t)})^T  \left[\frac{\sum_{i=1}^n Y_i\bm{X}_i L'\left(\frac{1-\bm{X}_i^T \bm{x}^{(t)}}{b^2} \right)}{\norm{\sum_{i=1}^n Y_i\bm{X}_i L'\left(\frac{1-\bm{X}_i^T \bm{x}^{(t)}}{b^2} \right)}}_2 \right]
	\end{split}
	\end{align}
	and $\bm{x}^{(t+1)} \gets \frac{\tilde{\bm{x}}^{(t+1)}}{\norm{\tilde{\bm{x}}^{(t+1)}}_2}$, where $\hat{\bm{v}}_{g,2}(\bm{x})$ is the last eigenvector of the estimated Riemannian Hessian 
	\begin{align}
	\label{Riem_Hess_wt_est}
	\begin{split}
	\mathcal{H} \hat{f}_g(\bm{x}) 
	&= \frac{C_L(b)}{b^2}\left(\bm{I}_3 -\bm{x}\bm{x}^T \right) \Bigg[\frac{1}{b^2}\sum_{i=1}^n Y_i\bm{X}_i\bm{X}_i^T\cdot L''\left(\frac{1-\bm{X}_i^T \bm{x}}{b^2} \right) \\
	& \quad + \sum_{i=1}^n Y_i\bm{X}_i^T \bm{x} \cdot \bm{I}_3\cdot L'\left(\frac{1-\bm{X}_i^T \bm{x}}{b^2} \right) \Bigg] \left(\bm{I}_3 -\bm{x}\bm{x}^T \right)
	\end{split}
	\end{align}
	inside the tangent space $T_{\bm{x}}$. Notice that the weighted \texttt{DirSCMS} algorithm \eqref{DirSCMS_wt} reduces to the usual \texttt{DirSCMS} algorithm \eqref{DirSCMS_iter} when all the weights $Y_1,...,Y_n$ are equal.

	\begin{figure*}
		\captionsetup[subfigure]{justification=centering}
		\centering
		\begin{subfigure}[t]{.49\textwidth}
			\centering
			\includegraphics[width=\linewidth]{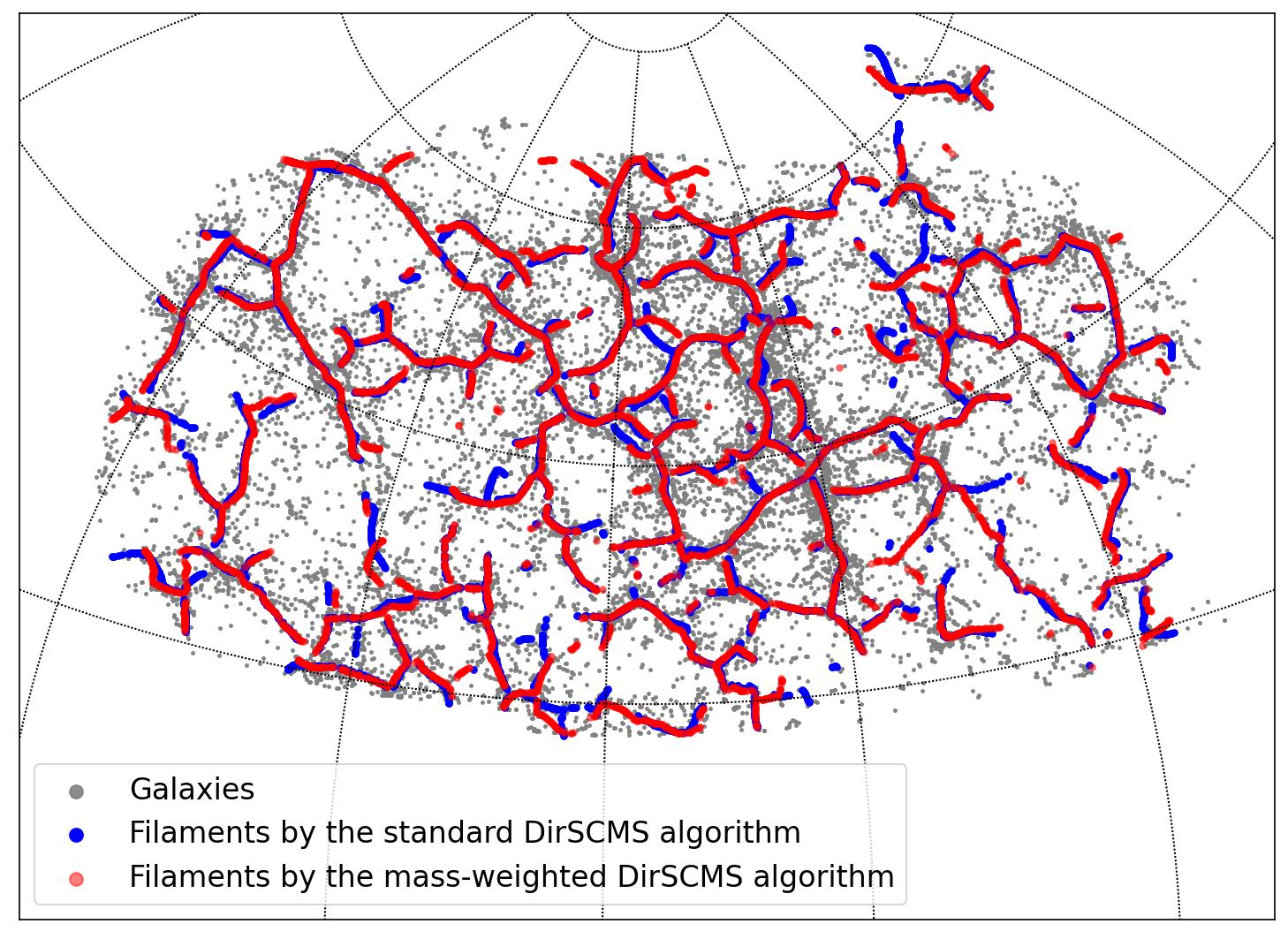}
		\end{subfigure}
		\hfil
		\begin{subfigure}[t]{.49\textwidth}
			\centering
			\includegraphics[width=\linewidth]{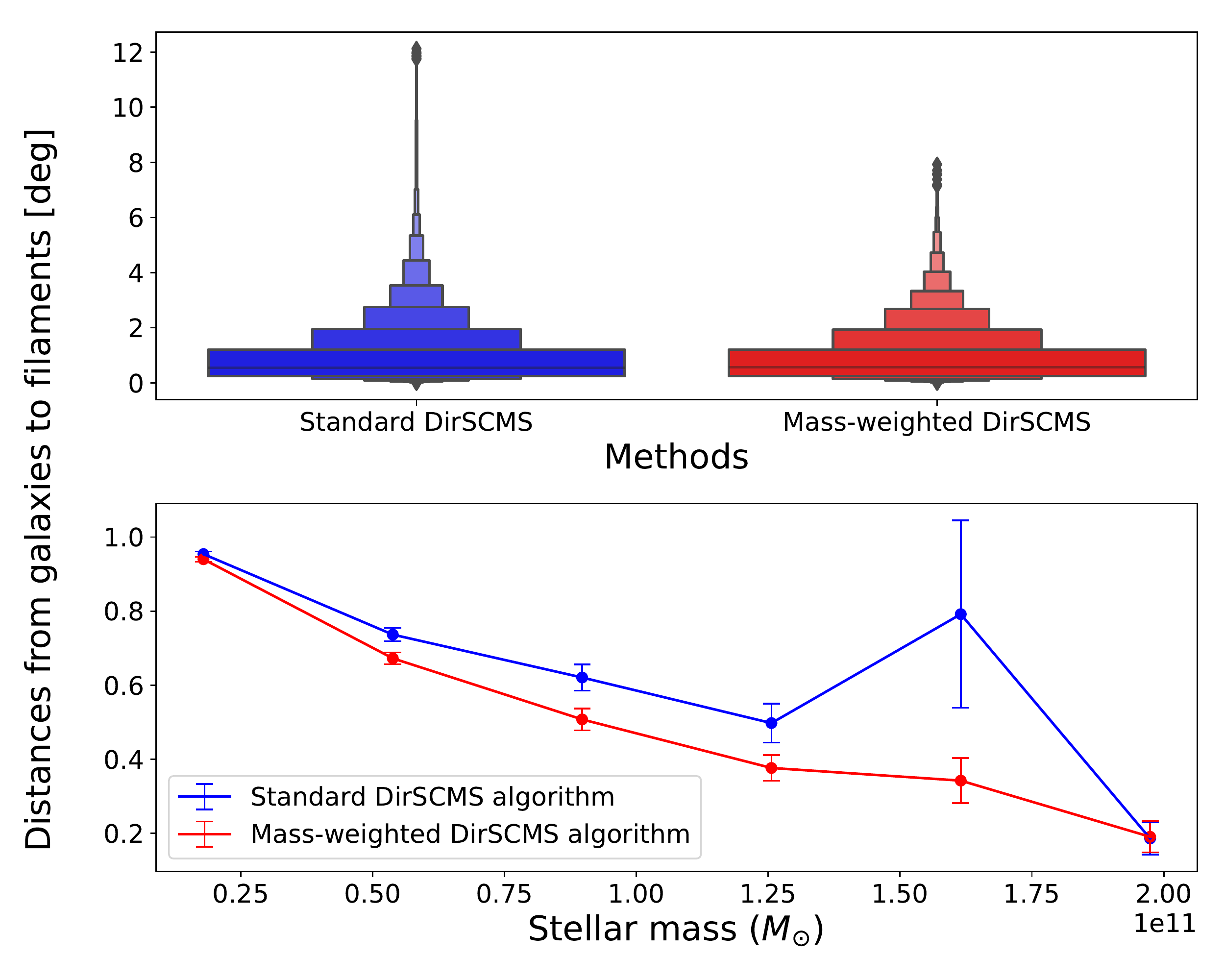}
		\end{subfigure}
		\caption{Comparisons of cosmic filaments detected by the usual \texttt{DirSCMS} (Algorithm~\ref{Algo:Dir_SCMS}) and mass-weighted \texttt{DirSCMS} (Section~\ref{Sec:GenDirEst}) algorithms on a thin redshift slice SDSS-IV galactic data ($0.06\leq z < 0.065$). \emph{Left}: The filaments detected by our \texttt{DirSCMS} algorithm on galaxies with and without the information of estimated stellar masses. \emph{Top Right}: The letter-value plots of the distances from galaxies to the two sets of filaments. \emph{Bottom Right}: The correlation between stellar masses of galaxies and their distances to the two sets of filaments on $\mathbb{S}^2$.}
		\label{fig:fila_mass_wt}
	\end{figure*} 
	
	\subsubsection{Comparison Between the Unweighted and Mass-Weighted {\tt DirSCMS} Algorithms on SDSS-IV Data}
	\label{Sec:std_mt_DirSCMS_comp}	
	
	Recent studies on simulation and survey data suggest that the galaxies tend to be redder and heavier in their stellar masses when they are closer to filaments \citep{Alpaslan2016,Malavasi2017VIMOS,chen2017detectSDSSIII}. While other stellar properties of galaxies, such as the luminosity and star formation rate, are also correlated with their proximity to the nearby filaments, we showcase how the stellar mass properties can influence the cosmic filament detection with our \texttt{DirSCMS} algorithm as an example here because the stellar mass property and its estimation function serve as a key measure for other stellar properties \citep{Weigel2016}. 
	
	To this end, we take another thin redshift slice $0.06\leq z < 0.065$ from the SDSS-IV galactic data described in \autoref{Sec:SDSS_data}. The stellar masses of selected galaxies are obtained from SDSS spectra by the FIREFLY (Fitting IteRativEly For Likelihood analYsis) stellar population model \citep{Firefly2017,Comparat2017}. This model estimates the stellar properties by leveraging the \cite{Chabrier2003} stellar initial mass function under the input stellar library MILES \citep{MILES2011}. We access the estimated stellar masses from the FIREFLY value added catalog\footnote{The data can be downloaded at \url{https://www.sdss.org/dr17/spectro/eboss-firefly-value-added-catalog}.}, among which a tiny portion ($\sim 6.36\%$) of stellar masses are missing in our selected redshift slice. For simplicity, we subset those galaxies with non-missing estimated stellar masses by the FIREFLY model in the North Galactic Cap ($100^{\circ} < \text{RA} < 270^{\circ}, -5^{\circ} < \text{DEC} < 70^{\circ}$), leading to a galactic dataset on $\mathbb{S}^2$ with 25,207 observations.
	
	Both the usual \texttt{DirSCMS} method (Algorithm~\ref{Algo:Dir_SCMS} with an equal weight for each observation) and mass-weighted \texttt{DirSCMS} algorithm are applied to the above galactic dataset on $\mathbb{S}^2$, whose bandwidth parameter is chosen using \eqref{bw_Dir} with $B_0=0.25$. The resulting filaments are presented in Figure~\ref{fig:fila_mass_wt}, in which they do exhibit non-negligible differences in terms of their structures and distances from the galaxies. On the one hand, the galaxies within the selected redshift slice are, on average, closer to the filaments detected by the mass-weighted \texttt{DirSCMS} algorithm than the standard one. Specifically, the galaxies within mass groups from $5\times 10^{10} M_{\odot}$ to $1.6 \times 10^{11} M_{\odot}$ are statistically closer to the mass-weighted filaments than its equally weighted counterparts. It provides some additional supporting evidence based on survey data other than simulation data in \cite{Outskirts2020} that it is beneficial to detect cosmic filaments with mass-weighted galaxies. Nevertheless, estimating the stellar mass properties is never easy on observed galaxies, so it is still more common to recover the filaments by assuming an equal weight for each galaxy on survey data. Furthermore, we observe from the decreasing trend in the bottom right panel of Figure~\ref{fig:fila_mass_wt} that the galaxies around the two filaments tend to be more massive than the ones that are farther away from the filaments. This result, to some extent, supports the previous claim about the correlation between the stellar mass and proximity to filaments.

	\subsection{{\tt DirLinSCMS} Algorithm on $\mathbb{S}^2\times \mathbb{R}$}
	\label{App:DirLinSCMS_detail}
	
	The full procedures of detecting cosmic filaments in $\mathbb{S}^2\times \mathbb{R}$ via our \texttt{DirLinSCMS} algorithm is essentially identical to Algorithm~\ref{Algo:Dir_SCMS}, except that we estimate the density function via \eqref{DLKDE} and compute its Riemannian Hessian $\mathcal{H}\hat{f}_{dl}(\bm{x},z)$ as:
	\begin{eqnarray}
	\label{Riem_Hess_est_DL}
	\mathcal{H} \hat{f}_{dl}(\bm{x},z) = \begin{pmatrix}
	\mathcal{H}_{\bm{x}\bm{x}} \hat{f}_{dl}(\bm{x},z) & \mathcal{H}_{\bm{x}z} \hat{f}_{dl}(\bm{x},z)\\
	\left[\mathcal{H}_{\bm{x}z} \hat{f}_{dl}(\bm{x},z)\right]^T & \mathcal{H}_{zz} \hat{f}_{dl}(\bm{x},z)
	\end{pmatrix} \in \mathbb{R}^{4\times 4},
	\end{eqnarray}
	where 
	\begin{align*}
	&\mathcal{H}_{\bm{x}\bm{x}} \hat{f}_{dl}(\bm{x},z) \\
	&= \frac{C_L(b_1)}{nb_1^2b_2}\left(\bm{I}_3 -\bm{x}\bm{x}^T \right) \sum_{i=1}^n\Bigg[ \frac{\bm{X}_i\bm{X}_i^T}{b_1^2} L''\left(\frac{1-\bm{X}_i^T \bm{x}}{b_1^2} \right) K\left(\norm{\frac{z-Z_i}{b_2}}_2^2 \right) \\
	& \quad + \sum_{i=1}^n \bm{x}^T\bm{X}_i\cdot \bm{I}_3\cdot L'\left(\frac{1-\bm{X}_i^T \bm{x}}{b_1^2} \right) K\left(\norm{\frac{z-Z_i}{b_2}}_2^2 \right) \Bigg] \left(\bm{I}_3 -\bm{x}\bm{x}^T \right),\\
	&\mathcal{H}_{\bm{x}z} \hat{f}_{dl}(\bm{x},z) \\
	&= -\frac{2C_L(b_1)}{nb_1^2b_2^3} \sum_{i=1}^n (z-Z_i)\bm{X}_i \cdot L'\left(\frac{1-\bm{X}_i^T \bm{x}}{b_1^2} \right) K'\left(\norm{\frac{z-Z_i}{b_2}}_2^2 \right),\\
	&\mathcal{H}_{\bm{x}z} \hat{f}_{dl}(z,z) = \frac{2C_L(b_1)}{nb_2^3}\sum_{i=1}^n\Bigg[ L\left(\frac{1-\bm{X}_i^T \bm{x}}{b_1^2} \right) K'\left(\norm{\frac{z-Z_i}{b_2}}_2^2 \right) \\
	& \hspace{17mm} + \sum_{i=1}^n \frac{2(z-Z_i)^2}{b_2^2}\cdot L\left(\frac{1-\bm{X}_i^T \bm{x}}{b_1^2} \right) K''\left(\norm{\frac{z-Z_i}{b_2}}_2^2 \right) \Bigg].
	\end{align*}
	Steps 3-3 and 3-4 in Algorithm~\ref{Algo:Dir_SCMS} will be replaced by \eqref{DirLinSCMS} as well.
	
	Different from our \texttt{DirSCMS} algorithm on the celestial sphere $\mathbb{S}^2$, which is controlled by a single bandwidth parameter, the performance of our \texttt{DirLinSCMS} algorithm on the 3D light cone $\mathbb{S}^2 \times \mathbb{R}$ relies on the proper tuning of bandwidth parameters for both the directional and linear data components. Here, we demonstrate how the estimated filaments by our \texttt{DirLinSCMS} algorithm change under different combinations of the directional and linear bandwidth parameters $b_1,b_2$ through the Illustris-3 FoF halo data at redshift $z=0$ in \autoref{Sec:illustris}. 
	
\begin{figure*}
		\captionsetup[subfigure]{justification=centering}
		\centering
		\begin{subfigure}[t]{.135\textwidth}
			\centering
			\includegraphics[width=\linewidth]{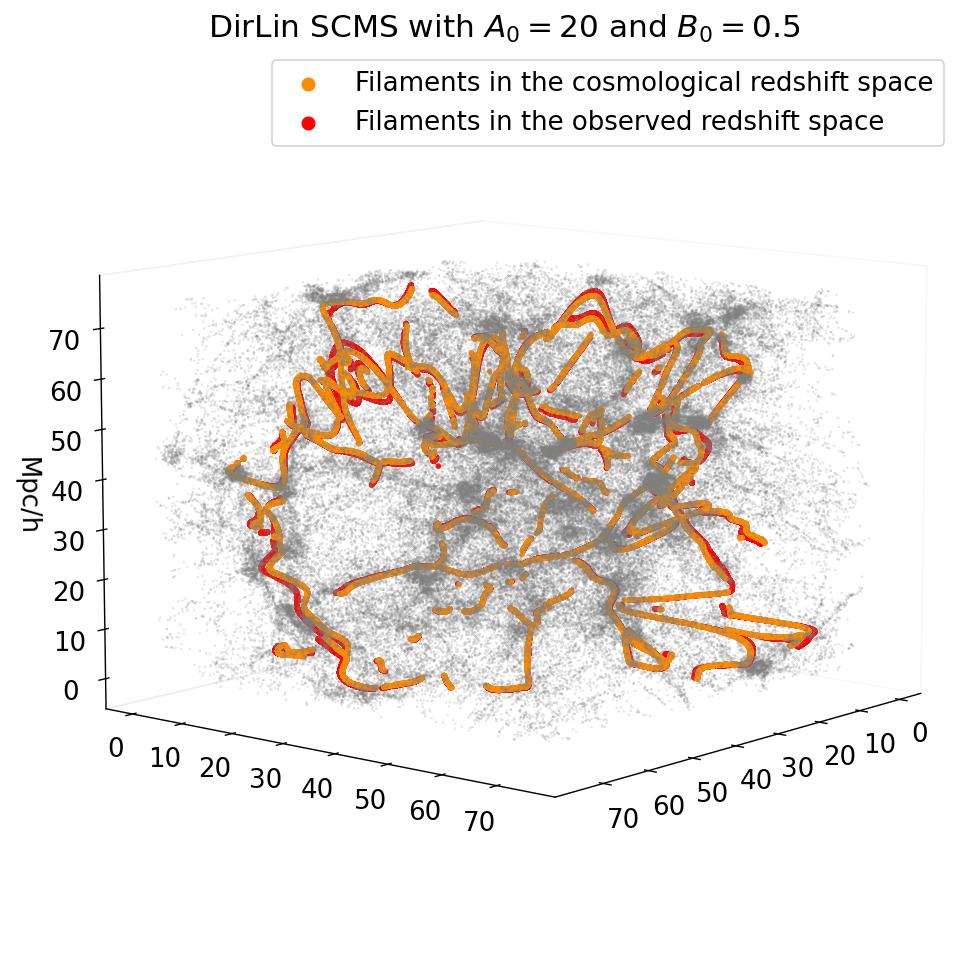}
		\end{subfigure}
		\hfil
		\begin{subfigure}[t]{.135\textwidth}
			\centering
			\includegraphics[width=\linewidth]{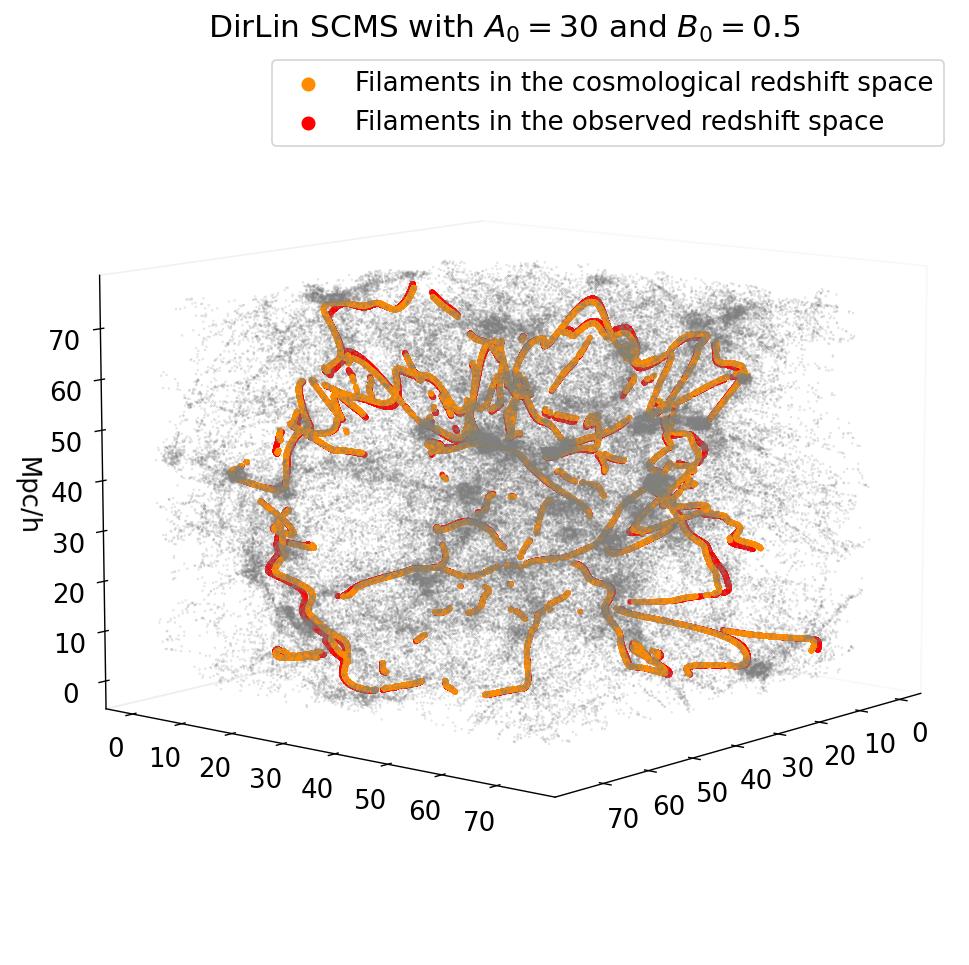}
		\end{subfigure}
		\hfil
		\begin{subfigure}[t]{.135\textwidth}
			\centering
			\includegraphics[width=\linewidth]{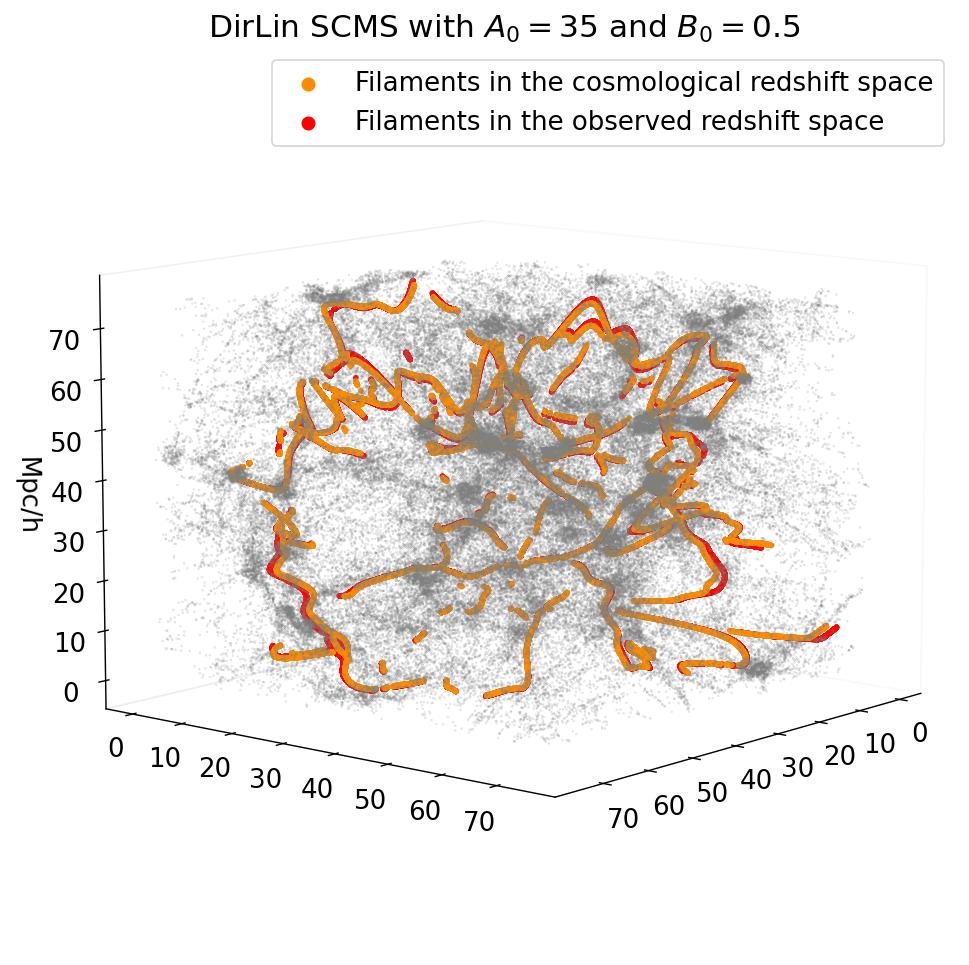}
		\end{subfigure}
		\hfil
		\begin{subfigure}[t]{.135\textwidth}
			\centering
			\includegraphics[width=\linewidth]{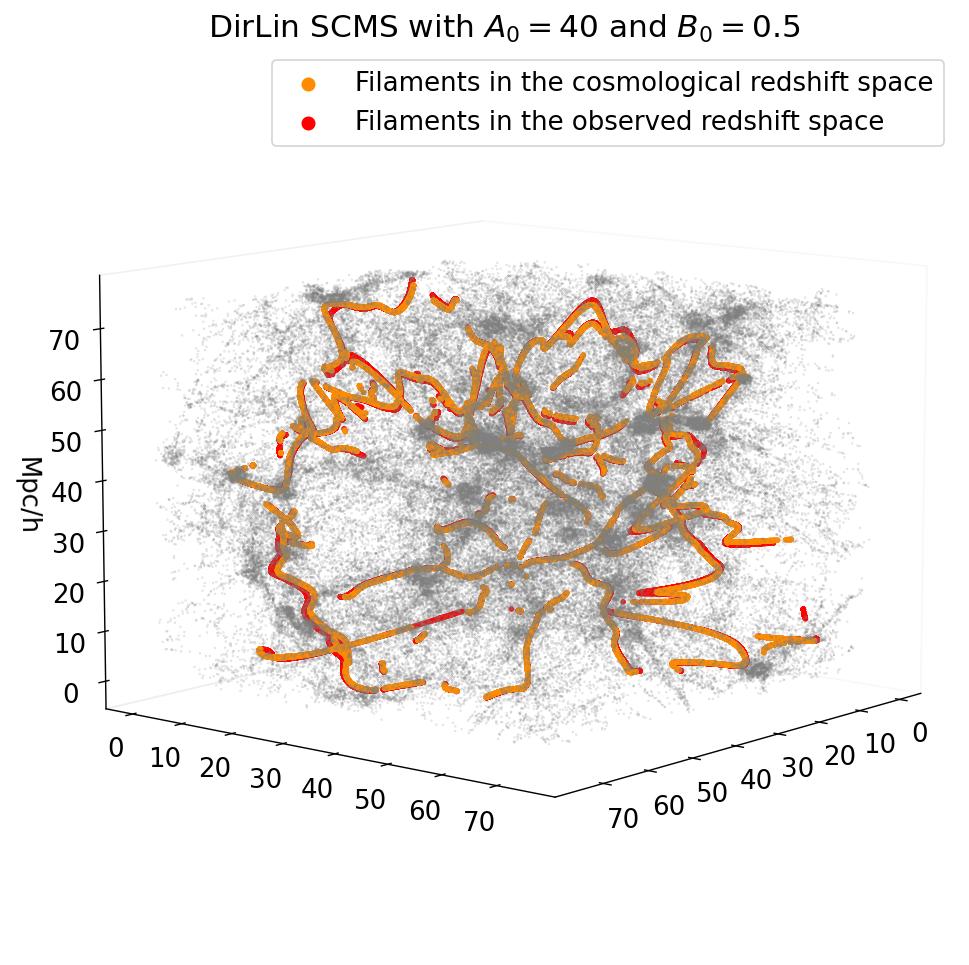}
		\end{subfigure}
		\hfil
		\begin{subfigure}[t]{.135\textwidth}
			\centering
			\includegraphics[width=\linewidth]{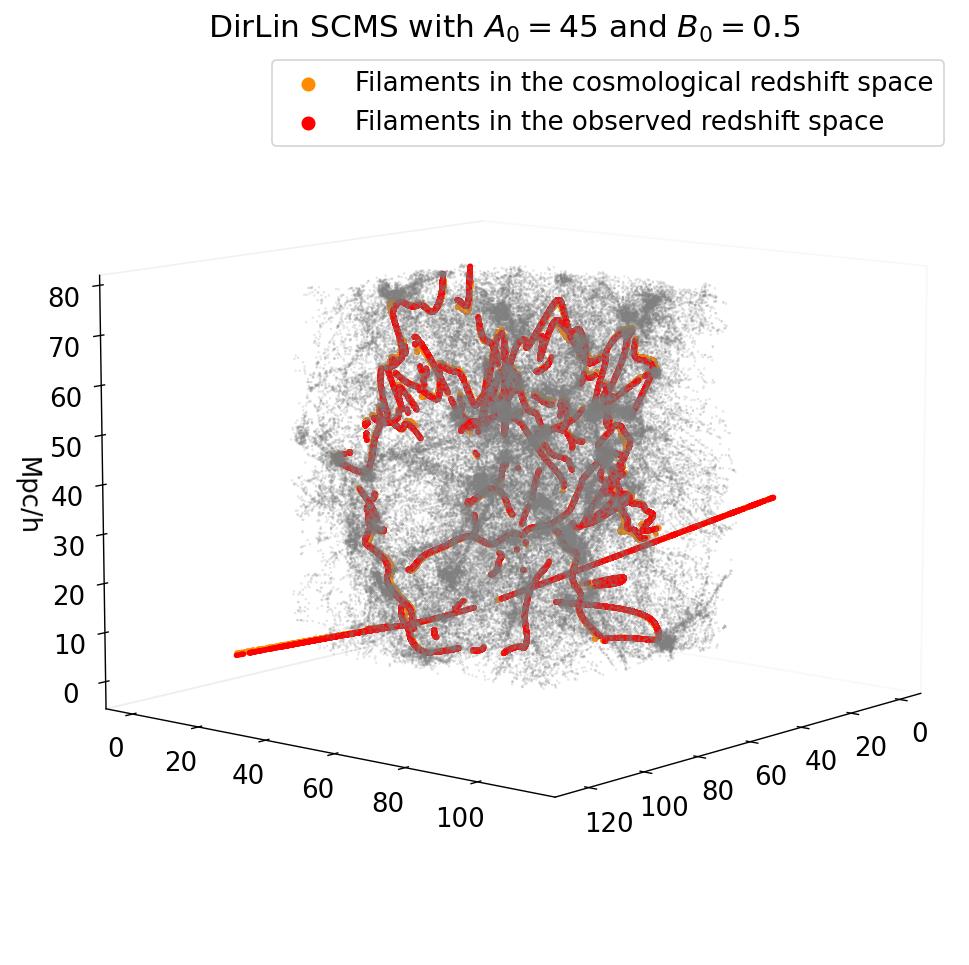}
		\end{subfigure}
		\hfil
		\begin{subfigure}[t]{.135\textwidth}
			\centering
			\includegraphics[width=\linewidth]{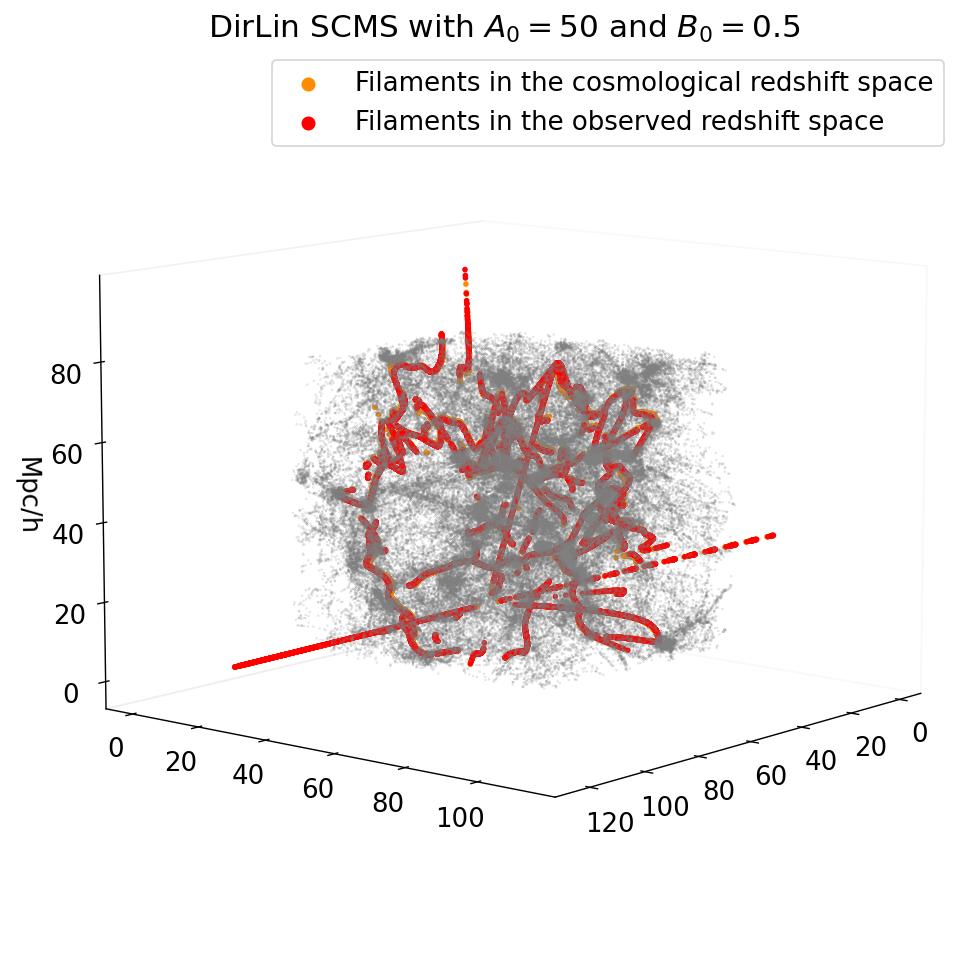}
		\end{subfigure}
		\hfil
		\begin{subfigure}[t]{.135\textwidth}
			\centering
			\includegraphics[width=\linewidth]{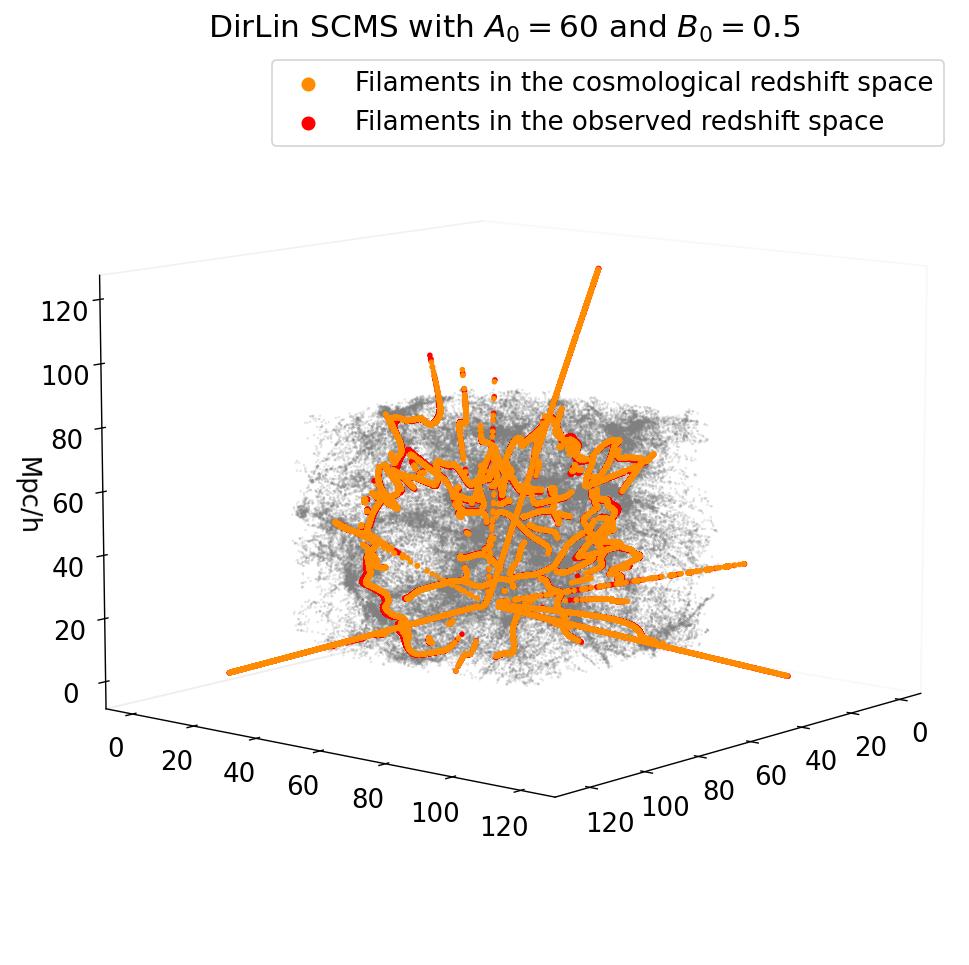}
		\end{subfigure}
		
		\begin{subfigure}[t]{.135\textwidth}
			\centering
			\includegraphics[width=\linewidth]{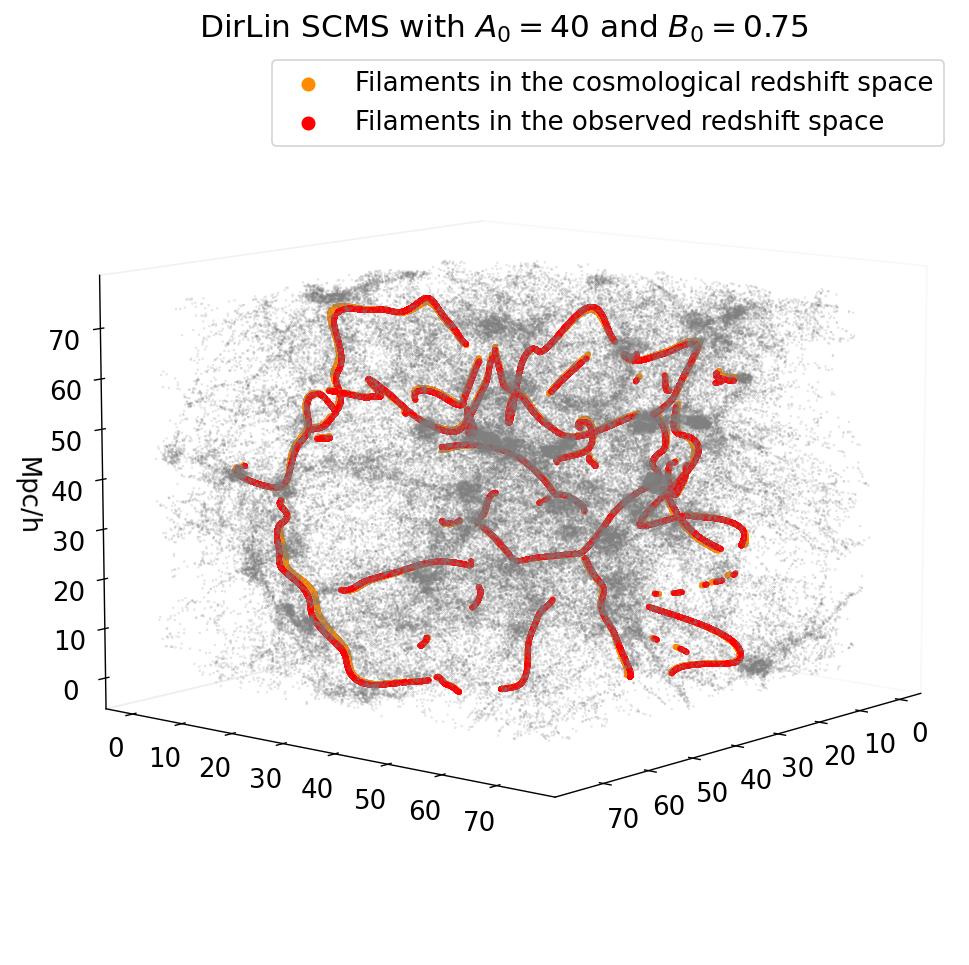}
		\end{subfigure}
		\hfil
		\begin{subfigure}[t]{.135\textwidth}
			\centering
			\includegraphics[width=\linewidth]{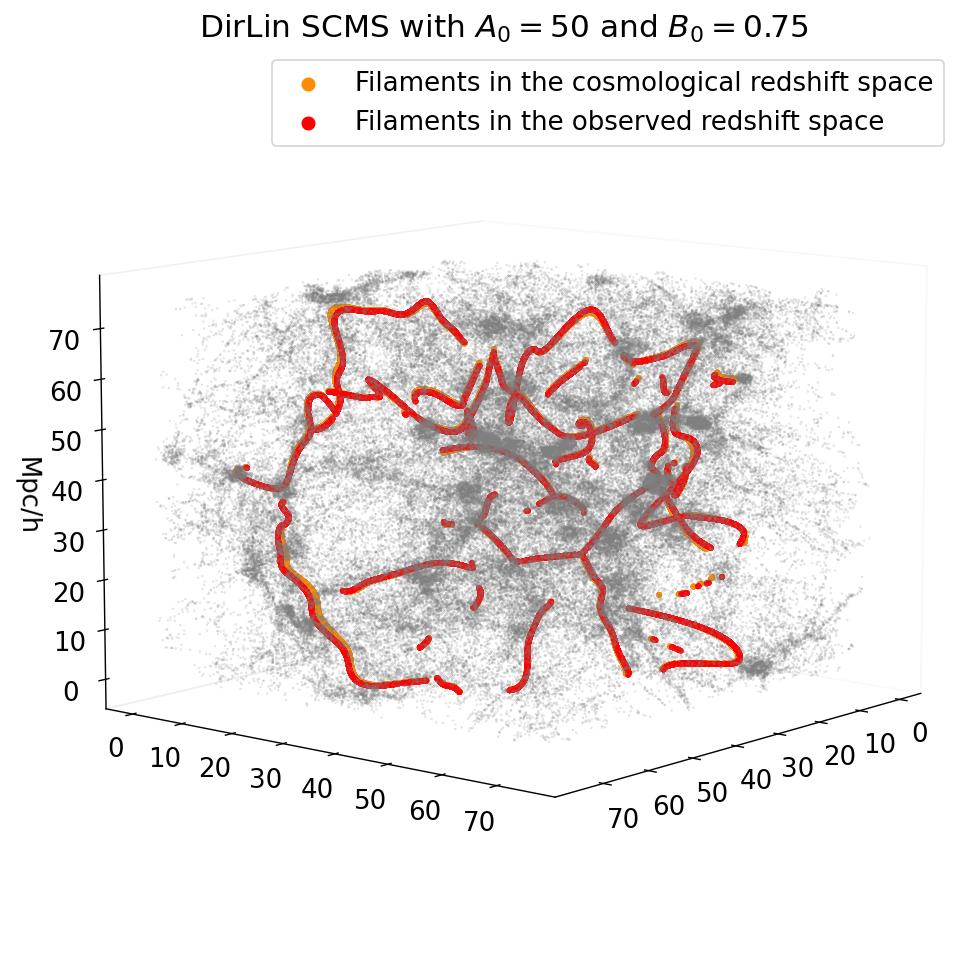}
		\end{subfigure}
		\hfil
		\begin{subfigure}[t]{.135\textwidth}
			\centering
			\includegraphics[width=\linewidth]{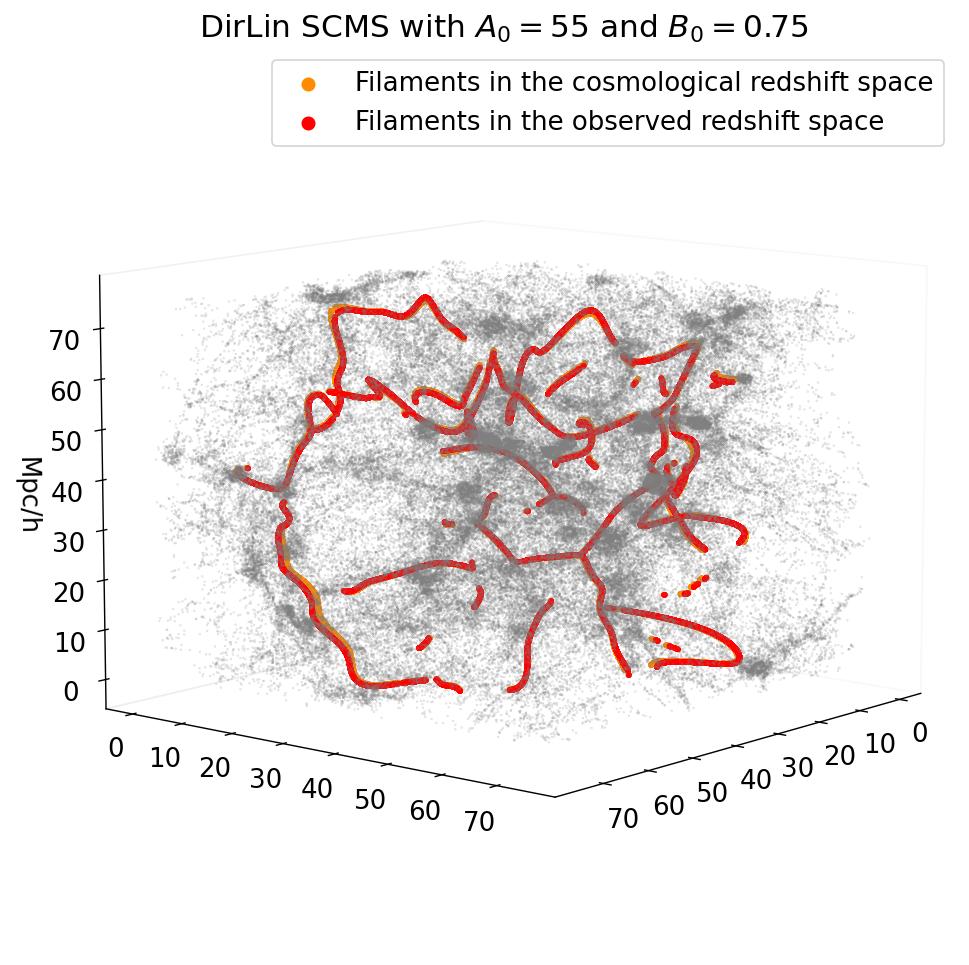}
		\end{subfigure}
		\hfil
		\begin{subfigure}[t]{.135\textwidth}
			\centering
			\includegraphics[width=\linewidth]{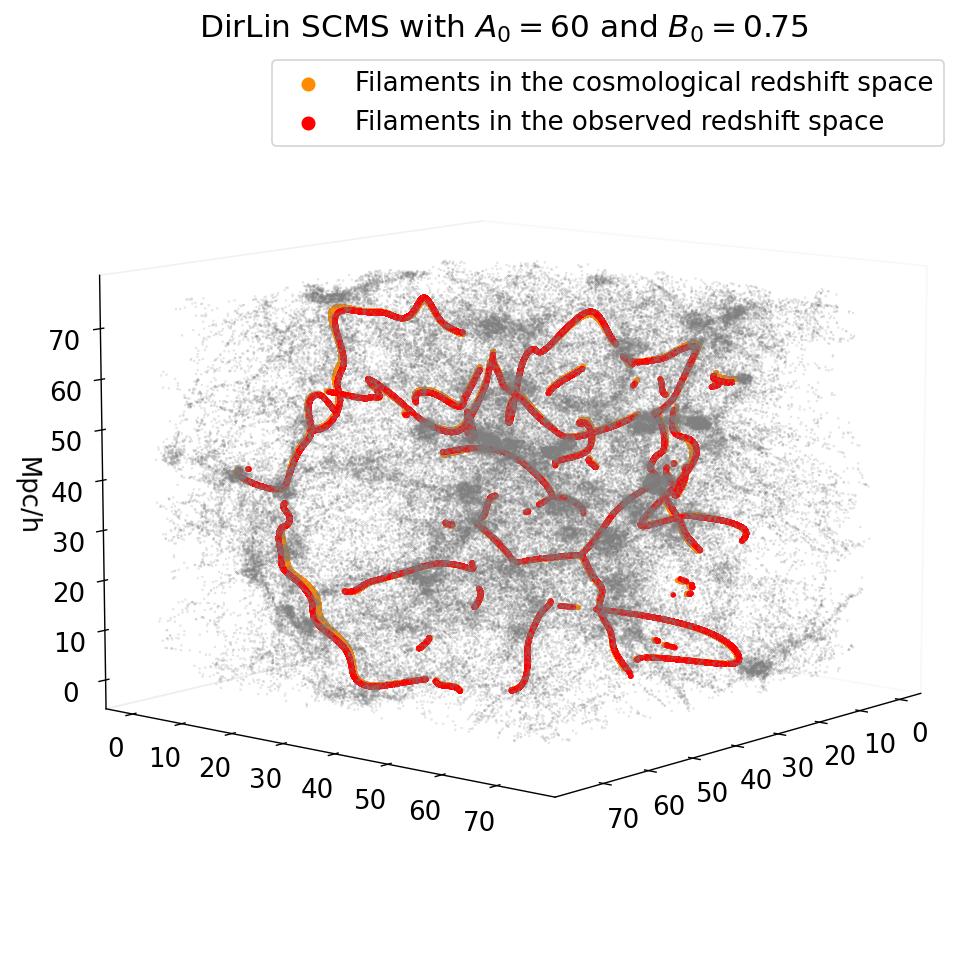}
		\end{subfigure}
		\hfil
		\begin{subfigure}[t]{.135\textwidth}
			\centering
			\includegraphics[width=\linewidth]{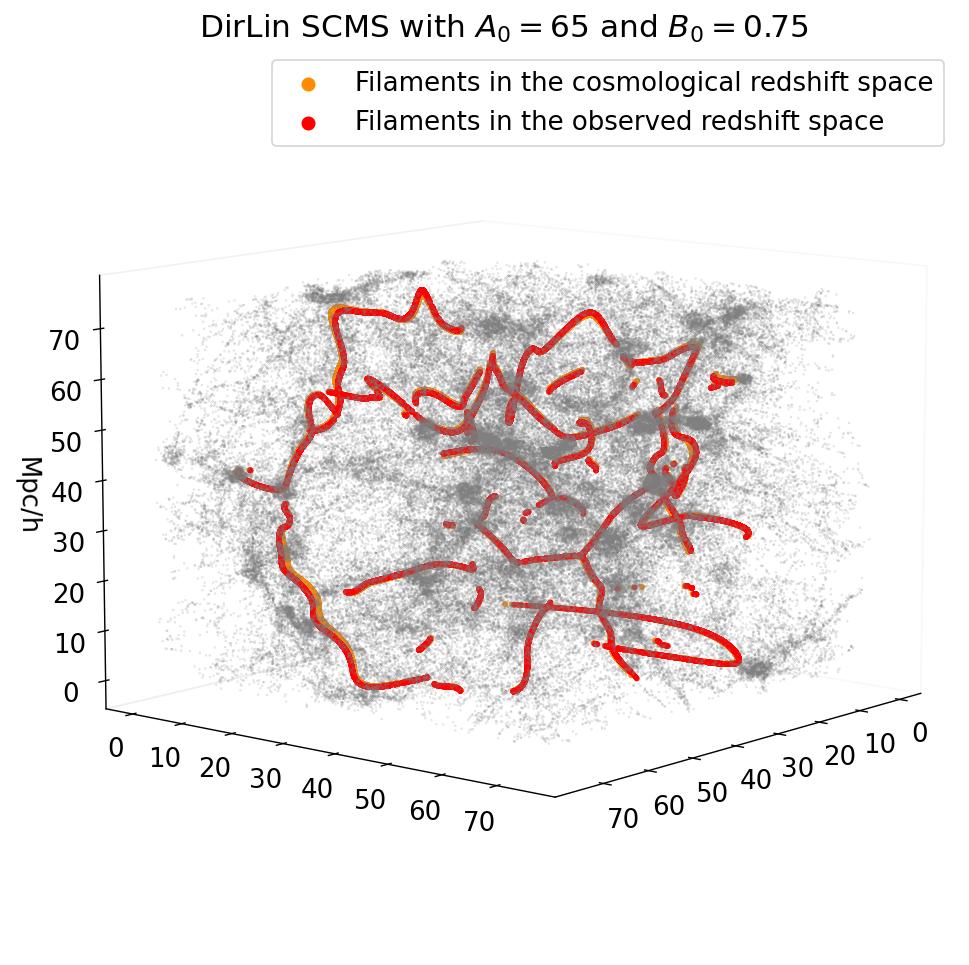}
		\end{subfigure}
		\hfil
		\begin{subfigure}[t]{.135\textwidth}
			\centering
			\includegraphics[width=\linewidth]{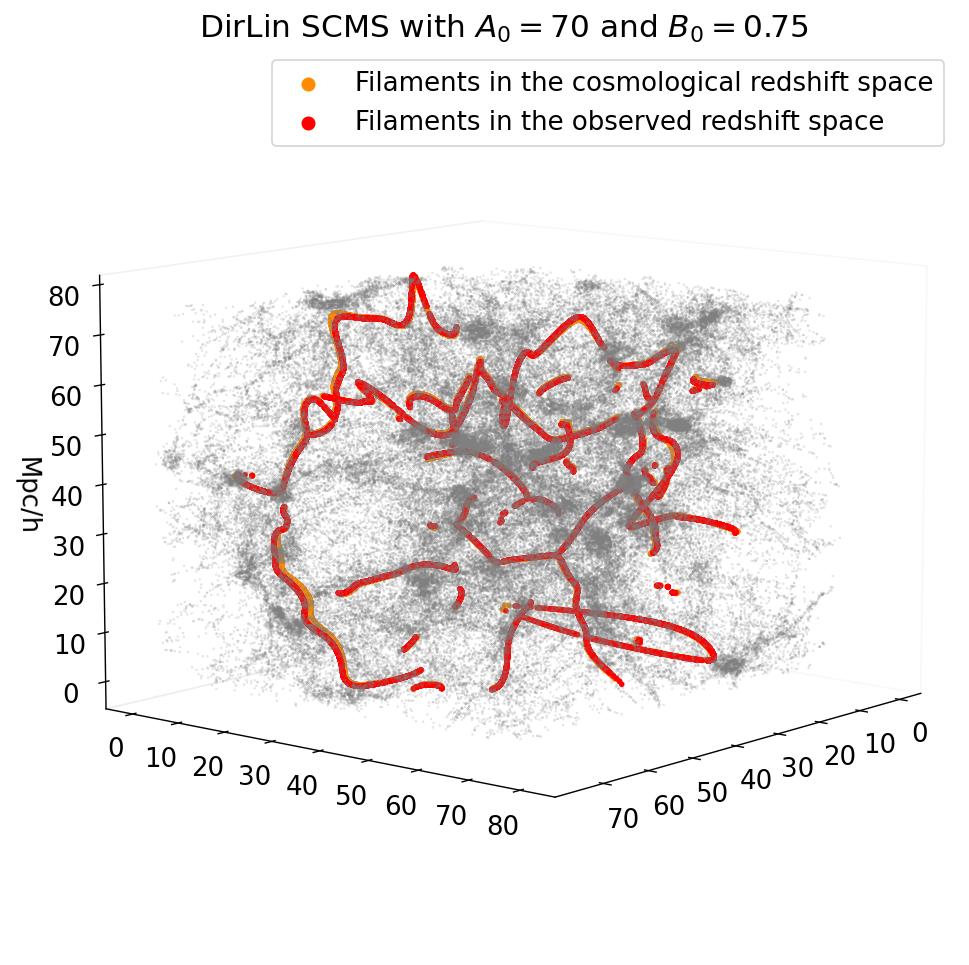}
		\end{subfigure}
		\hfil
		\begin{subfigure}[t]{.135\textwidth}
			\centering
			\includegraphics[width=\linewidth]{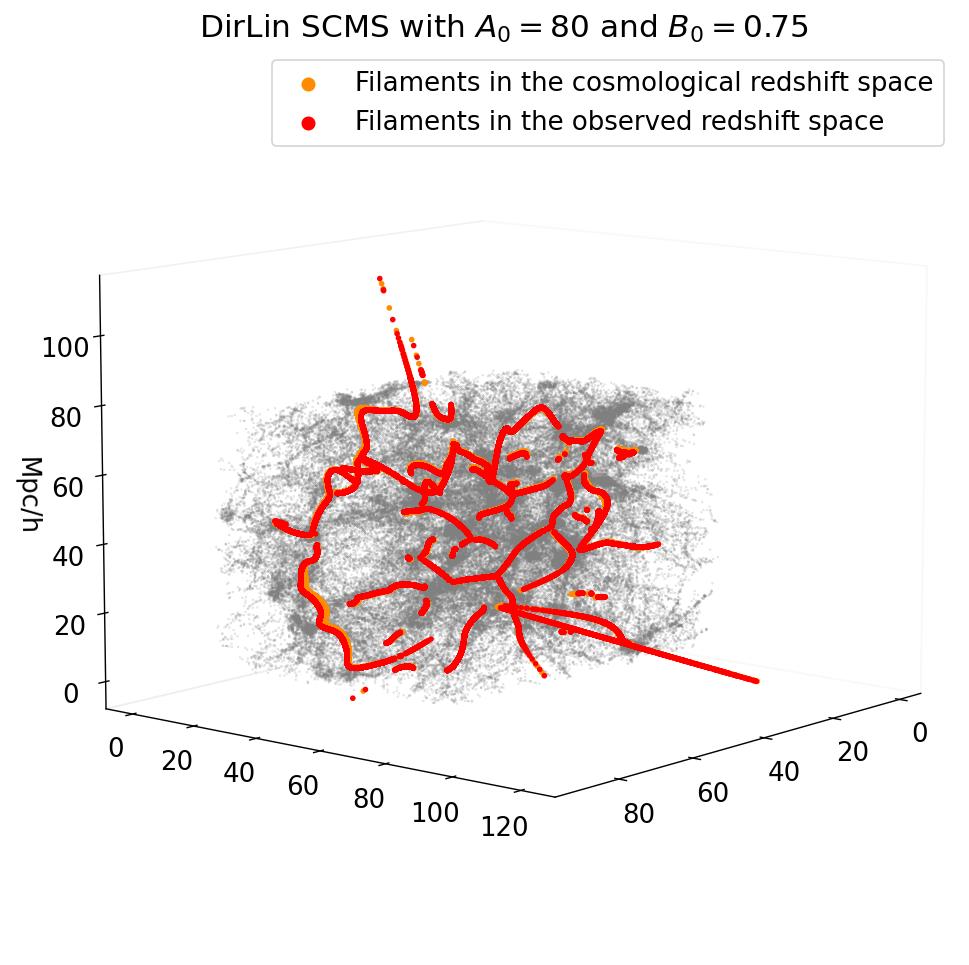}
		\end{subfigure}
		
		\begin{subfigure}[t]{.135\textwidth}
			\centering
			\includegraphics[width=\linewidth]{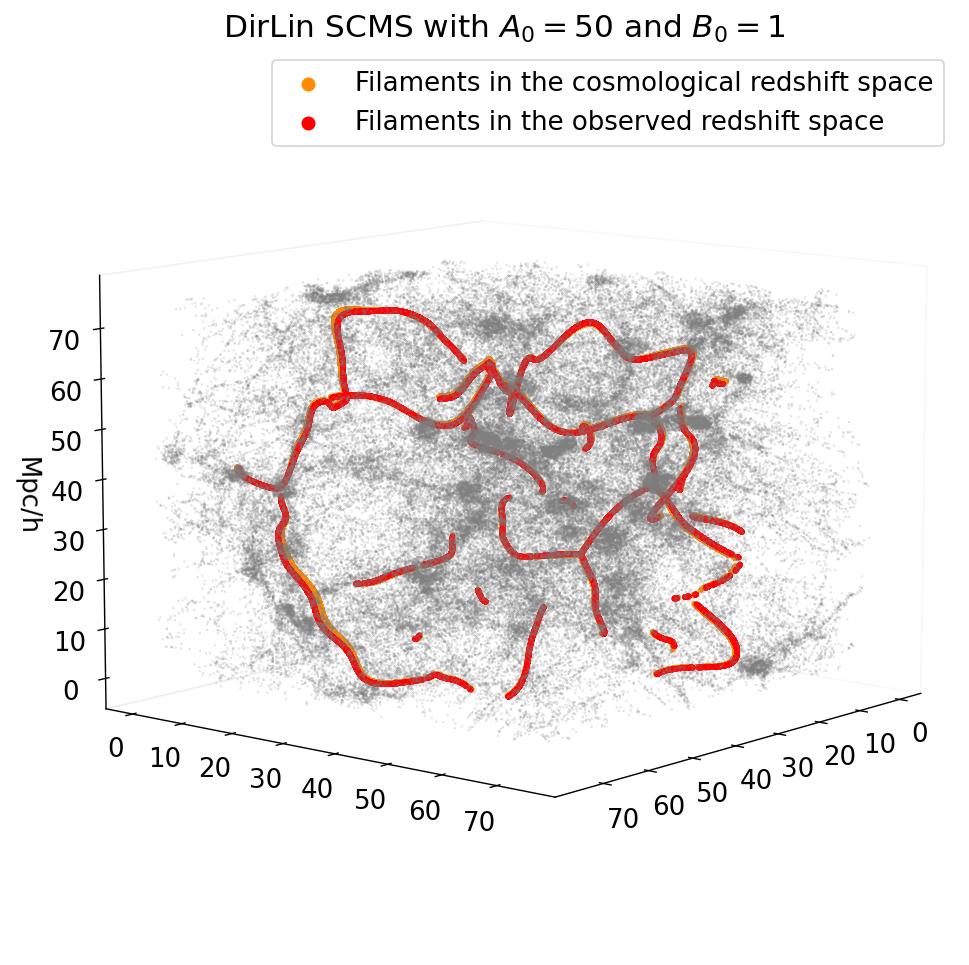}
		\end{subfigure}
		\hfil
		\begin{subfigure}[t]{.135\textwidth}
			\centering
			\includegraphics[width=\linewidth]{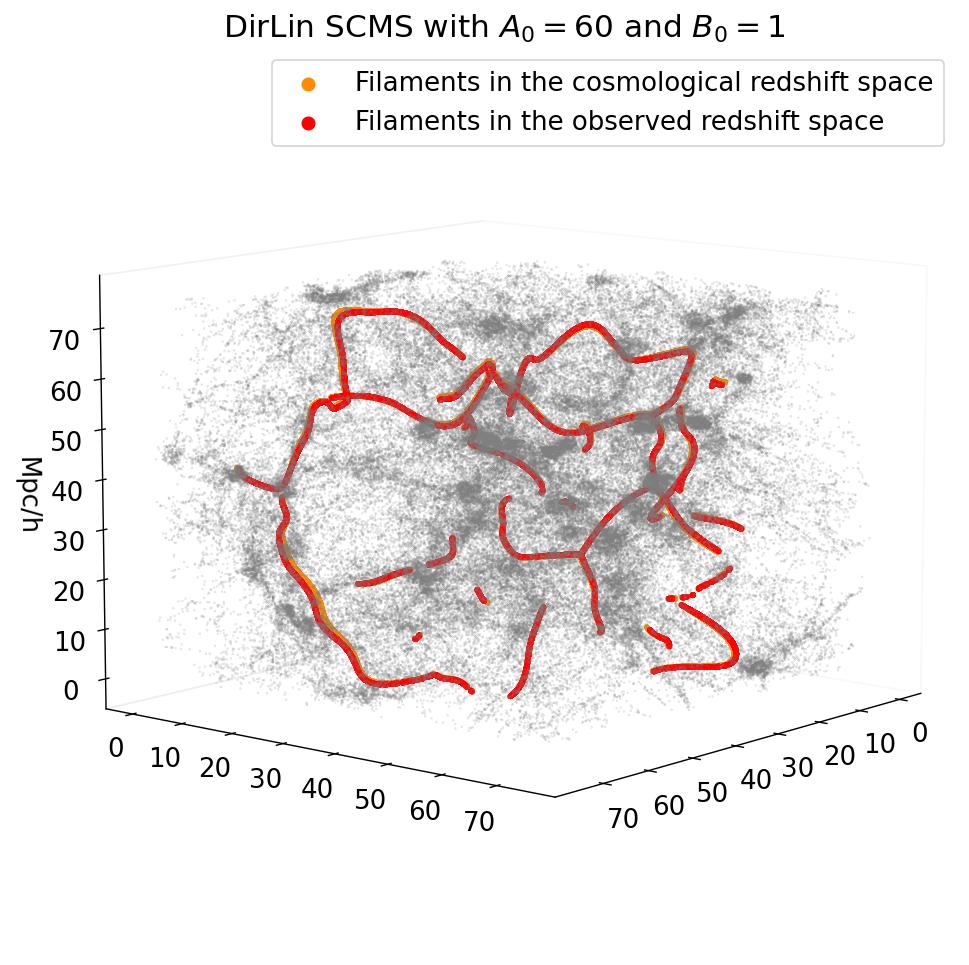}
		\end{subfigure}
		\hfil
		\begin{subfigure}[t]{.135\textwidth}
			\centering
			\includegraphics[width=\linewidth]{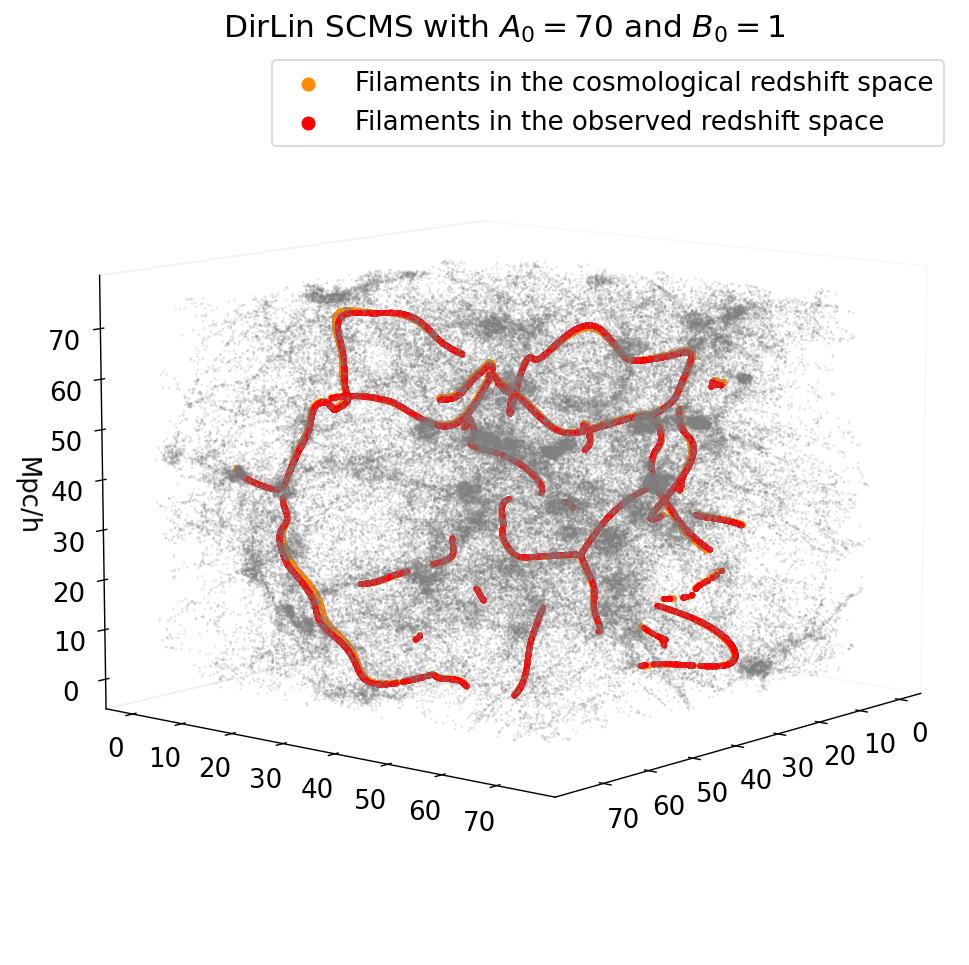}
		\end{subfigure}
		\hfil
		\begin{subfigure}[t]{.135\textwidth}
			\centering
			\includegraphics[width=\linewidth]{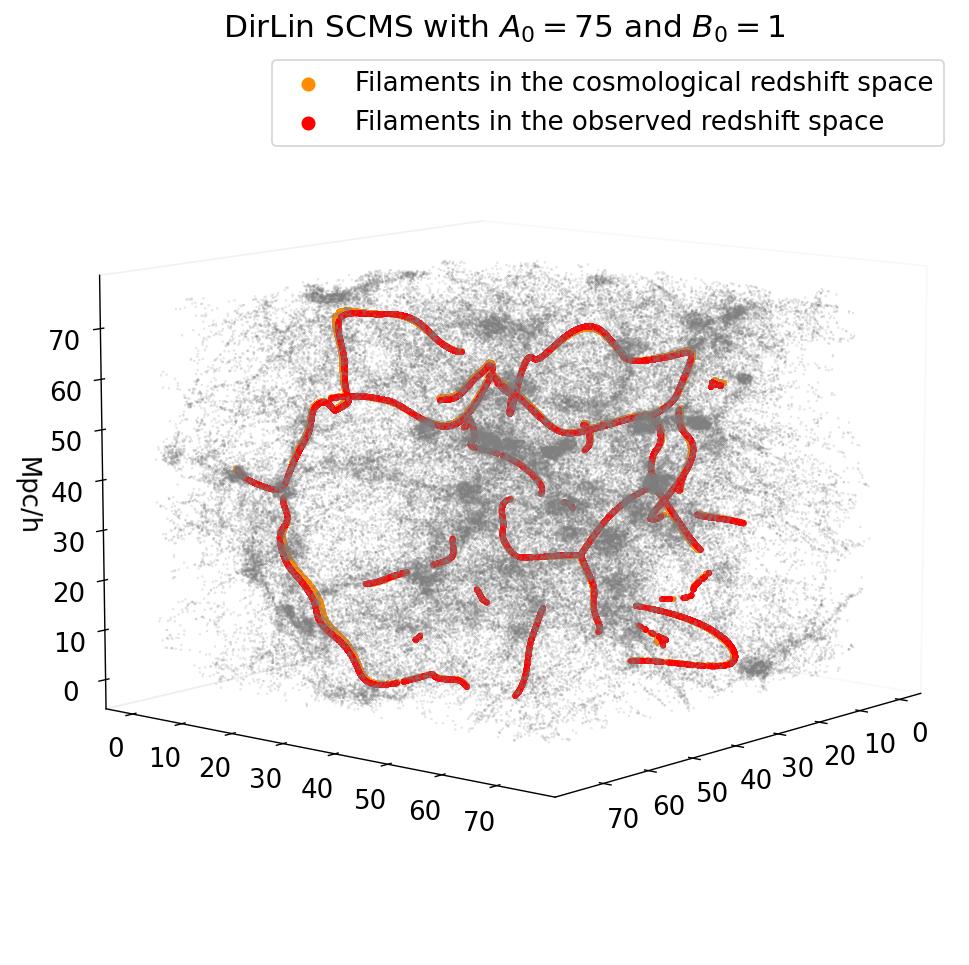}
		\end{subfigure}
		\hfil
		\begin{subfigure}[t]{.135\textwidth}
			\centering
			\includegraphics[width=\linewidth]{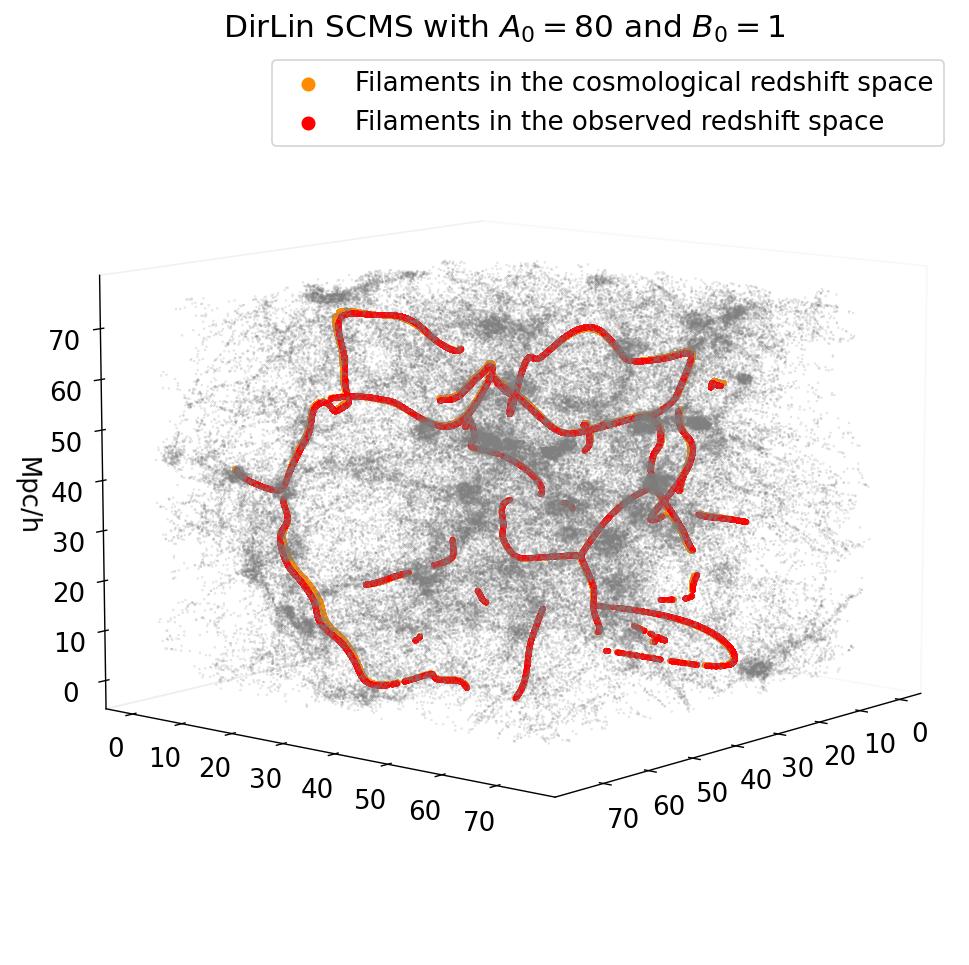}
		\end{subfigure}
		\hfil
		\begin{subfigure}[t]{.135\textwidth}
			\centering
			\includegraphics[width=\linewidth]{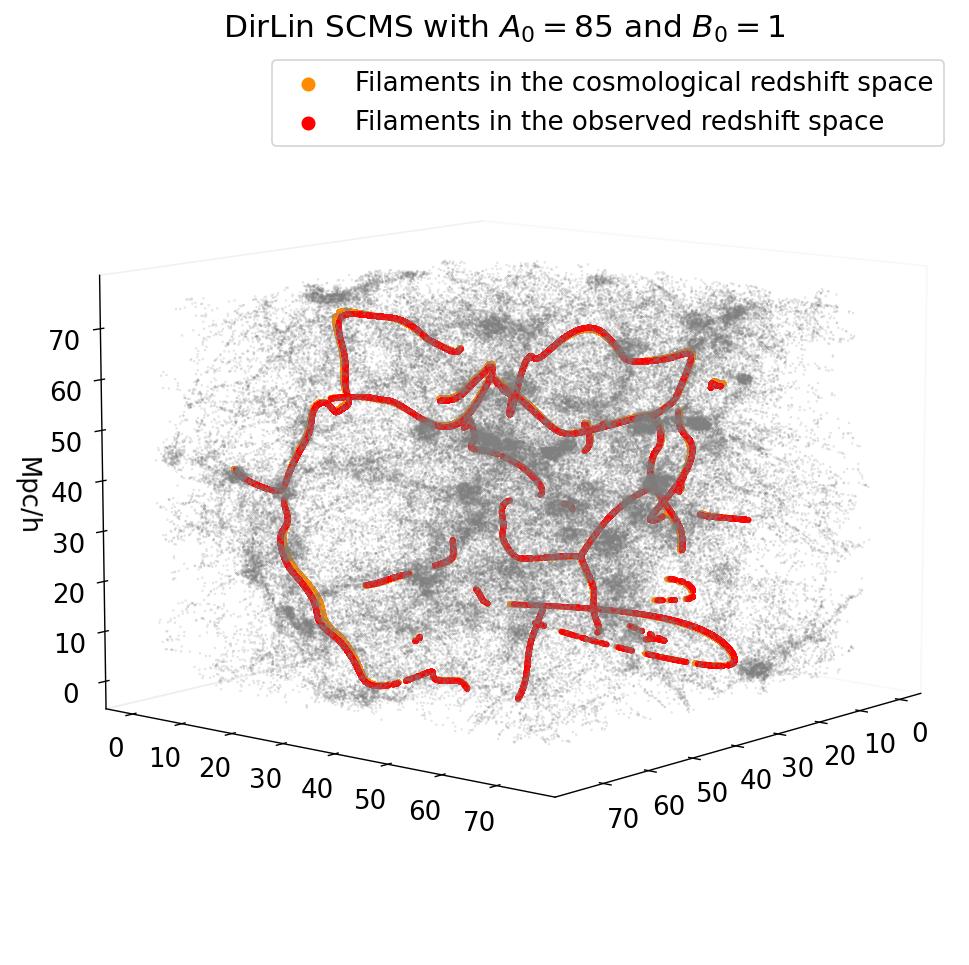}
		\end{subfigure}
		\hfil
		\begin{subfigure}[t]{.135\textwidth}
			\centering
			\includegraphics[width=\linewidth]{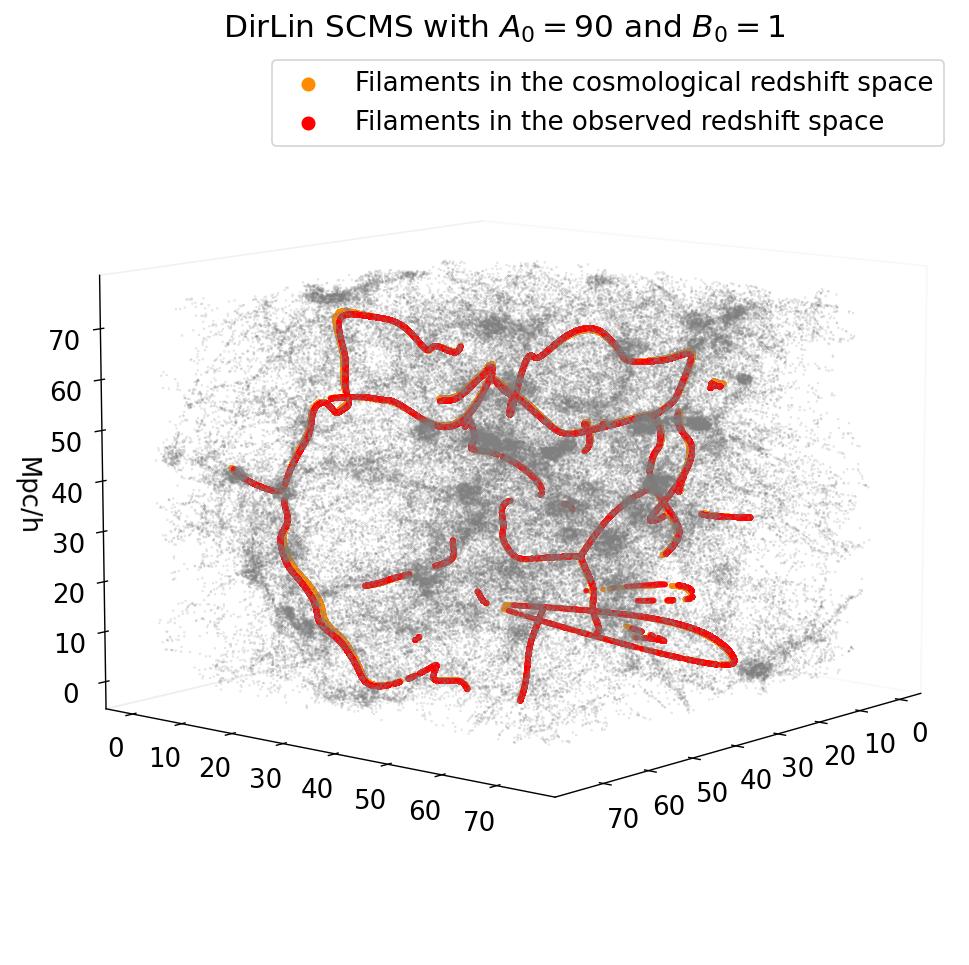}
		\end{subfigure}
		
		\begin{subfigure}[t]{\textwidth}
			\centering
			\includegraphics[width=\linewidth]{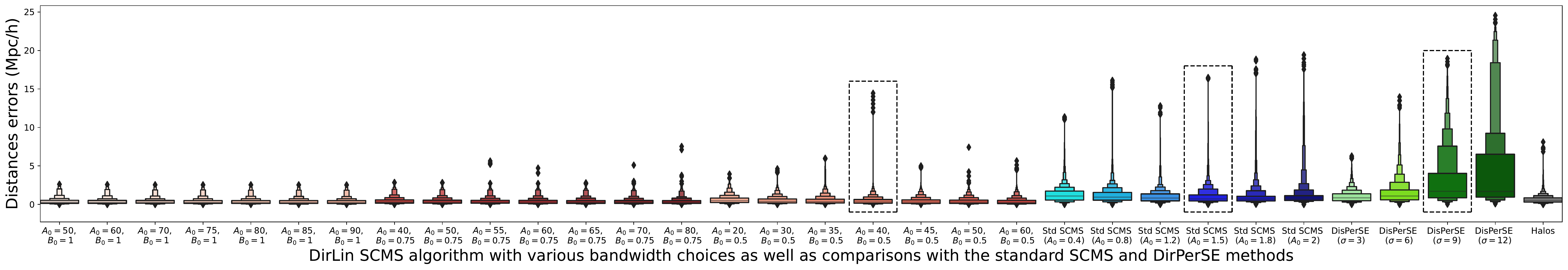}
		\end{subfigure}
		\begin{subfigure}[t]{\textwidth}
			\centering
			\includegraphics[width=\linewidth]{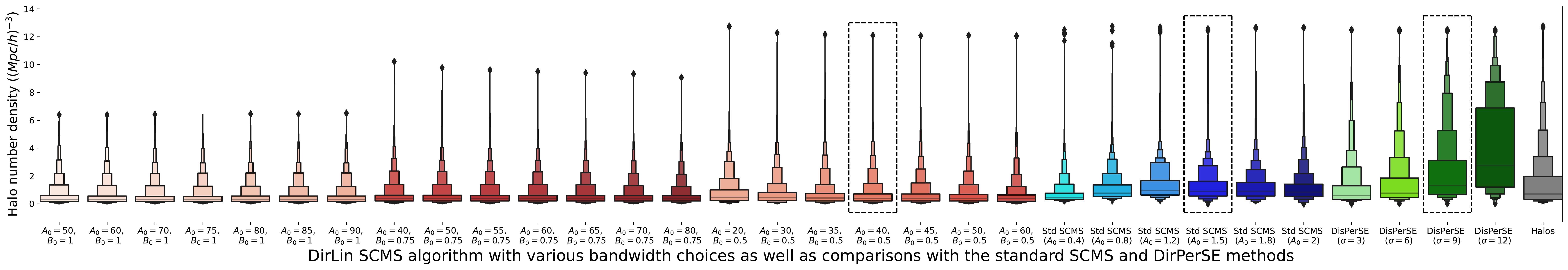}
		\end{subfigure}
		\caption{Comparisons of the filamentary structures detected by our \texttt{DirLinSCMS} algorithm in the observed and cosmological redshift spaces under different choices of the bandwidth parameters on the FoF halos of Illustris-3 at redshift $z=0$. \emph{Fourth Row}: The comparative letter-value plots of distance errors from the filaments detected by each method in the observed redshift space to the ones in the cosmological redshift space. \emph{Fifth row}: The comparative letter-value plots of the halo number densities for the filamentary points detected by each method in the observed redshift space. At the fourth and fifth rows, the letter-value plots within dashed rectangles corresponds to the results in \autoref{fig:Illustris_DirLin_Eu_comp}.}
		\label{fig:Illustris_diff_bw}
	\end{figure*}

	The bandwidth parameter $b_1$ for the directional data part is selected via \eqref{bw_Dir} and $b_2$ for the linear data part is chosen by \eqref{bw_Eu} with $d=1$. We vary their values by changing the scale factors $B_0,A_0$ in \eqref{bw_Dir} and \eqref{bw_Eu}. To ensure that $b_1,b_2$ are properly balanced on the Illustris FoF halo data, we consider 21 combinations of $(A_0,B_0)$ as specified in \autoref{fig:Illustris_diff_bw}, under which the bandwidth parameters $b_1,b_2$ are roughly of the same value when $(A_0,B_0)=(80,1)$, $(60, 0.75)$, and $(40, 0.5)$. As in \autoref{Sec:illustris}, we compute the distance errors from the filament by our \texttt{DirLinSCMS} algorithm in the observed redshift space to the one in the cosmological redshift space under each combination of the directional and linear bandwidth parameters. The 3D plots in \autoref{fig:Illustris_diff_bw} reveal that the filamentary structures by our \texttt{DirLinSCMS} algorithm are sensitive to the choices of its two bandwidth parameters. Their distance errors under the redshift distortions, however, remain steadily small under all these combinations of the directional and linear bandwidths; see the fourth row of \autoref{fig:Illustris_diff_bw}. For better comparison, we also compute the distance error distributions under the redshift distortions for the FoF halo themselves, the standard \texttt{SCMS} algorithm under different choices of its single bandwidth parameter, and \texttt{DisPerSE} with various persistence ratio thresholds. Compared with our \texttt{DirLinSCMS} algorithm, the distance error distributions for the standard \texttt{SCMS} algorithm and \texttt{DisPerSE} change more dramatically under different choices of their tuning parameters and have average values larger than the mean distance errors embraced by the FoF halos themselves.
	
	Another criterion to judge the quality of an estimated filamentary structure is whether the filament traces out the high-density regions of the observed data. We quantify this criterion by calculating the number of FoF halos, $N_h$, within a $3\Mpc/h$ neighborhood sphere centered at each point of the estimated filament in the observed redshift space. The quantity $N_h$ is further normalized by the volume of the neighborhood sphere $V_n=\frac{4\pi}{3} (3\Mpc/h)^3$ to obtain the halo number density $N_H/V_n$ in unit of $(\Mpc/h)^{-3}$ for each filamentary point. The results are presented at the fifth row of \autoref{fig:Illustris_diff_bw}. Compared with the standard SCMS and \texttt{DisPerSE} algorithms, our \texttt{DirLinSCMS} algorithm produced some filamentary structure with relatively low halo number densities. This undesired result for our \texttt{DirLinSCMS} algorithm may be due to the fact that FoF halos of the Illustris simulation are designed in a regular 3D cube and do not inherently embrace a conic geometry. In the future, we are planning to apply our \texttt{DirLinSCMS} algorithm to other survey data, where each object is originally observed in the (RA,DEC,$z$) light cone space, and further access the utility of our \texttt{DirLinSCMS} algorithm.
	
	\subsection{Computation of Cosmic Nodes}
	\label{App:mode_seek}
	
	We know from \autoref{Sec:cosmic_nodes} that our filament model based on density ridges is a natural host of two types of cosmic nodes, local modes and knots. Here, we delineate the practical algorithms to seek out these two types of cosmic nodes on both the celestial sphere $\mathbb{S}^2$ and the 3D light cone $\mathbb{S}^2\times \mathbb{R}$.
	
	$\bullet$ {\bf Identifying local modes}. To obtain the local modes of directional \eqref{DirKDE} or directional-linear KDE \eqref{DLKDE}, the mean shift method \citep{oba2005multi,kafai2010directional,DirMS2020,DirLinProd2021}, the prototype of the \texttt{SCMS} algorithm, is an ideal choice. The mean shift algorithm updates an iterative point $\bm{x}^{(t)} \in \mathbb{S}^2$ or $(\bm{x}^{(t)},z^{(t)})\in \mathbb{S}^2\times \mathbb{R}$ at step $t$ with a fixed-point equation as:
	\begin{align}
	\label{MS_iter}
	&\bullet\,\text{Spherical case}: \bm{x}^{(t+1)} \gets -\frac{\sum_{i=1}^n \bm{X}_i L'\left(\frac{1-\bm{X}_i^T \bm{x}^{(t)}}{b^2} \right)}{\norm{\sum_{i=1}^n \bm{X}_i L'\left(\frac{1-\bm{X}_i^T \bm{x}^{(t)}}{b^2} \right)}}_2,\\
	&\bullet\, \text{Conic case}: \begin{pmatrix}
	\tilde{\bm{x}}^{(t+1)}\\
	z^{(t+1)}
	\end{pmatrix} \gets 
	\begin{pmatrix}
	\frac{\sum\limits_{i=1}^n \bm{X}_i\cdot L'\left(\frac{1-\bm{X}_i^T\bm{x}^{(t)}}{b_1} \right)  K\left(\norm{\frac{z^{(t)}-Z_i}{b_2}}_2^2 \right) }{\sum\limits_{i=1}^n L'\left(\frac{1-\bm{X}_i^T\bm{x}^{(t)}}{b_1} \right) K\left(\norm{\frac{z^{(t)}-Z_i}{b_2}}_2^2 \right)}\\ \frac{\sum\limits_{i=1}^n Z_i \cdot L\left(\frac{1-\bm{X}_i^T\bm{x}^{(t)}}{b_1} \right)   K'\left(\norm{\frac{z^{(t)}-Z_i}{b_2}}_2^2 \right) }{\sum\limits_{i=1}^n L\left(\frac{1-\bm{X}_i^T\bm{x}^{(t)}}{b_1} \right)  K'\left(\norm{\frac{z^{(t)}-Z_i}{b_2}}_2^2 \right)}
	\end{pmatrix},
	\end{align}
	given the Cartesian coordinates $\left\{\bm{X}_1,...,\bm{X}_n\right\}$ of observations on $\mathbb{S}^2$ and their redshift values $\left\{Z_1,...,Z_n\right\}$. The mean shift algorithm is guaranteed to converge to a local mode of the estimated density function for any initial point on $\mathbb{S}^2$ or $\mathbb{S}^2 \times \mathbb{R}$ if the smoothing bandwidth parameters are sufficiently small \citep{MSconsist2016,DMSEM2021}.
	
	$\bullet$ {\bf Identifying knots}. We adopt the metric graph reconstruction algorithm \citep{aanjaneya2011metric,lecci2014statistical} from Appendix A in \cite{Chen2016Catalog} to identify the knots (or intersections) from a given set $R$ of the filament. Specifically, the algorithm classifies each point $\bm{y}\in R$ as ``Knot'' or ``Non-Knot'' as follows:
	\begin{enumerate}
		\item Subset those points on the filaments whose distance to $\bm{y}$ is between $r_{\text{in}}$ and $r_{\text{out}}$.
		\item Partition this subset of points via hierarchical clustering, where the clusters will not be merged if the average distance between these two clusters is above the threshold value $r_{\text{sep}}$.
		\item Count the number of clusters in the last step. If this number is greater or equal to three, we assign label ``Knot'' to $\bm{y}$. Otherwise, $\bm{y}$ is a ``Non-Knot'' point.
	\end{enumerate}
	We set the tuning parameters in the above procedures as:
	\begin{align*}
	&\bullet\, \text{Spherical case}: r_{\text{in}}=\frac{2b}{3}, \quad r_{\text{out}} = 2r_{\text{in}}, \quad r_{\text{sep}} = \frac{r_{\text{in}} + r_{\text{out}}}{2};\\
	&\bullet\, \text{Conic case}: r_{\text{in}}=\frac{2(b_1+b_2)}{3}, \quad r_{\text{out}} = 2r_{\text{in}}, \quad r_{\text{sep}} = \frac{r_{\text{in}} + r_{\text{out}}}{2}.
	\end{align*}
	
	\section{Uncertainty Measure with (Smoothed) Bootstrap Method}
	\label{Sec:UncertMeasure}

	Besides detecting some valid filaments from discrete observations with our extended \texttt{SCMS} algorithms in \texttt{SCONCE}, we reveal the possibility of quantifying the uncertainty level of each filamentary point on the celestial sphere $\mathbb{S}^2$ and 3D light cone $\mathbb{S}^2\times \mathbb{R}$ via the bootstrap technique \citep{jackknife1979,efron1994introduction}. As in \cite{Chen2015methods}, we consider both the nonparametric and smoothed bootstraps \citep{efron1981nonparametric,silverman1987bootstrap} under our extended \texttt{SCMS} algorithms in \autoref{Sec:SCMS_sph} and \autoref{Sec:SCMS_cone}.
	
	\subsection{Outline of the Bootstrap Procedure}
	\label{App:boot_procedure}
	
	Recall that our observational data sample is $\mathcal{D}=\left\{\bm{U}_1,...,\bm{U}_n \right\}$, where $\bm{U}_i = \bm{X}_i \in \mathbb{S}^2$ for spherical data and $\bm{U}_i=(\bm{X}_i,Z_i) \in \mathbb{S}^2\times \mathbb{R}$ for conic data for $i=1,...,n$. Notice that in Appendix~\ref{Sec:GenDirEst}, the stellar property $Y_i$ for each observation $\bm{X}_i\in \mathbb{S}^2$ is an auxiliary variable assisting the filament detection, so the observational data $\bm{U}_i=(\bm{X}_i,Y_i), i=1,...,n$ in that scenario should also be treated as spherical data as well. At a high level, the bootstrap procedure of measuring the uncertainties on the estimated filaments $\hat{R}$ consists of three main steps: 
	\begin{enumerate}
		\item resample $B$ different datasets from the original data sample,
		\item estimate the bootstrapped filaments $\hat{R}^{*(j)}, j=1,...,B$ on each resampled dataset via the \texttt{DirSCMS} algorithm on $\mathbb{S}^2$ or \texttt{DirLinSCMS} algorithm on $\mathbb{S}^2\times \mathbb{R}$, and
		\item measure the uncertainty of each point $\bm{u}\in \hat{R}$ as:
		\begin{eqnarray}
		\label{uncertain_measure}
		\rho(\bm{u}) = \sqrt{\frac{1}{B} \sum_{j=1}^B d_g\left(\bm{u},\hat{R}^{*(j)}\right)^2},
		\end{eqnarray}
		where $d_g\left(\bm{u},\hat{R}^{*(j)}\right)=\min\left\{d_g\left(\bm{u},\bm{u}^*\right): \bm{u}^*\in \hat{R}^{*(j)} \right\}$ for $j=1,...,B$.
	\end{enumerate}
	This framework of measuring the filamentary uncertainties is adopted from \cite{Chen2015methods,novel_cat2021}. 
	The only difference is that we use the geodesic distance metric to calculate the distances from each point on the estimated filament $\hat{R}$ with the original data to the bootstrapped filaments $\hat{R}^{*(j)}, j=1,...,B$ as:
	\begin{equation}
	\label{geo_dist}
	d_g(\bm{u}_1,\bm{u}_2) =
	\begin{cases}
	\arccos(\bm{u}_1^T\bm{u}_2) & \text{ if } \quad \bm{u}_i\in \mathbb{S}^2,\\
	\norm{z_1\bm{x}_1 - z_2\bm{x}_2}_2 & \text{ if } \quad \bm{u}_i=(\bm{x}_i,z_i) \in \mathbb{S}^2\times \mathbb{R},
	\end{cases}
	\end{equation} 
	for $i=1,2$ so as to take into account the spherical or conic geometry.
	
	
	\subsection{Two Types of the Bootstrap Scheme}
	\label{App:two_bootstrap}
	
	With regards to the bootstrap resampling mechanism, we consider two different versions available in the literature:
	\begin{itemize}
		\item {\bf Nonparametric Bootstrap}: It was the very first bootstrap version initially proposed by \cite{jackknife1979}, which has its own simplicity and is widely used in all scientific fields. Each bootstrap sample $\left\{\bm{U}_1^{*(j)},...,\bm{U}_n^{*(j)} \right\}$ with $j=1,...,B$ is sampled \emph{with replacement} from the original data sample $\mathcal{D}=\left\{\bm{U}_1,...,\bm{U}_n \right\}$.
		
		\item {\bf Smoothed Bootstrap:} It is a variant of the nonparametric bootstrap in which each bootstrap sample is generated not directly from the original data sample $\mathcal{D}$ but from the estimated density function (c.f., Equations \eqref{DirKDE}, \eqref{GenDirKDE}, or \eqref{DLKDE}) . Given the original data sample $\mathcal{D}=\left\{\bm{U}_1,...,\bm{U}_n \right\}$, it repeats the following two-step process $n$ times ($i=1,...,n$) in order to obtain one smoothed bootstrap data sample $\left\{\bm{U}_1^{*(j)},...,\bm{U}_n^{*(j)} \right\}$:
		\begin{enumerate}
			\item We first sample a candidate point $\tilde{\bm{U}}$ uniformly from $\mathcal{D}$.
			\item Depending on the data type, we draw $\bm{U}_i^{*(j)}$ as follows:
			\begin{itemize}
				\item If $\tilde{\bm{U}}=\tilde{\bm{X}} \in \mathbb{S}^2$, sample $\bm{X}^*$ from the distribution with density $\tilde{f}(\bm{x}) = C_L(b) \cdot L\left(\frac{1-\bm{x}^T\tilde{\bm{X}}}{b^2} \right)$ and take $\bm{U}_i^{*(j)} = \bm{X}^*$.
				
				\item If $\tilde{\bm{U}} =(\tilde{\bm{X}},\tilde{Z}) \in \mathbb{S}^2\times \mathbb{R}$, sample $\bm{X}^*$ from the distribution with density $\tilde{f}_1(\bm{x})=C_L(b_1) \cdot L\left(\frac{1-\bm{x}^T\tilde{\bm{X}}}{b_1^2} \right)$ and $Z^*$ from the distribution with density $\tilde{f}_2(z)=\frac{1}{b_2} K\left(\norm{\frac{z- \tilde{Z}}{b_2}}_2^2 \right)$. Then, take $\bm{U}_i^{*(j)} = (\bm{X}^*,Z^*)$.
			\end{itemize}
		\end{enumerate}
		Sampling data points from the above densities (or equivalently, some arbitrary kernel functions) could be difficult in practice. However, the problem can be resolved with our applications of the von Mises kernel $L(r)=e^{-r}$ and Gaussian kernel $K(r)=\frac{1}{\sqrt{2\pi}} e^{-\frac{r}{2}}$. We discuss how to analytically sample $\bm{X}^*$ from $\tilde{f}_1(\bm{x})$ on $\mathbb{S}^2$ in Appendix~\ref{Sec:vMF_sam}, and sampling $Z^*$ from $\tilde{f}_2(z)$ is the same as adding Gaussian noise with mean $0$ and variance $\frac{1}{b_2^2}$ to $\tilde{Z}$.
	\end{itemize}

	Compared to the smoothed bootstrap, the nonparametric bootstrap is more comprehensible and independent of the density function estimator. The smoothed bootstrap, on the other hand, requires a sampling scheme from the directional kernel function $L$ in \eqref{DirKDE} (as well as the linear kernel function $K$ in \eqref{DLKDE}) a priori, which may be computationally intensive for some kernel functions. Although one can always design a rejection sampling scheme \citep{flury1990acceptance} to generate data points from any kernel function, the rejection sampling may be quite inefficient depending on the proposal density. Fortunately, we do have an analytic sampling approach for the von Mises kernel on $\mathbb{S}^2$ and $\mathbb{S}^2\times \mathbb{R}$ in Appendix~\ref{Sec:vMF_sam}. Despite the complexity in its design, the smoothed bootstrap resolves the issue of only having (repeated) observations from the original data as in the nonparametric bootstrap and can reduce the mean squared error in the estimation of uncertainty measures under a proper choice of the smoothing bandwidth parameter \citep{silverman1987bootstrap}.
	
	\subsection{Analytic Sampling from the von Mises Kernel on $\mathbb{S}^2$}
	\label{Sec:vMF_sam}
	
	One interesting fact about the von Mises kernel $L(r)=e^{-r}$ is that the resulting directional KDE \eqref{DirKDE} becomes a $n$-mixture of von-Mises Fisher (vMF) distribution on $\mathbb{S}^2$. The density function of a vMF distribution is defined as:
	\begin{eqnarray}
	\label{vMF_density}
	f_{\text{vMF}}(\bm{x};\bm{\mu},\kappa) = C_q(\kappa) \cdot \exp\left(\kappa \bm{\mu}^T\bm{x} \right)
	\end{eqnarray}
	with $C_q(\kappa) = \frac{\kappa^{\frac{q-1}{2}}}{(2\pi)^{\frac{q+1}{2}} \mathcal{I}_{\frac{q-1}{2}}(\kappa)}$. Here, $\bm{\mu} \in \mathbb{S}^2$ is the directional mean of the vMF distribution, $\kappa \geq 0$ is the concentration parameter, and 
	$$\mathcal{I}_{\alpha}(\kappa) = \frac{\left(\frac{\kappa}{2} \right)^{\alpha}}{\pi^{\frac{1}{2}} \Gamma(\alpha + \frac{1}{2})} \int_{-1}^1 (1-t^2)^{\alpha -\frac{1}{2}} \cdot e^{\kappa t} dt$$
	is the modified Bessel function of the first kind of order $\alpha$, where $\Gamma(\cdot)$ is the Gamma function. We denote this distribution by $\text{vMF}(\bm{\mu}, \kappa)$. Under this notation, the directional KDE \eqref{DirKDE} under the von Mises kernel $L(r)=e^{-r}$ can be expressed as:
	$$f(\bm{x}) = \frac{1}{n}\sum_{i=1}^n \text{vMF}\left(\bm{X}_i, \frac{1}{b^2}\right),$$
	and thus, sampling a data point from the density $\tilde{f}(\bm{x})=C_L(b)\cdot L\left(\frac{1-\bm{x}^T\tilde{\bm{X}}}{b^2}\right)$ on $\mathbb{S}^2$ 
	in the smoothed bootstrap of Appendix~\ref{App:two_bootstrap} is equivalent to generating a random observation from $\text{vMF}\left(\tilde{\bm{X}}, \frac{1}{b^2}\right)$ 

	Now, we illuminate an analytic approach to randomly sampling data points from a general $\text{vMF}(\bm{\mu},\kappa)$ on $\mathbb{S}^2$, which is far more efficient than the naive rejection sampling. We first consider generating $\bm{X}$ from $\text{vMF}(\bm{\mu}_0, \kappa)$ with $\bm{\mu}_0=(0,0,1)^T \in \mathbb{S}^2\subset \mathbb{R}^3$. Based on the results in \cite{Gen_dist_sph1984,Sim_vMF1994,Kurz2015}, we know that $\bm{X}$ follows $\text{vMF}(\bm{\mu}_0, \kappa)$ if and only if
	$$\bm{X} = \left(\sqrt{1-W^2}\cdot \bm{U}, W \right)^T,$$
	where $\bm{U}$ is uniformly distributed on $\mathbb{S}^1=\left\{\bm{x}\in \mathbb{R}^2: \norm{\bm{x}}_2=1\right\}$ and $W \in [-1,1]$ has its probability density function as:
	\begin{eqnarray}
	\label{W_density}
	f_W(w) = \frac{\kappa}{e^{\kappa} -e^{-\kappa}} \cdot \exp(\kappa w) \quad \text{ when } \quad w \in [-1,1],
	\end{eqnarray}
	whose corresponding cumulative distribution function (CDF) is
	$$F_W(t) = \frac{e^{\kappa t} - e^{-\kappa}}{e^{\kappa}-e^{-\kappa}} \quad \text{ when } \quad t\in [-1,1].$$
	Some algebra show that the inverse of the above CDF is given by
	\begin{eqnarray}
	\label{W_CDF_inv}
	F_W^{-1}(y) = \frac{1}{\kappa} \log\left[y(e^{\kappa} - e^{-\kappa}) + e^{-\kappa} \right].
	\end{eqnarray}
	To avoid numerical overflow for large values of $\kappa$, one can replace \eqref{W_CDF_inv} with the following equivalent expression
	\begin{eqnarray}
	\label{W_CDF_inv2}
	F_W^{-1}(y) = 1+ \frac{1}{\kappa} \log\left[y + (1-y) e^{-2\kappa} \right].
	\end{eqnarray}
	The random variable $F_W^{-1}(Y)$ with $Y\sim \text{Uniform}[0,1]$ will have a density function \eqref{W_density} and CDF \eqref{W_CDF_inv} or \eqref{W_CDF_inv2}. The sampling of $\bm{U}$ that is uniformly distributed on $\mathbb{S}^1$ is relatively easy. For instance, one can sample $U\sim \text{Uniform}[0,1]$ and take $\bm{U}= (\cos 2\pi U, \sin 2\pi U)$. Thus, the analytic strategy for sampling $\bm{X} \sim \text{vMF}(\bm{\mu}_0, \kappa)$ on $\mathbb{S}^2$ is now clear.
	
	\begin{figure*}
		\captionsetup[subfigure]{justification=centering}
		\centering
		\begin{subfigure}[t]{.49\textwidth}
			\centering
			\includegraphics[width=\linewidth]{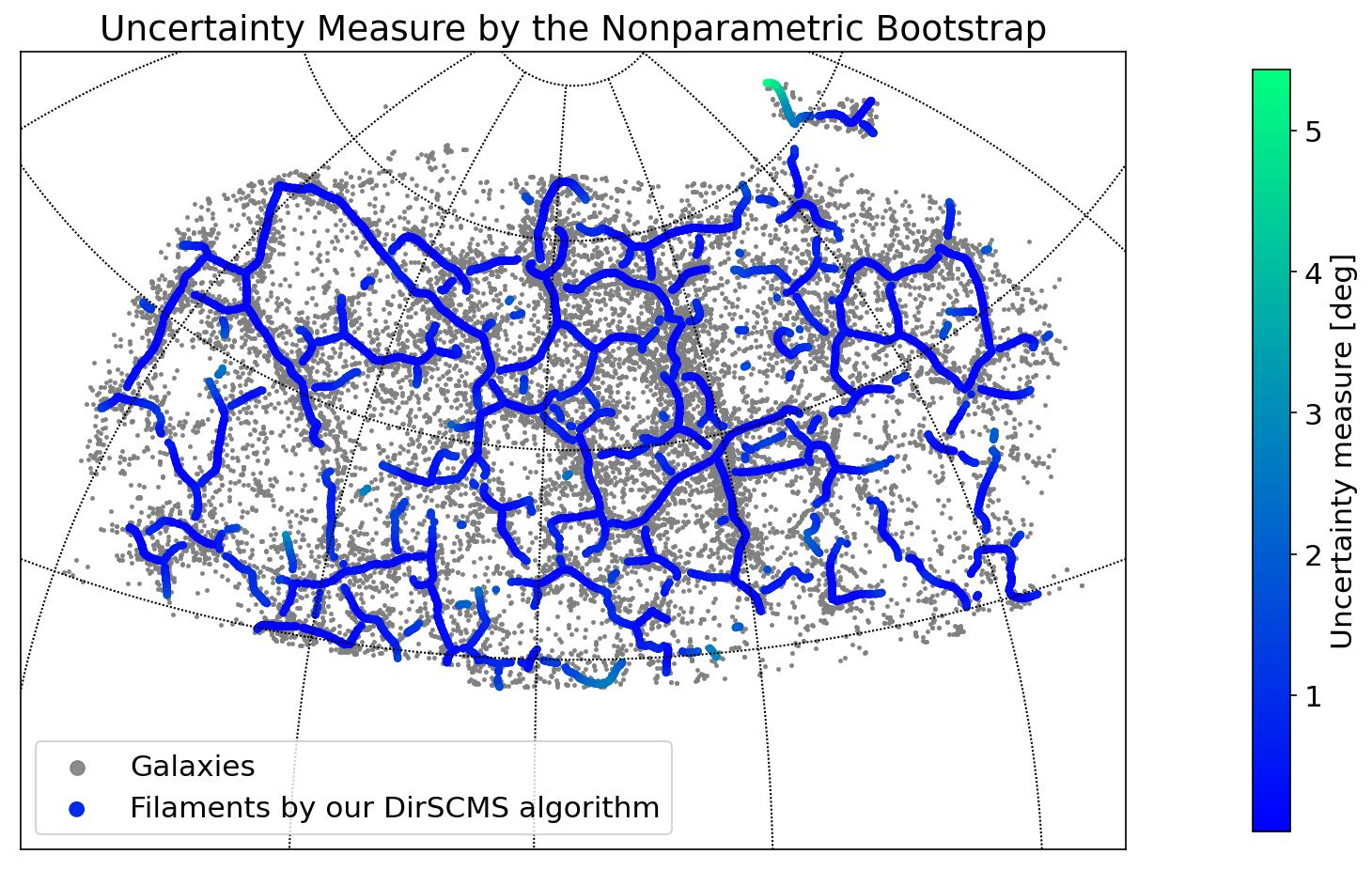}
		\end{subfigure}
		\hfil
		\begin{subfigure}[t]{.49\textwidth}
			\centering
			\includegraphics[width=\linewidth]{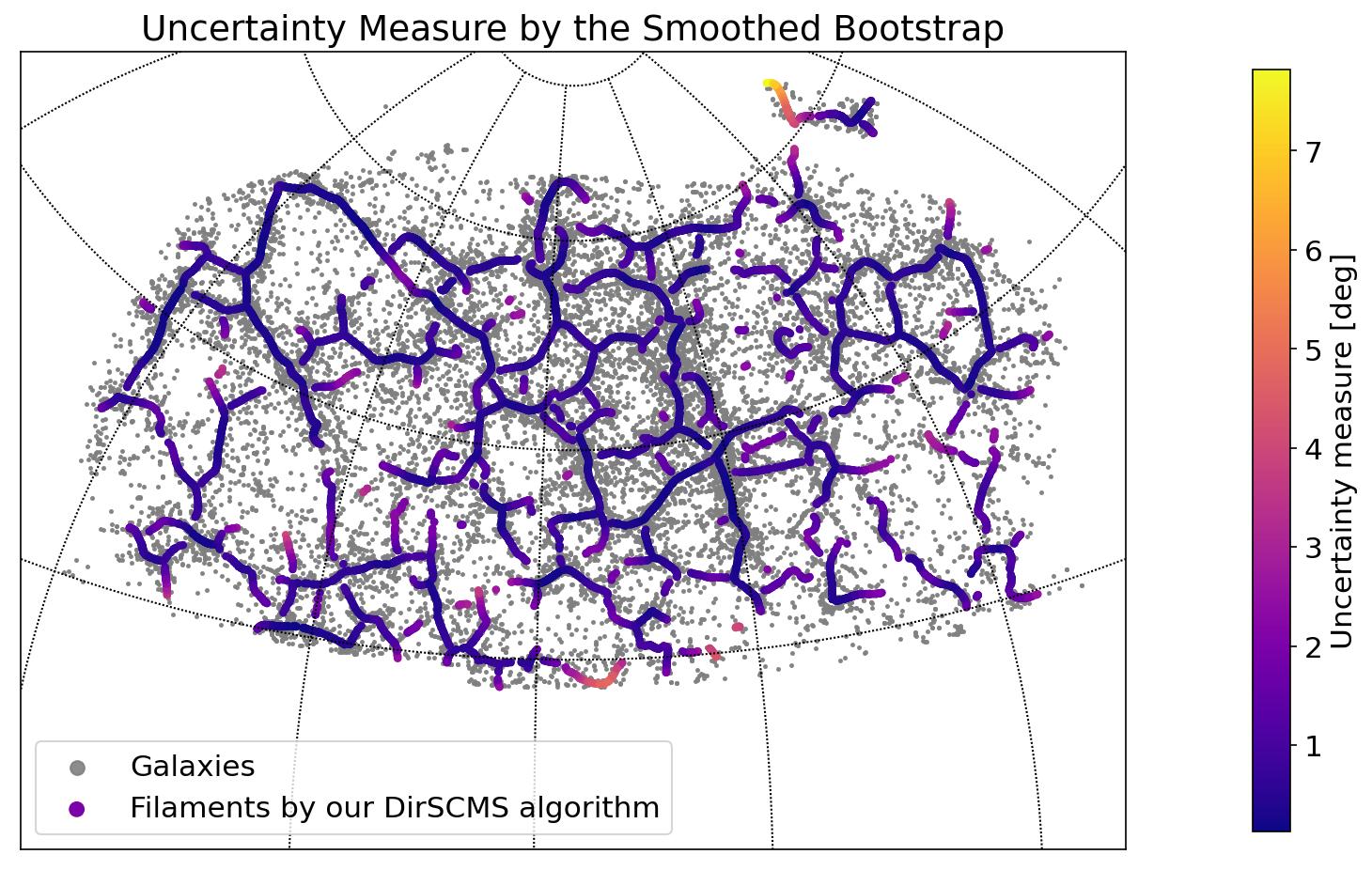}
		\end{subfigure}
		\begin{subfigure}[t]{.47\textwidth}
			\centering
			\includegraphics[width=\linewidth]{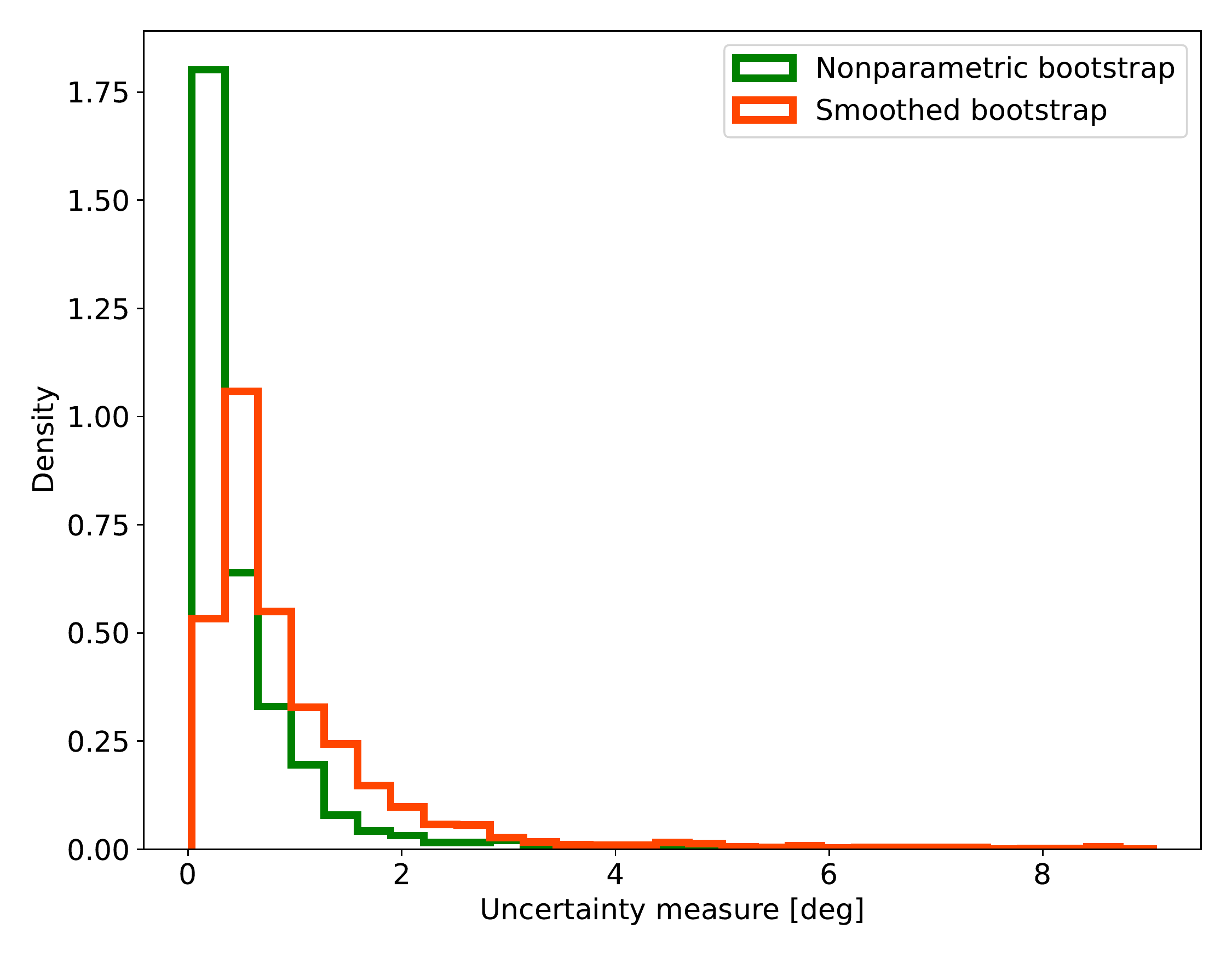}
		\end{subfigure}
		\hfil
		\begin{subfigure}[t]{.47\textwidth}
			\centering
			\includegraphics[width=\linewidth]{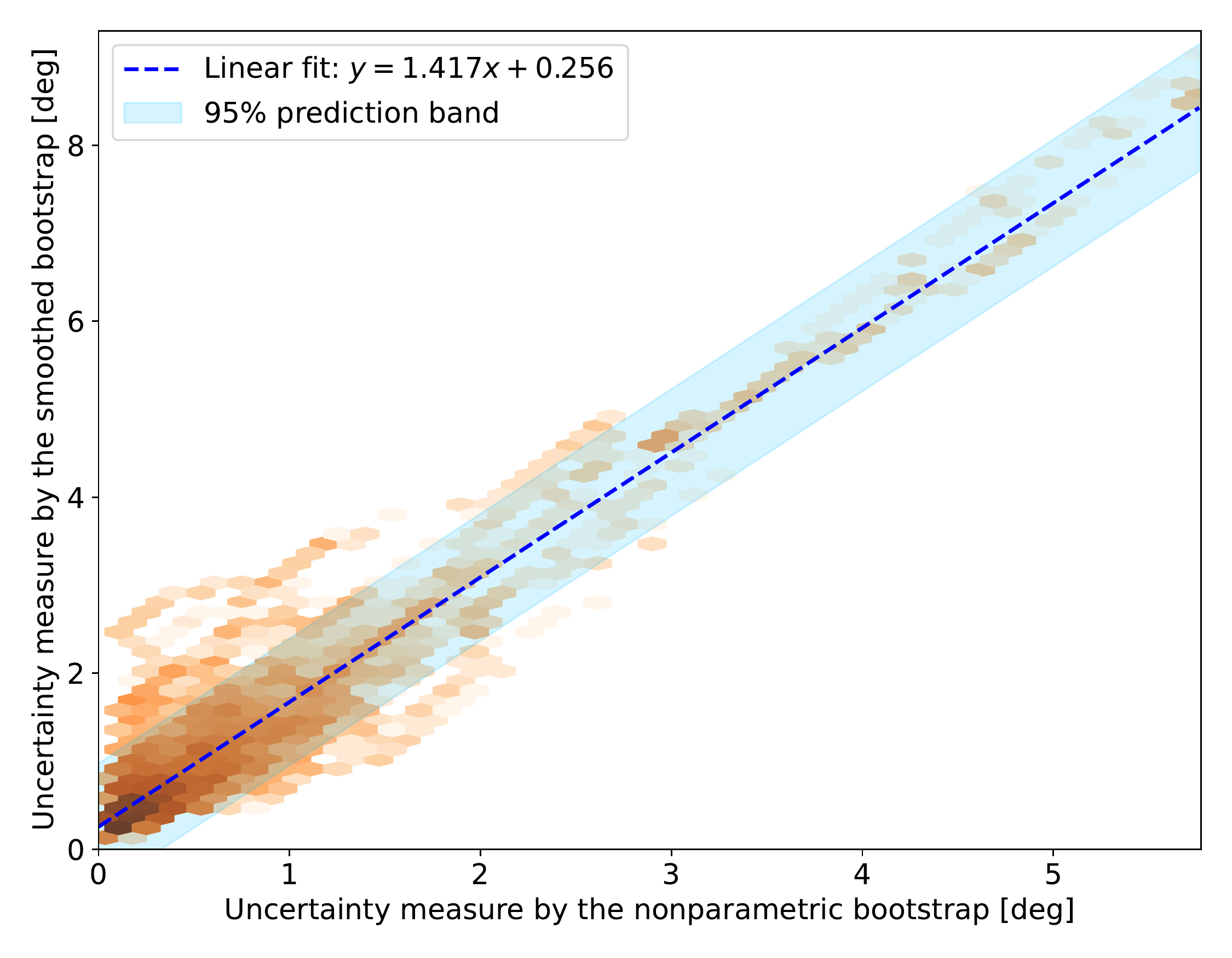}
		\end{subfigure}
		\caption{Comparisons of the uncertainty measures on the estimated filaments by the nonparametric and smoothed bootstraps with our \texttt{DirSCMS} algorithm on a thin redshift slice ($0.06\leq z < 0.065$) of the SDSS-IV galactic data. \emph{Upper Row}: The cosmic filaments detected by our \texttt{DirSCMS} algorithm with their uncertainty measures by the two bootstrap methods. \emph{Bottom Left}: Histograms of the uncertainty measures for the filaments by the two bootstrap methods. \emph{Bottom Right}: The 2D histogram of the uncertainty measures for filamentary points by the two bootstrap methods, where the blue dashed line is a linear fit on the uncertainty measure data.}
		\label{fig:boot_comp}
	\end{figure*}
	
	To obtain data points from $\text{vMF}(\bm{\mu}, \kappa)$ with an arbitrary mean direction $\bm{\mu} \in \mathbb{S}^2$, one can apply a rotation matrix $R_{\text{rot}} \in \mathbb{R}^{3\times 3}$ to the data sample from $\text{vMF}(\bm{\mu}_0, \kappa)$. One remarkable fact related to the expression of $R_{\text{rot}}$ is that, in order to move a normalized vector $\bm{\mu}_0 \in \mathbb{S}^2$ to another normalized vector $\bm{\mu} \in \mathbb{S}^2$, one only needs to rotate $\bm{\mu}_0$ with respect to the axis $\bm{k} = \frac{\bm{\mu}_0 + \bm{\mu}}{2}$ by the angle $\pi$. With the matrix form of the Rodrigues's rotation formula \citep{maritz2021rotations}, one gets the beautiful formula for the rotation matrix as:
	\begin{eqnarray}
	\label{Rodrigues_rot}
	R_{\text{rot}} = 2\frac{(\bm{\mu}_0 + \bm{\mu})(\bm{\mu}_0 + \bm{\mu})^T}{(\bm{\mu}_0 + \bm{\mu})^T (\bm{\mu}_0 + \bm{\mu})} - \bm{I}_3.
	\end{eqnarray}
	Note that this rotation matrix will indeed rotate the data point with respect to $\bm{\mu}_0$ by the angle $\pi$ when $\bm{\mu}_0=\bm{\mu}$.\\
	
	In short, the analytic algorithm for randomly sampling a data point $\bm{X}_{\text{vMF}} \sim \text{vMF}(\bm{\mu}, \kappa)$ with $\bm{\mu}\in \mathbb{S}^2$ is given by
	\begin{enumerate}
		\item Sample two independent data points $Y$ and $U$ from $\text{Uniform}[0,1]$.
		\item Compute $W=1+ \frac{1}{\kappa} \log\left[Y + (1-Y) e^{-2\kappa} \right]$ and $\bm{U} = (\cos 2\pi U, \sin 2\pi U)$.
		\item Obtain $\bm{X} = \left(\sqrt{1-W^2}\bm{U}, W \right)^T$ and rotate it as $\bm{X}_{\text{vMF}} = R_{\text{rot}} \bm{X}$.
	\end{enumerate}
	
	This efficient sampling scheme will be applied to our smoothed bootstrap procedure in Appendix~\ref{App:two_bootstrap} when we deal with astronomical observations on $\mathbb{S}^2$ or $\mathbb{S}^2\times \mathbb{R}$.
	
	\subsection{Comparison between the nonparametric and smoothed bootstraps on SDSS data}
	\label{App:boot_comp}
	
	We conduct a quantitative comparison between the nonparametric and smoothed bootstraps on $\mathbb{S}^2$ by applying them to the SDSS-IV galactic data in \autoref{Sec:SDSS_data} within the thin redshift slice $0.06 \leq z< 0.065$ and the North Galactic Cap ($100^{\circ}< \text{RA}<270^{\circ}, -5^{\circ}< \text{DEC}< 70^{\circ}$). The total number of galaxies is 26,919 in this spherical slice, including those with missing stellar masses by the FIREFLY model. Both the nonparametric and smoothed bootstrap procedures follow from the procedures in Appendix~\ref{App:boot_procedure} with our \texttt{DirSCMS} algorithm and bootstrapping time $B=100$.  
	
	The results are shown in \autoref{fig:boot_comp}, where, on average, the smoothed bootstrap produces a higher value of the uncertainty measure for each point on the filament estimated by our \texttt{DirSCMS} algorithm than the nonparametric bootstrap. Nevertheless, the two bootstrap methods do not exhibit too much difference in quantifying the uncertainties on the estimated filament as a whole. Those points on the filament with high uncertainties by the nonparametric bootstrap are also less robust under the smoothed bootstrap. Indeed, as revealed by the bottom right panel of \autoref{fig:boot_comp}, the uncertainty measures yielded by the nonparametric and smoothed bootstraps are linearly correlated with the Pearson's correlation coefficient as 0.936. Hence, in the cosmic filament detection process, one can freely choose to use either the nonparametric or smoothed bootstrap methods to measure the uncertainties of the estimated filaments, as long as the method is consistent along the entire filament detection pipeline.
	


	\bsp	
	\label{lastpage}
\end{document}